\documentstyle[bezier,pictex,editedvolume,numreferences,epsfig]{crckapb} 

\newcommand{\tr}[1]{\,{\rm tr}\,#1}
\def\e{{\,\rm e}\,}
\def\eol{\hspace*{\fill}\linebreak}

\def\be{\begin{equation}}
\def\ee{\end{equation}}
\def\bea{\begin{eqnarray}}
\def\eea{\end{eqnarray}}
\def\LA{\left\langle}
\def\RA{\right\rangle}
\newcommand{\rf}[1]{(\ref{#1})}
\newcommand{\eq}[1]{Eq.~(\ref{#1})}

\def\d{\partial}

\def\l{\lambda}

\def\s{\sigma}

\def\A{{A}}
\def\F{{F}}
\def\Beta{{\cal B}}

\def\j{\mbox{\boldmath $j$}}
\def\J{\mbox{\boldmath $J$}}
\newcommand{\ie}{{\it i.e.}\ }
\newcommand{\p}{{\prime}}
\newcommand{\ra}{\rightarrow}
\newcommand{\case}[2]{{\textstyle {#1 \over #2}}}
\newcommand{\bbox}[1]{\mbox{\boldmath $#1$}}
\def\lesssim{\mathrel{\mathpalette\fun <}}
\def\gtrsim{\mathrel{\mathpalette\fun >}}
\def\fun#1#2{\lower3.6pt\vbox{\baselineskip0pt\lineskip.9pt
\ialign{$\mathsurround=0pt#1\hfil##\hfil$\crcr#2\crcr\sim\crcr}}}

\newcommand{\eps}{\varepsilon}
\newcommand{\non}{\nonumber \\*}

\font\thinlinefont=cmr5
\begingroup\makeatletter\ifx\SetFigFont\undefined
\def\x#1#2#3#4#5#6#7\relax{\def\x{#1#2#3#4#5#6}}%
\expandafter\x\fmtname xxxxxx\relax \def\y{splain}%
\ifx\x\y   
\gdef\SetFigFont#1#2#3{%
  \ifnum #1<17\tiny\else \ifnum #1<20\small\else
  \ifnum #1<24\normalsize\else \ifnum #1<29\large\else
  \ifnum #1<34\Large\else \ifnum #1<41\LARGE\else
     \huge\fi\fi\fi\fi\fi\fi
  \csname #3\endcsname}%
\else
\gdef\SetFigFont#1#2#3{\begingroup
  \count@#1\relax \ifnum 25<\count@\count@25\fi
  \def\x{\endgroup\@setsize\SetFigFont{#2pt}}%
  \expandafter\x
    \csname \romannumeral\the\count@ pt\expandafter\endcsname
    \csname @\romannumeral\the\count@ pt\endcsname
  \csname #3\endcsname}%
\fi
\fi\endgroup

\begin{opening}

\title{Large-N Gauge Theories}

\subtitle{Lectures presented at the 1999 NATO-ASI on ``Quantum Geometry'' 
in Akureyri, Iceland. \\ Preprint: ITEP-TH-80/99, hep-th/0001047}

\author{Yuri Makeenko}

\institute{Institute of Theoretical and Experimental Physics, \\
           B. Cheremushkinskaya 25,  \\
           117218 Moscow, Russia \\
           email: makeenko@itep.ru}
\end{opening}

\runningtitle{Large-N Gauge Theories}

\begin{document}

\begin{abstract}
Four pedagogical Lectures at the NATO-ASI on ``Quantum Geometry'' 
in Akureyri, Iceland, August 1999. 

\vspace*{.1cm}	

\hspace*{3.5cm}	Contents: \\
\hspace*{3cm}		1. O$(N)$ Vector Models \\
\hspace*{3cm}	2. Large-$N$ QCD\\
\hspace*{3cm}	3. QCD in Loop Space\\
\hspace*{3cm}	4. Large-$N$ Reduction\\
\end{abstract}

\section{Introduction}

The large-$N$ limit of Gauge Theory was proposed in 1974
by 't~Hooft~\cite{Hoo74} for
quantum chromodynamics (QCD). The dimensionality of the gauge
group \mbox{SU$(N_c)$} was used \sloppy
as a parameter, considering 
$N_c$ as a large number and performing an expansion in $1/N_c$.
The motivation was an expansion in the inverse number of 
field-components $N$ in statistical mechanics where it is known as 
the $1/N$-expansion, and is a standard method for non-perturbative 
investigations. 

The expansion of QCD in $1/N_c$ rearranges
diagrams of perturbation theory in a way which is consistent 
with a string picture of strong interaction, whose
phenomenological consequences agree with experiment. The accuracy of
the leading-order term, which is often called  large-$N$ 
QCD or multicolor QCD, is expected to be of the order of the ratios 
of meson widths to their masses, \ie about 10--15\%. 

While QCD is simplified in the large-$N_c$ limit, it is still not
yet solved. Generically, it is a problem of infinite matrices, rather than
of infinite vectors as in the theory of second-order phase
transitions in statistical mechanics.

We start these Lectures by showing how the $1/N$-expansion
works for the O$(N)$-vector models, and describing some applications
to the four-Fermi interaction and the nonlinear sigma model. 
Then we concentrate on the Yang--Mills theory at large $N_c$.

The methods described in these Lectures are developed mostly
in the seventies and the beginning of the eighties.  
They are used over and over again when discussing a relation
between Gauge Theory and String Theory.

The conformal invariance, described in
Section~2 for the four-Fermi interaction, 
is used in the modern approaches based on the 
AdS/CFT correspondence~\cite{Mal98}
(discussed in the lectures by L.~Thorlacius at this School).
The notion of the large-$N$ limit,
described in Section~3, is widely used, in particular,
in Matrix Theory~\cite{BFSS97}
(discussed in  the lectures by W.~Taylor at this School).
The loop equation, described in Section~4,
 has been recently applied~\cite{DGO99}
for studying the AdS/CFT correspondence.
The large-$N$ reduced models, described in Section~5, are applied 
to a nonperturbative formulation of superstring~\cite{IKKT97} and 
very recently to noncommutative gauge theory~\cite{Aok99}.

The content of my fifth lecture at this School, which was devoted
to an application of the large-$N$ methods to Matrix Theory
at finite temperature, is not included in the text.
It is mostly contained in Ref.~\cite{Mak99}.


\section{O$(N)$ Vector Models}

The simplest models, which become solvable in the limit of a large number 
of field components, deal with a field which has $N$ components
forming an O$(N)$ vector in an internal symmetry space.
A model of this kind was first considered by Stanley~\cite{Sta68} 
in statistical mechanics and is known as the spherical model.
The extension to quantum field theory was done by Wilson~\cite{Wil73}
both for the four-Fermi and $\varphi^4$ theories.

In the framework of perturbation theory, the four-Fermi interaction is
renormalizable only in $d=2$ dimensions and is non-renormalizable for 
\mbox{$d>2$}. The $1/N$-expansion resums perturbation-theory diagrams after 
which the four-Fermi interaction becomes renormalizable to each order in 
$1/N$ for \mbox{$2\leq d<4$}. 
An analogous expansion exists for the nonlinear O$(N)$ sigma model.

The $1/N$ expansion of the vector models is associated with a resummation of
Feynman diagrams. A very simple class of diagrams --- the bubble graphs ---
survives to the leading order in $1/N$. This is why the large-$N$ limit
of the vector models is solvable. Alternatively, the large-$N$ solution 
is nothing but a saddle-point solution in the path-integral approach.
The existence of the saddle point is due to the fact that $N$ is large. This
is to be distinguished from a perturbation-theory saddle point which is due
to the fact that the coupling constant is small. Taking into account
fluctuations around the saddle-point results in the $1/N$-expansion of the 
vector models.

We begin this Section with a description of the $1/N$-expansion  
of the $N$-component four-Fermi theory analyzing the bubble graphs.
Then we introduce functional methods and construct the 
$1/N$-expansion of the O$(N)$-symmetric nonlinear sigma model.
At the end we discuss the factorization in the O$(N)$ vector models
at large $N$. 

\subsection{Four-Fermi Interaction \label{ss:f-F}}

The action of the O$(N)$-symmetric four-Fermi 
interaction in a $d$-dimensional Euclidean space
is defined by
\be
S\left[ \bar \psi ,\psi \right]  =  \int d^dx\left( \bar \psi\, \hat
\partial \, \psi +m\,\bar \psi \psi -\frac G2\left( \bar \psi \psi \right)
^2\right).
\label{FFaction}
\ee
Here
$\hat \partial  = \gamma _\mu \partial _\mu $
and
$
\psi  =  \left( \psi _1,\ldots ,\psi _N\right) 
$
is a spinor field which forms 
an $N$-component vector in an internal symmetry space so that
\be
\bar \psi \psi  =  \sum_{i=1}^N\bar \psi _i\psi _i \,.
\ee
In $d=2$ this model was studied in the large-$N$ limit in 
Ref.~\cite{GN74} and is often called the Gross--Neveu model. 

The dimension of the four-Fermi coupling constant $G$ is
$
\dim \left[ G\right]  =  m^{2-d} \,.
$
For this reason, the perturbation theory for the 
four-Fermi interaction is renormalizable in $d=2$ 
but is non-renormalizable for $d>2$ (and, in particular, in $d=4$).
This is why the old Fermi theory of weak interactions 
was replaced by the modern electroweak theory, 
where the interaction is mediated by the $W^\pm$ and $Z$ bosons.

The action~\rf{FFaction} can be equivalently rewritten as
\be
S\left[ \bar \psi ,\psi,\chi\right]  =  \int d^dx\left( \bar \psi \,\hat
\partial\, \psi +m\,\bar \psi \psi -\chi \,\bar \psi \psi 
+\frac{\chi ^2}{2G} \right) ,
\label{auxiliary}
\ee
where $\chi$ is an auxiliary field. The two forms of the action,
\rf{FFaction} and \rf{auxiliary},
are equivalent due to the equation of motion 
\be
{\chi}  =  G \bar{\psi} {\psi} 
\label{chi}
\ee
which can be derived by varying the action~\rf{auxiliary}
with respect to $\chi$.

In the path-integral quantization, where the partition function
is defined by
\be
Z=\int D\chi D\bar\psi D\psi \e^{-S\left[ \bar \psi ,\psi,\chi\right]}
\label{chiFZ}
\ee
with $S\left[ \bar \psi ,\psi,\chi\right]$ given by \eq{auxiliary},
the action~\rf{FFaction} appears after performing the Gaussian integral
over $\chi$. Therefore, one alternatively gets 
\be
Z=\int D\bar\psi D\psi \e^{-S\left[ \bar \psi ,\psi\right]}
\ee
with $S\left[ \bar \psi ,\psi\right]$ given by \eq{FFaction}.

The perturbative expansion of the O$(N)$-symmetric four-Fermi theory
can be conveniently represented using the formulation~\rf{auxiliary} via the 
auxiliary field $\chi$. Then the diagrams are of the type of those in
Yukawa theory, and resemble the ones for QED with
$\bar\psi$ and $\psi$ being an analog of the electron-positron field and 
$\chi$ being an analog of the photon field. However,
the auxiliary field $\chi(x)$ does not propagate, since
it follows from the action~\rf{auxiliary} that
\be
D_0\left( x-y \right) \equiv
\LA\, \chi(x) \chi(y)\,\RA_{\rm Gauss}=G\,\delta^{(d)}(x-y)
\label{chipropagatorx}
\ee
or
\be
D_0\left( p \right) \equiv
\LA\, \chi(-p) \chi(p)\,\RA_{\rm Gauss}~=~G
\label{chipropagatorp}
\ee
in momentum space.

It is convenient to represent the four-Fermi vertex as 
the sum of two terms
\bea
\unitlength=1mm
\linethickness{0.6pt}
\begin{picture}(100.00,14.00)(-3,9)
\put(7.00,7.00){\vector(1,1){5.00}}
\put(11.00,11.00){\line(1,1){4.00}}
\put(15.00,15.00){\vector(1,1){5.00}}
\put(19.00,19.00){\line(1,1){4.00}}
\put(15.00,15.00){\vector(-1,1){5.00}}
\put(11.00,19.00){\line(-1,1){4.00}}
\put(23.00,7.00){\vector(-1,1){5.00}}
\put(19.00,11.00){\line(-1,1){4.00}}
\put(30.00,15.00){\makebox(0,0)[cc]{$=$}}
\put(37.00,7.00){\vector(1,1){5.00}}
\put(41.00,11.00){\line(1,1){4.00}}
\put(47.00,15.00){\vector(1,1){5.00}}
\put(51.00,19.00){\line(1,1){4.00}}
\put(45.00,15.00){\vector(-1,1){5.00}}
\put(41.00,19.00){\line(-1,1){4.00}}
\put(55.00,7.00){\vector(-1,1){5.00}}
\put(51.00,11.00){\line(-1,1){4.00}}
\put(61.00,15.00){\makebox(0,0)[cc]{$-$}}
\put(67.00,7.00){\vector(1,1){5.00}}
\put(71.00,11.00){\line(1,1){4.00}}
\put(75.00,15.00){\vector(1,1){5.00}}
\put(79.00,19.00){\line(1,1){4.00}}
\put(74.00,16.00){\vector(-1,1){4.00}}
\put(71.00,19.00){\line(-1,1){4.00}}
\put(83.00,7.00){\vector(-1,1){5.00}}
\put(80.00,10.00){\line(-1,1){4.00}}
\bezier{20}(76.00,14.00)(76.00,16.00)(74.00,16.00)
\end{picture}
\eea
where the empty space inside the vertex is associated with the 
propagator~\rf{chipropagatorx} (or \rf{chipropagatorp} in momentum 
space). The relative minus sign makes the vertex antisymmetric in
both incoming and outgoing fermions as is prescribed by the Fermi
statistics.

\begin{figure}[tb]
\unitlength=1mm
\linethickness{0.6pt}
\begin{picture}(20.00,37.00)(0,4)
\put(7.00,12.00){\vector(1,1){5.00}}
\put(11.00,16.00){\line(1,1){4.00}}
\put(15.00,25.00){\oval(10.00,10.00)[l]}
\put(15.00,30.00){\vector(-1,1){5.00}}
\put(11.00,34.00){\line(-1,1){4.00}}
\put(17.00,30.00){\vector(1,1){5.00}}
\put(21.00,34.00){\line(1,1){4.00}}
\put(25.00,12.00){\vector(-1,1){5.00}}
\put(21.00,16.00){\line(-1,1){4.00}}
\put(17.00,25.00){\oval(10.00,10.00)[r]}
\put(10.10,24.80){\vector(0,1){1.00}}
\put(21.90,24.80){\vector(0,1){1.00}}
\put(16.00,7.00){\makebox(0,0)[cc]{\large a)}}
\end{picture}
\begin{picture}(30.00,37.00)(6,4)
\put(7.00,17.00){\vector(1,1){5.00}}
\put(11.00,21.00){\line(1,1){4.00}}
\put(29.00,25.00){\vector(1,1){5.00}}
\put(33.00,29.00){\line(1,1){4.00}}
\put(15.00,25.00){\vector(-1,1){5.00}}
\put(11.00,29.00){\line(-1,1){4.00}}
\put(37.00,17.00){\vector(-1,1){5.00}}
\put(33.00,21.00){\line(-1,1){4.00}}
\put(22.00,25.00){\circle{10.00}}
\put(21.80,29.90){\vector(1,0){1.00}}
\put(22.20,20.10){\vector(-1,0){1.00}}
\put(22.00,7.00){\makebox(0,0)[cc]{\large b)}}
\end{picture}
\begin{picture}(30.00,37.00)(8,4)
\put(20.00,26.00){\oval(10.00,10.00)[t]}
\put(20.00,24.00){\oval(10.00,10.00)[b]}
\put(25.00,24.00){\line(0,1){2.00}}
\put(7.00,16.00){\vector(1,1){5.00}}
\put(11.00,20.00){\line(1,1){4.00}}
\put(27.00,25.00){\vector(1,1){5.00}}
\put(31.00,29.00){\line(1,1){4.00}}
\put(15.00,26.00){\vector(-1,1){5.00}}
\put(11.00,30.00){\line(-1,1){4.00}}
\put(35.00,17.00){\vector(-1,1){5.00}}
\put(31.00,21.00){\line(-1,1){4.00}}
\put(19.80,30.90){\vector(-1,0){1.00}}
\put(20.20,19.10){\vector(1,0){1.00}}
\put(21.00,7.00){\makebox(0,0)[cc]{\large c)}}
\end{picture}
\begin{picture}(61.00,37.00)(-13,4)
\put(22.00,26.00){\oval(10.00,10.00)[t]}
\put(22.00,24.00){\oval(10.00,10.00)[b]}
\put(17.00,24.00){\line(0,1){2.00}}
\put(7.00,17.00){\vector(1,1){5.00}}
\put(11.00,21.00){\line(1,1){4.00}}
\put(27.00,26.00){\vector(1,1){5.00}}
\put(31.00,30.00){\line(1,1){4.00}}
\put(15.00,25.00){\vector(-1,1){5.00}}
\put(11.00,29.00){\line(-1,1){4.00}}
\put(35.00,16.00){\vector(-1,1){5.00}}
\put(31.00,20.00){\line(-1,1){4.00}}
\put(21.80,30.90){\vector(1,0){1.00}}
\put(22.20,19.10){\vector(-1,0){1.00}}
\put(22.00,7.00){\makebox(0,0)[cc]{\large d)}}
\end{picture}
\begin{picture}(30.00,37.00)(0,4)
\put(20.00,26.00){\oval(10.00,10.00)[t]}
\put(20.00,24.00){\oval(10.00,10.00)[b]}
\put(7.00,16.00){\vector(1,1){5.00}}
\put(11.00,20.00){\line(1,1){4.00}}
\put(25.00,24.00){\vector(1,1){7.00}}
\put(31.00,30.00){\line(1,1){4.00}}
\put(15.00,26.00){\vector(-1,1){5.00}}
\put(11.00,30.00){\line(-1,1){4.00}}
\put(35.00,16.00){\vector(-1,1){5.00}}
\put(31.00,20.00){\line(-1,1){4.00}}
\put(19.80,30.90){\vector(-1,0){1.00}}
\put(20.20,19.10){\vector(1,0){1.00}}
\bezier{20}(27.00,24.00)(27.00,26.00)(25.00,26.00)
\put(21.00,7.00){\makebox(0,0)[cc]{\large e)}}
\end{picture}
\caption[One-loop diagrams for the four-vertex]   
{Diagrams of the second order of perturbation theory for
   the four-Fermi vertex. The diagram b) involves the sum over
   the O$(N)$ indices.}
   \label{fi:somegraphs}
   \end{figure}
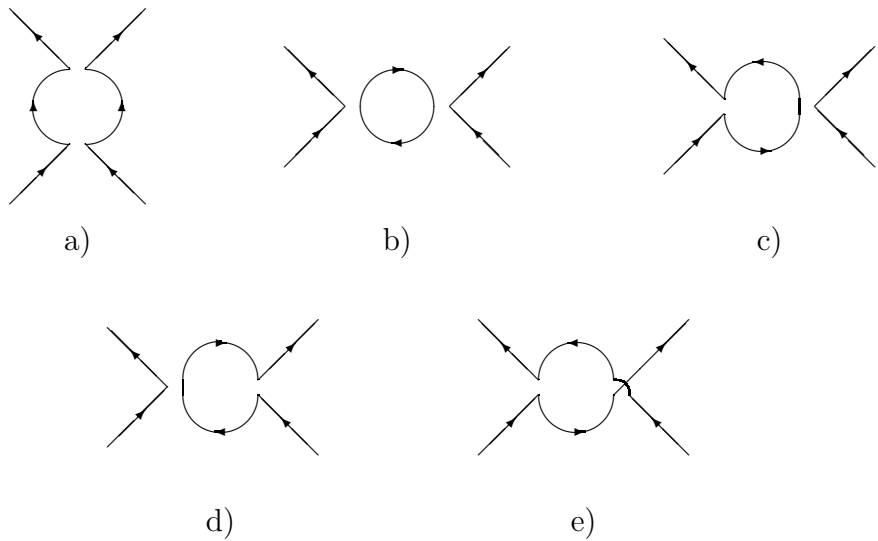

\begin{figure}[tb]
\unitlength=1mm
\linethickness{0.6pt}
\begin{picture}(60.00,30.00)(-20,3)
\put(5.00,15.00){\vector(1,0){5.00}}
\put(10.00,15.00){\line(1,0){5.00}}
\put(15.00,20.00){\oval(10.00,10.00)[l]}
\put(17.00,20.00){\oval(10.00,10.00)[r]}
\put(10.10,20.20){\vector(0,1){1.00}}
\put(21.90,19.80){\vector(0,-1){1.00}}
\put(16.00,7.00){\makebox(0,0)[cc]{\large a)}}
\put(15.00,25.00){\line(1,0){2.00}}
\put(17.00,15.00){\vector(1,0){7.00}}
\put(23.00,15.00){\line(1,0){4.00}}
\end{picture}
\begin{picture}(20.00,30.00)(0,3)
\put(5.00,15.00){\vector(1,0){5.00}}
\put(10.00,15.00){\line(1,0){5.00}}
\put(15.00,7.00){\makebox(0,0)[cc]{\large b)}}
\put(15.00,22.00){\circle{10.00}}
\put(10.10,22.20){\vector(0,1){1.00}}
\put(19.90,21.80){\vector(0,-1){1.00}}
\put(15.00,15.00){\vector(1,0){7.00}}
\put(21.00,15.00){\line(1,0){4.00}}
\end{picture}
\caption[One-loop diagrams for the propagator of $\psi$-field]   
{ One-loop diagrams for the propagator of the $\psi$-field. 
   The diagram b) involves the sum over the O$(N)$ indices.
     }
   \label{fi:prop1loop}
   \end{figure}
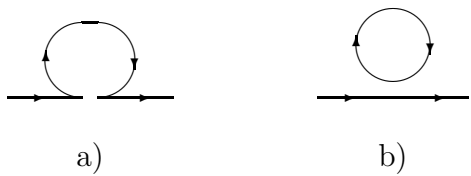

The diagrams that contribute to second order in $G$ for the four-Fermi vertex
are depicted, in these notations, in Figure~\ref{fi:somegraphs}.
The O$(N)$ indices propagate through the solid lines so that the closed line
in the diagram in Figure~\ref{fi:somegraphs}b corresponds to the sum over 
the O$(N)$ indices which results in a factor of $N$. 
Analogous one-loop diagrams for the propagator of the $\psi$-field
are depicted in Figure~\ref{fi:prop1loop}.

\subsubsection*{Remark on the one-loop Gell-Mann--Low function 
of four-Fermi theory} 

Evaluating the diagrams in Figure~\ref{fi:somegraphs} 
which are logarithmically divergent in $d=2$,
and noting that the diagrams in Figure~\ref{fi:prop1loop} do 
not contribute to the wave-function renormalization of the $\psi$-field,
which emerges to the next order in $G$, one gets for 
the one-loop Gell-Mann--Low function
\be
\Beta\left( G \right)=- \frac{(N-1)\,G^2}{2\pi} \,.
\label{betaFF2d}
\ee
The four-Fermi theory in $2$ dimensions is asymptotically free as 
was first noted by Anselm~\cite{Ans59} and rediscovered in Ref.~\cite{GN74}.

The vanishing of the one-loop Gell-Mann--Low function 
in the Gross--Neveu model for $N=1$ is related
to the same phenomenon in the Thirring model. 
The latter model is associated with the vector-like \sloppy
interaction $\left( \bar\psi \gamma_\mu \psi \right)^2$ of one species of
fermions with $\gamma_\mu$
being the $\gamma$-matrices in $2$ dimensions. Since a bispinor has in $d=2$
only two components $\psi_1$ and $\psi_2$, both the vector-like and the
scalar-like interaction~\rf{FFaction} for $N=1$ reduce to 
$\bar \psi_1 \psi_1 \bar \psi_2 \psi_2$ since the square of a Grassmann  
variable vanishes. Therefore, these two models coincide. 
For the Thirring model, the vanishing of the Gell-Mann--Low function 
for any $G$ was shown by Johnson~\cite{Joh61} to all loops.

\subsection{Bubble graphs as zeroth order in $1/N$} 

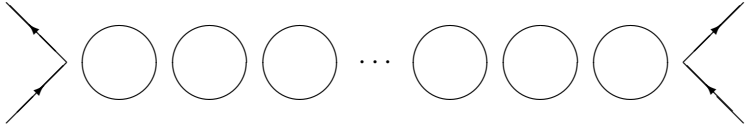
\begin{figure}[tb]
\unitlength=1mm
\linethickness{0.6pt}
\begin{picture}(100.00,22.00)(-3,0)
\put(7.00,2.00){\vector(1,1){5.00}}
\put(11.00,6.00){\line(1,1){4.00}}
\put(97.00,10.00){\vector(1,1){5.00}}
\put(101.00,14.00){\line(1,1){4.00}}
\put(15.00,10.00){\vector(-1,1){5.00}}
\put(11.00,14.00){\line(-1,1){4.00}}
\put(105.00,2.00){\vector(-1,1){5.00}}
\put(101.00,6.00){\line(-1,1){4.00}}
\put(22.00,10.00){\circle{10.00}}
\put(34.00,10.00){\circle{10.00}}
\put(46.00,10.00){\circle{10.00}}
\put(56.00,10.00){\makebox(0,0)[cc]{$\cdots$}}
\put(66.00,10.00){\circle{10.00}}
\put(78.00,10.00){\circle{10.00}}
\put(90.00,10.00){\circle{10.00}}
\end{picture}
\caption[Bubble diagram which survives the large-$N$ limit]   
{ Bubble diagram which survives the large-$N$ limit of the O$(N)$ vector
   models.
     }
   \label{fi:bubble}
   \end{figure}

The perturbation-theory expansion of the O$(N)$-symmetric four-Fermi theory 
contains, in particular, the diagrams of the 
type depicted in Figure~\ref{fi:bubble} 
which are called {\em bubble graphs}.
Since each bubble has a factor of $N$, the contribution of the $n$-bubble 
graph 
is $\propto G^{n+1} N^n$ which is of the order of
\be
G^{n+1} N^n\sim G
\ee
as $N\ra\infty$ since 
\be
G\sim\frac 1N \,.
\label{orderG}
\ee
Therefore, all the bubble graphs are essential to the leading order in $1/N$.

Let us denote
\bea
\unitlength=1.00mm
\linethickness{0.6pt}
\begin{picture}(100,15)(5,2)
\multiput(1.30,10.00)(4.00,0.00){5}{
  \bezier{28}(0.00,0.00)(1.00,1.00)(2.00,0.00)}
\multiput(3.30,10.00)(4.00,0.00){5}{
  \bezier{28}(0.00,0.00)(1.00,-1.00)(2.00,0.00)}
\put(26.00,9.80){\makebox(0,0)[cc]{$=$}}
\put(31.00,10.00){\makebox(0,0)[cc]{$G$}}
\put(43.00,10.00){\makebox(0,0)[cc]{$+\ldots+\;G^2$}}
\put(57.50,10.00){\circle{10.00}}
\put(70.50,10.00){\makebox(0,0)[cc]{$+\;\,G^{n+1}$}}
\put(83.00,10.00){\circle{10.00}}
\put(91.00,10.00){\makebox(0,0)[cc]{$\cdots$}}
\put(99.00,10.00){\circle{10.00}} 
\put(91.00,17.00){\makebox(0,0)[cc]{$n$ {\small loops} }}
\put(111.00,10.00){\makebox(0,0)[cc]{$+\ldots$}}
\end{picture}
\label{wavy}
\eea
In fact the wavy line is nothing but the propagator $D$ 
of the $\chi$ field 
with the bubble corrections included. The first term $G$ 
on the RHS of \eq{wavy} is 
nothing but the free propagator~\rf{chipropagatorp}.

Summing the geometric series of the fermion-loop chains
on the RHS of \eq{wavy}, one gets analytically%
\footnote{Recall that the free Euclidean fermionic propagator is given
by $S_0(p)=\left(i\hat{p}+m \right)^{-1}$ due to Eqs.~\rf{auxiliary}, 
\rf{chiFZ} and the additional minus sign is associated with the fermion
loop.} 
\be
D^{-1} (p)=\frac{1}{G}
- N\int\frac{d^dk}{(2\pi)^d}\, \frac{
\hbox{sp}\, \left(\hat{k}+im\right)
\left(\hat{k}+\hat{p}+im \right)}{\left(k^2+m^2\right)
\left((k+p)^2+m^2\right)}\,.
\label{F1}
\ee
This determines the exact propagator of the $\chi$ field
at large $N$. It is ${\cal O}\left( N^{-1}\right)$ since the coupling
$G$ is included in the definition of the propagator.

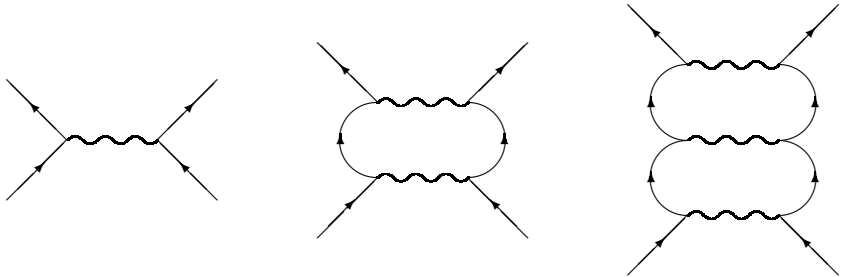
\begin{figure}[tb]
\unitlength=1mm
\linethickness{0.6pt}
\begin{picture}(40.00,41.00)(-3,5)
\put(0.00,17.00){\vector(1,1){5.00}}
\put(4.00,21.00){\line(1,1){4.00}}
\put(8.00,25.00){\vector(-1,1){5.00}}
\put(4.00,29.00){\line(-1,1){4.00}}
\multiput(8.00,25.00)(4.00,0.00){3}{
  \bezier{28}(0.00,0.00)(1.00,1.00)(2.00,0.00)}
\multiput(10.00,25.00)(4.00,0.00){3}{
  \bezier{28}(0.00,0.00)(1.00,-1.00)(2.00,0.00)}
\put(20.00,25.00){\vector(1,1){5.00}}
\put(24.00,29.00){\line(1,1){4.00}}
\put(28.00,17.00){\vector(-1,1){5.00}}
\put(24.00,21.00){\line(-1,1){4.00}}
\end{picture}
\begin{picture}(40.00,41.00)(-3,5)
\put(0.00,12.00){\vector(1,1){5.00}}
\put(4.00,16.00){\line(1,1){4.00}}
\put(8.00,25.00){\oval(10.00,10.00)[l]}
\put(8.00,30.00){\vector(-1,1){5.00}}
\put(4.00,34.00){\line(-1,1){4.00}}
\put(20.00,30.00){\vector(1,1){5.00}}
\put(24.00,34.00){\line(1,1){4.00}}
\put(28.00,12.00){\vector(-1,1){5.00}}
\put(24.00,16.00){\line(-1,1){4.00}}
\put(20.00,25.00){\oval(10.00,10.00)[r]}
\put(3.10,24.80){\vector(0,1){1.00}}
\put(24.90,24.80){\vector(0,1){1.00}}
\multiput(8.00,20.00)(4.00,0.00){3}{
  \bezier{28}(0.00,0.00)(1.00,1.00)(2.00,0.00)}
\multiput(10.00,20.00)(4.00,0.00){3}{
  \bezier{28}(0.00,0.00)(1.00,-1.00)(2.00,0.00)}
\multiput(8.00,30.00)(4.00,0.00){3}{
  \bezier{28}(0.00,0.00)(1.00,1.00)(2.00,0.00)}
\multiput(10.00,30.00)(4.00,0.00){3}{
  \bezier{28}(0.00,0.00)(1.00,-1.00)(2.00,0.00)}
\end{picture}
\begin{picture}(30.00,41.00)(-3,5)
\put(0.00,7.00){\vector(1,1){5.00}}
\put(4.00,11.00){\line(1,1){4.00}}
\put(8.00,35.00){\vector(-1,1){5.00}}
\put(4.00,39.00){\line(-1,1){4.00}}
\multiput(8.00,15.00)(4.00,0.00){3}{
  \bezier{28}(0.00,0.00)(1.00,1.00)(2.00,0.00)}
\multiput(10.00,15.00)(4.00,0.00){3}{
  \bezier{28}(0.00,0.00)(1.00,-1.00)(2.00,0.00)}
\multiput(8.00,25.00)(4.00,0.00){3}{
  \bezier{28}(0.00,0.00)(1.00,1.00)(2.00,0.00)}
\multiput(10.00,25.00)(4.00,0.00){3}{
  \bezier{28}(0.00,0.00)(1.00,-1.00)(2.00,0.00)}
\multiput(8.00,35.00)(4.00,0.00){3}{
  \bezier{28}(0.00,0.00)(1.00,1.00)(2.00,0.00)}
\multiput(10.00,35.00)(4.00,0.00){3}{
  \bezier{28}(0.00,0.00)(1.00,-1.00)(2.00,0.00)}
\put(20.00,35.00){\vector(1,1){5.00}}
\put(24.00,39.00){\line(1,1){4.00}}
\put(28.00,7.00){\vector(-1,1){5.00}}
\put(24.00,11.00){\line(-1,1){4.00}}
\put(8.00,20.00){\oval(10.00,10.00)[l]}
\put(8.00,30.00){\oval(10.00,10.00)[l]}
\put(20.00,20.00){\oval(10.00,10.00)[r]}
\put(20.00,30.00){\oval(10.00,10.00)[r]}
\put(3.10,19.80){\vector(0,1){1.00}}
\put(24.90,19.80){\vector(0,1){1.00}}
\put(3.10,29.80){\vector(0,1){1.00}}
\put(24.90,29.80){\vector(0,1){1.00}}
\end{picture}
\caption[Diagrams of the $1/N$-expansion]   
{ Some diagrams of the $1/N$-expansion for
   the O$(N)$ four-Fermi theory. The wavy line represents the
   (infinite) sum of the bubble graphs~\rf{wavy}.
     }
   \label{fi:newpt}
   \end{figure}

The idea is now to change the order of summation of diagrams of
perturbation theory using $1/N$ rather than $G$ as the expansion
parameter. Therefore, the zeroth-order propagator of the expansion
in $1/N$ is defined as the sum over the bubble graphs~\rf{wavy}
which is given by \eq{F1}.
Some of the diagrams of the new expansion 
for the four-Fermi vertex are depicted in Figure~\ref{fi:newpt}. 
The first diagram is proportional to $G$ while
the second and third ones are proportional to $G^2$ or $G^3$,
respectively, and therefore are of order ${\cal O}(N^{-1})$ 
or ${\cal O}(N^{-2})$ with respect to the first diagram.
The perturbation theory is thus rearranged as the $1/N$-expansion.

The general
structure of the $1/N$-expansion is the same for all vector models,
say, for the $N$-component nonlinear sigma model which is considered
in Subsection~\ref{ss:n.s.m.}.

The main advantage of the expansion in $1/N$ for the four-Fermi 
interaction, over the perturbation theory, is that it is
renormalizable in $d<4$
while the perturbation-theory expansion in $G$ is 
renormalizable only in $d=2$.  
Moreover, the $1/N$-expansion of the four-Fermi theory
in $2<d<4$ demonstrates~\cite{Wil73} an existence of
an ultraviolet-stable fixed point, \ie a nontrivial zero
of the Gell-Mann--Low function.

In order to
show that the $1/N$-expansion of the four-Fermi theory is renormalizable
in $2\leq d <4$, let us 
analyze indices of the diagrams of the $1/N$-expansion.
First of all, we shall get rid of an ultraviolet divergence of the integral
over the $d$-momentum $k$ in \eq{F1}. The divergent part of the integral
is proportional to $\Lambda^{d-2}$ 
(logarithmically divergent in $d=2$)
with $\Lambda$ being an
ultraviolet cutoff. It can be canceled by choosing 
\be
G= \frac{g^2}{N} \, \Lambda^{2-d} \,,
\label{Gg}
\ee
where $g^2$ is a proper dimensionless constant which is not necessarily
positive since the four-Fermi theory is stable with either sign of $G$. 
The power of $\Lambda$ in \eq{Gg} is consistent with the dimension of $G$.
This prescription works for $2<d<4$ where there is only one
divergent term while another divergency $\propto p^2 \ln{\Lambda}$ emerges
additionally in $d=4$. This is why the consideration is not applicable
in $d=4$.

The propagator $D(p)$ is therefore finite, and behaves at large
momenta $|p|\gg m$ as
\be
D(p)\propto\frac{1}{|p|^{d-2}} \,.
\label{D3p}
\ee
The standard power-counting arguments then show that the only divergent
diagrams appear in the propagators of the $\psi$ and $\chi$ fields, and
in the $\bar\psi$-$\chi$-$\psi$ three-vertex. These divergencies can be
removed by a renormalization of the coupling $g$, mass, and wave functions
of $\psi$ and $\chi$. 

This completes a demonstration of renormalizability
of the $1/N$-expan\-sion for the four-Fermi interaction in $2\leq d <4$.
For more detail, see Ref.~\cite{Par75}.

\subsection{Scale and Conformal Invariance of Four-Fermi Theory}

The coefficient in \eq{Gg} can easily be
calculated in $d=3$. 
To evaluate the divergent part of the integral in \eq{F1}, we put
$p=0$ and $m=0$. Remembering that the $\gamma$-matrices 
are $2\times 2$ matrices in $d=3$, we get
\be
\int^\Lambda\frac{d^3k}{(2\pi)^3}\, \frac{
\hbox{sp}\, \hat{k}\hat{k} }{k^2
k^2} =2\int^\Lambda\frac{d^3k}{(2\pi)^3}\, 
\frac{1}{k^2} 
= \frac{1}{\pi^2} \int^\Lambda d|k|= \frac{\Lambda}{\pi^2} \,. 
\ee 
Note that the integral is linearly divergent in $d=3$ and $\Lambda$
is the cutoff for the integration over $|k|$.
This divergence can be canceled by choosing $G$ according to \eq{Gg} with
$g$ equal to
\be
g_*=\pi\,.
\label{Gg3}
\ee

To calculate in $d=3$ the coefficient of proportionality in \eq{D3p},
let us choose $G=\pi^2/N\Lambda$ as is prescribed by Eqs.~\rf{Gg}, \rf{Gg3}
and put in \eq{F1} $m=0$ since we are interested in the asymptotics 
$|p|\gg m$. Then the RHS of \eq{F1} can be rearranged as 
\be
D^{-1} (p)=
- 2N\int\frac{d^3k}{(2\pi)^3}\left[ \frac{
 k^2 +kp}{k^2 (k+p)^2}-\frac{1}{k^2} \right]
= \frac{N\,|p|}{8} \,.
\label{8overp}
\ee
The integral is obviously convergent.

Equation~\rf{8overp} (or \rf{D3p} in $d$ dimensions) is remarkable since
it shows that the scale dimension of the field $\chi$ changes its value from
$l_\chi=d/2$ in perturbation theory to 
$l_\chi=1$ in the zeroth order of the $1/N$
expansion (remember that the momentum-space propagator of a field
with the scale dimension $l$ is proportional to $|p|^{2l-d}$). 
This appearance of scale invariance in the $1/N$-expansion of the
four-Fermi theory at $2<d<4$ was first pointed out by Wilson~\cite{Wil73}
and implies that the Gell-Mann--Low function $\Beta(g)$ has a zero
at $g=g_*$ which is given in $d=3$ by \eq{Gg3}.

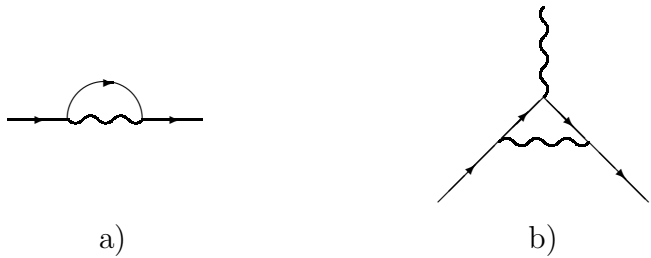
\begin{figure}[tb]
\unitlength=1mm
\linethickness{0.6pt}
\begin{picture}(70.00,37.00)(-17,9)
\put(0.00,28.00){\vector(1,0){5.00}}
\put(4.00,28.00){\line(1,0){4.00}}
\put(22.00,28.00){\line(1,0){4.00}}
\put(18.00,28.00){\vector(1,0){5.00}}
\multiput(8.00,28.00)(4.00,0.00){3}{
  \bezier{28}(0.00,0.00)(1.00,-1.00)(2.00,0.00)}
\multiput(10.00,28.00)(4.00,0.00){2}{
  \bezier{28}(0.00,0.00)(1.00,1.00)(2.00,0.00)}
\put(13.00,28.00){\oval(10.00,10.00)[t]}
\put(14.00,12.00){\makebox(0,0)[cc]{\large a)}}
\put(13.20,32.90){\vector(1,0){1.00}}\end{picture}
\begin{picture}(40.00,37.00)(-3,9)
\put(0.00,17.00){\vector(1,1){5.00}}
\put(4.00,21.00){\line(1,1){4.00}}
\put(8.00,25.00){\vector(1,1){4.00}}
\put(10.00,27.00){\line(1,1){4.00}}
\multiput(8.00,25.00)(4.00,0.00){3}{
  \bezier{28}(0.00,0.00)(1.00,1.00)(2.00,0.00)}
\multiput(10.00,25.00)(4.00,0.00){3}{
  \bezier{28}(0.00,0.00)(1.00,-1.00)(2.00,0.00)}
\put(14.00,31.00){\vector(1,-1){4.00}}
\put(16.00,29.00){\line(1,-1){4.00}}
\put(20.00,25.00){\vector(1,-1){5.00}}
\put(24.00,21.00){\line(1,-1){4.00}}
\multiput(14.00,31.00)(0.00,4.00){3}{
  \bezier{28}(0.00,0.00)(1.00,1.00)(0.00,2.00)}
\multiput(14.00,33.00)(0.00,4.00){3}{
  \bezier{28}(0.00,0.00)(-1.00,1.00)(0.00,2.00)}
\put(14.00,12.00){\makebox(0,0)[cc]{\large b)}}
\end{picture}
\caption[$1/N$-corrections to $\psi$-propagator and three-vertex]   
{Diagrams for the $1/N$-correction to
   the $\psi$-field propagator (a) and the three-vertex (b). 
    }
   \label{fi:1overN}
   \end{figure}

The (logarithmic) anomalous dimensions of the fields $\psi$, $\chi$,
and of the $\bar\psi$-$\chi$-$\psi$ three-vertex 
in $d=3$ to order $1/N$ can be found as follows. 
The $1/N$-correction to the propagator of the $\psi$-field is 
given by the diagram depicted in Figure~\ref{fi:1overN}a). 
Since we are interested in an ultraviolet behavior, we can put
again $m=0$. Analytically, we have
\begin{eqnarray}
S^{-1}(p) ~=~ i\hat p + \frac{8i}{N}
\int^\Lambda\frac{d^3 k}{(2\pi)^3}
\frac{\hat{k}+\hat{p}}{|k|(k+p)^2} \,.
\end{eqnarray}
The (logarithmically) divergent contribution emerges from the
domain of integration $|k|\gg|p|$ so we can expand the integrand in $p$.
The $p$-indepen\-dent term vanishes after integration over the directions
of $k$ so that we get
\be
S^{-1}(p) =  
i\hat{p} \left [ 1+
\frac{8}{N}
\int^\Lambda\frac{d^3 k}{(2\pi)^3}
\frac{1}{|k|^3}\right]=
i\hat{p} \left [ 1+\frac{2}{3\pi^2N}\ln\frac{\Lambda^2}{p^2} +
\hbox{finite}\right]\,.
\label{psi0}
\ee

The diagram, which gives a non-vanishing contribution to the three-vertex
in order $1/N$, is depicted in Figure~\ref{fi:1overN}b. It reads analytically
\be
\Gamma(p_1,p_2) = 1 + \frac{8}{N}
\int^\Lambda\frac{d^3 k}{(2\pi)^3}
\frac{(\hat{k}+\hat{p}_1)(\hat{k}+\hat{p}_2)}{|k|(k+p_1)^2(k+p_2)^2} \,,
\label{gamma00}
\ee
where $p_1$ and $p_2$ the incoming and outgoing fermion momenta,
respectively. The logarithmic domain is $|k|\gg|p|_{\rm max}$ 
with $|p|_{\rm max}$ being the largest of $|p_1|$ and $|p_2|$.
This gives
\be
\Gamma(p_1,p_2) =  
1-\frac{2}{\pi^2N}\ln\frac{\Lambda^2}{p^2_{\rm max}}
+\hbox{finite} \,.
\label{gamma0}
\ee

An analogous calculation of the $1/N$ correction for the field $\chi$
is a bit more complicated since involves three two-loop diagrams
(see Ref.~\cite{CMS93}).
The resulting expression for $D^{-1}(p)$ reads
\be
\Big(N\,D(p)\Big)^{-1}=\frac{\Lambda}{g^2} +\left[-\frac{\Lambda}{\pi^2}
+\frac {|p|}8 \right]+ 
\frac{1}{\pi^2N}\left[ 2\Lambda
-|p| \left(\frac{2}{3}\ln{\frac{\Lambda^2}{p^2}} +\hbox{finite} 
\right)\right].
\label{chi0}
\ee
The linear divergence is canceled to order $1/N$ providing $g$ is equal to
\be
g_*=\pi \left(1+\frac{1}{N}\right)\,,
\label{g1N}
\ee
which determines $g_*$ to order $1/N$. After this $D^{-1}(p)$
takes the form
\be
 D^{-1}(p)
= \frac{N\,|p|}{8}\left(1-\frac{16}{3\pi^2N}
\ln{\frac{\Lambda^2}{p^2}}\right)\,.
\label{chi00}
\ee

To make all three expressions~\rf{psi0}, \rf{gamma0}, and \rf{chi00} finite,
we need logarithmic renormalizations of the wave functions of $\psi$- and 
$\chi$-fields and of the vertex $\Gamma$. This can be achieved 
by multiplying them by the renormalization constants
\be
Z_i(\Lambda)=1-\gamma_i \ln\frac{\Lambda^2}{\mu^2}
\ee
where $\mu$ stands for a reference mass scale and $\gamma_i$ are
anomalous dimensions. The index $i$ stands for $\psi$, $\chi$, or $\rm{v}$
for the $\psi$- and $\chi$-propagators or the three-vertex $\Gamma$,
respectively. We have, therefore, calculated
\bea
\gamma_\psi&=& \frac{2}{3\pi^2N}\,, \non
\gamma_{\rm v}&=& -\frac{2}{\pi^2N} \,, \non
\gamma_\chi&=& -\frac{16}{3\pi^2N} 
\label{indices}
\eea
to order $1/N$. Due to \eq{chi} $\gamma_\chi$ coincides with
the anomalous dimension of the composite fields $\bar\psi\psi$
\be
\gamma _{\bar \psi \psi } =\gamma_\chi \,. 
\ee

Note, that
\be
Z_\psi^2 Z_{\rm v}^{-2} Z_\chi=1 \,.
\label{ZZZ}
\ee
This implies that the effective charge is not renormalized and is given
by \eq{g1N}. Thus, the nontrivial zero of the Gell-Mann--Low function
persists to order $1/N$ (and, in fact, to all orders of the 
$1/N$-expansion).

If $g$ is chosen exactly at the critical point $g_*$, then the
renormalization-group equations 
\be
\frac{\mu\, d\ln \Gamma_i}{d\mu}=\gamma_i \left( g^2 \right)\,,
\ee
where $\Gamma_i$ stands generically either for vertices or for 
inverse propagators, possess the scale invariant solutions
\be
\Gamma_i\propto \mu^{\gamma_i \left( g^2_* \right)} \,.
\label{scalesolutions}
\ee

For the four-Fermi theory in $d=3$, \eq{scalesolutions} yields
\bea
S(p) &=& 
\frac{1}{ i\hat{p} }\left(\frac{p^2}{\mu^2} \right)^{\gamma_\psi} \,,
\label{psiscale} \\
 D(p)&
= &\frac{8}{N\,|p|}\left(\frac{p^2}{\mu^2} \right)^{\gamma_\chi} \,,
\label{chiscale} \\
\Gamma (p_1,p_2)&=& \left(\frac{\mu^2}{p_1^2} \right)^{\gamma_{\rm v}} 
f\left( \frac{p_2^2}{p_1^2},\frac{p_1p_2}{p_1^2}\right)\,,
\label{gammascale} 
\end{eqnarray}
where $f$ is an arbitrary function of the dimensionless ratios
which is not determined by scale invariance.
The indices here obey the relation
\be
\gamma_{\rm v} = \gamma_\psi + \frac{1}{2}\gamma_\chi 
\label{vertexindex}
\ee
which guarantees that \eq{ZZZ}, implied by scale invariance, is satisfied.

The indices $\gamma_i$ are given to order $1/N$ by Eqs.~\rf{indices}.
When expanded in $1/N$, Eqs.~\rf{psiscale} and \rf{chiscale} obviously
reproduces Eqs.~\rf{psi0} and \rf{chi00}. Therefore, one gets 
the exponentiation of the logarithms which 
emerge in the $1/N$-expansion. The calculation of the next
terms of the $1/N$-expansion for the indices $\gamma_i$ is contained in
Ref.~\cite{Gra91}.

Scale invariance implies, in a renormalizable quantum field theory,
more general conformal invariance as is first pointed out in 
Refs.~\cite{MS69,GW70}. The conformal group in a $d$-dimensional 
space-time has $(d+1)(d+2)/2$ parameters as is illustrated by
Table~\ref{t:1}.
\begin{table}[t] 
\vspace*{.2cm}
\centerline{
\begin{tabular}{||c|l|c|c||} \hline
Group & 
\multicolumn{2}{c|} {Transformations}   & \# Parameters \\ \hline\hline  
\mbox{} & & & \\
Lorentz& $ ~~\frac{d(d-1)}{2}~\hbox{rotations} $&    
$ x_\mu^\prime= \Omega_{\mu\nu} x_\nu $
  & $\frac{d(d-1)}{2}$ \\ 
\mbox{} & & & \\\hline 
\mbox{} & & & \\
Poincar\'{e}& $+~d~\hbox{translations}$ & 
$x_\mu^\prime =x_\mu+a_\mu $ & $\frac{d(d+1)}{2}$ \\ 
\mbox{} & & & \\\hline 
\mbox{} & & & \\
Weyl& $ +~1~\hbox{dilatation}$ & 
$x_\mu^\prime =\rho\, x_\mu$ & $\frac{d^2+d+2}{2}$ \\ 
\mbox{} & & & \\\hline 
\mbox{} & & & \\
Conformal&  $+~d~\hbox{special conformal} $& 
$\frac{x_\mu^\prime} {\left(x^\prime\right)^2}=\frac{x_\mu}{x^2}+\alpha_\mu   
 $ & $\frac{(d+1)(d+2)}{2}$ \\ 
\mbox{} & & & \\\hline
\end{tabular} }
\caption[Groups of space-time symmetry]   
{ Contents and the number of parameters of groups of space-time symmetry.
     }
\label{t:1}
\end{table} 
More about the conformal group can be found in the lecture by 
Jackiw~\cite{Jac72}.

A heuristic proof~\cite{MS69} 
of the fact that scale invariance implies conformal invariance  
is based on the explicit form of the conformal current $K_\mu^\alpha$,
which is associated with the special conformal transformation, via
the energy-momentum tensor:
\be
K_\mu^\alpha=\left( 2x_\nu x^\alpha - x^2 \delta^\alpha_\nu \right)
\theta_{\mu\nu} \,.
\ee 
Differentiating, we get 
\be
\partial_\mu K_\mu^\alpha=2 x^\alpha \theta_{\mu\mu} \,,
\ee
which is proportional to divergence of the
dilatation current. Therefore, both the dilatation and conformal 
currents vanish simultaneously when $\theta_{\mu\nu} $ is traceless
which is provided, in turn, by the vanishing of the Gell-Mann--Low function.

Conformal invariance completely fixes three-vertices as was first
shown by Polyakov~\cite{Pol70} for scalar theories. The proper formula
for the four-Fermi theory reads~\cite{Mig71}
\bea
\lefteqn{\Gamma(p_1,p_2)~=~ \mu^{2\gamma_{\rm v} }
\frac{\Gamma(\frac d2)
\Gamma(\frac d2-\gamma_{\rm v})}{\Gamma(\gamma_{\rm v})} } \non & &\times
\int \frac{d^dk}{\pi^{d/2}}\frac{\hat k +\hat p_1}{[(k+p_1)^2]^{1+\gamma_\chi/2}}
\frac{\hat k +\hat p_2}{[(k+p_2)^2]^{1+\gamma_\chi/2}}
\frac{1}{|k|^{d-2+2\gamma_\psi-\gamma_\chi/2}}\,,~~~~~~~~
\label{gamap}
\eea
where the coefficient in the form of the ratio of the $\Gamma$-functions
is prescribed by the normalization~\rf{psiscale} and \rf{chiscale}
and the indices are related by \eq{vertexindex} but can be arbitrary
otherwise%
\footnote{The only restriction $\gamma_\psi\geq0$ is imposed by the 
K\"{a}ll\'{e}n--Lehmann
representation of the propagator while there is no such restriction on
$\gamma_\chi$ since it is a composite field.}.

Equation~\rf{gamap}, which results from conformal invariance, 
unambiguously fixes the function $f$ in \eq{gammascale}.
In contrast to infinite-dimensional conformal symmetry in $d=2$,
the conformal group in $d>2$ is less restrictive. 
It fixes only the tree-point vertex while, say, the four-point
vertex remains an unknown function of two variables.

The integral on the RHS of \eq{gamap} looks in $d=3$ very much like that in 
\eq{gamma00} and can easily be calculated to the leading order in $1/N$
when only the region of integration over large momenta
with $|k|\gtrsim |p|_{\rm max}\equiv \hbox{max}\{|p_1|, |p_2|\}$
is essential to this accuracy.

Let us first note that the coefficient in front of the integral is
$\propto\gamma_{\rm v}\sim 1/N$, so that one has to peak up  
the term $\sim 1/\gamma_{\rm v}$ in the integral
for the vertex to be of order $1$. 
This term comes from the region of integration with 
$|k| \gtrsim |p|_{\rm max}$.
Recalling that $|p_1-p_2| \lesssim |p|_{\rm max}$ in
Euclidean space, one gets
\be
\int \frac{d^3k}{2\pi}\frac{\hat k +\hat p_1}{[(k+p_1)^2]^{1+\gamma_\chi/2}}
\frac{\hat k +\hat p_2}{[(k+p_2)^2]^{1+\gamma_\chi/2}}
\frac{1}{|k|^{1+2\gamma_\psi-\gamma_\chi/2}}   =
\frac{1}{\gamma_{\rm v}\left(p^2_{\rm max}\right)^{\gamma_{\rm v}}}\,,
\label{gamapmax}
\ee
where \eq{vertexindex} has been used and
\be
\Gamma(p_1,p_2)= 
\left( \frac{\mu^2}{p^2_{\rm max} } \right)^{\gamma_{\rm v}}\,.
\label{maxvertex}
\ee
While the integral in \eq{gamapmax} is divergent in the ultraviolet
for $\gamma_{\rm v}<0$, this divergence disappears after the renormalization.

Equation~\rf{gamma0} is reproduced by \eq{gamapmax} when expanding in $1/N$.
This dependence of the three-vertex solely on the largest momentum is
typical for logarithmic theories in the ultraviolet region where one can put,
say, $p_1=0$ without changing the integral with logarithmic accuracy. This is
valid if the integral is fast convergent
in infrared regions which is our case. 

\subsubsection*{Remark on broken scale invariance}

Scale (and conformal) invariance at a fixed point $g=g_*$ holds only for
large momenta $|p|\gg m$. For smaller values of momenta, scale invariance 
is broken by masses.
In fact, any dimensional parameter $\mu$ breaks scale invariance.
If the bare coupling $g$ is chosen in the vicinity of $g_*$, 
then scale invariance holds even in the
massless case only for $|p|\gg \mu$ where $g(p)$ approaches $g_*$
while it is broken if 
$|p|\lesssim \mu$. 

\subsection{Nonlinear sigma model}
 \label{ss:n.s.m.}

The nonlinear O$(N)$ sigma model%
\footnote{The name comes from elementary particle physics where
a nonlinear sigma model in $4$ dimensions is used as an effective 
Lagrangian for describing low-energy scattering of the Goldstone
$\pi$-mesons.}
in $2$ Euclidean dimensions is defined by the partition function
\begin{eqnarray}
Z & = & \int D\vec n\,\delta \left( \vec n^2-\frac 1{g^2}\right) 
\e^{-\frac 12\int
d^2x\left( \partial _\mu \vec n\right) ^2} 
\label{sigmapartition}
\end{eqnarray}
where
$
\vec n  =  \left( n_1,\ldots ,n_N\right) 
$
is an O$(N)$ vector. While the action in \eq{sigmapartition} is
pure Gaussian, the model is not free due to the constraint
\be
\vec n^2(x)=\frac 1{g^2}\,,
\label{nconstraint}
\ee
which is imposed on the $\vec{n}$ field via the (functional)
delta-function.

The sigma model in $d=2$
 is sometimes considered as a toy model for QCD since it possesses:
\begin{itemize}
\item[1)] asymptotic freedom~\cite{Pol75};
\item[2)] instantons for $N=3$~\cite{BP75}.
\end{itemize}

The action in \eq{sigmapartition} is $\sim N$ as $N\ra \infty$ but
the entropy, \ie a contribution from the measure of integration, 
is also $\sim N$ so that a straightforward saddle point is not 
applicable.

To overcome this difficulty, we introduce an auxiliary field $u(x)$,
which is $\sim 1$ as $N\ra\infty$, and rewrite the partition 
function~\rf{sigmapartition} as
\be
Z  \propto  \int_\uparrow Du\left( x\right) \int D\vec n\left( x\right) 
\e^{-\frac
12\int d^2x\,\left[ \left( \partial _\mu \vec n\right) ^2-u\left( \vec
n^2-\frac 1{g^2}\right) \right] } ,
\ee
where the contour of integration over $u(x)$ is parallel to imaginary
axis. 

Doing the Gaussian integration over $\vec n$, we get
\be
Z  \propto   \int_\uparrow Du\left( x\right) 
\e^{-\frac N2{\rm Sp}\ln \left(
-\partial _\mu ^2+u\left( x\right) \right) +\frac 1{2g^2}\int
d^2x\,u\left( x\right) }\,.
\label{usigmapartition}
\ee
The first term in the exponent is
nothing but the sum of one-loop diagrams in $2$ dimensions
\be
\frac N2{\rm Sp}\ln \left(- \partial _\mu ^2+u\left( x\right) \right) ~=~
\sum_n \frac 1n
\unitlength=.7mm
\linethickness{0.6pt}
\begin{picture}(37.00,14.50)(1,18.6)
\put(20.00,20.00){\circle{10.00}}
\multiput(16.46,23.535)(-2.00,3.33){3}{
  \bezier{28}(0.00,0.00)(-1.333,0.333)(-1.00,1.665)}
\multiput(15.46,25.20)(-2.00,3.33){2}{
  \bezier{28}(0.00,0.00)(0.333,1.167)(-1.00,1.665)}
\multiput(23.535,23.535)(2.00,3.33){3}{
  \bezier{28}(0.00,0.00)(1.333,0.333)(1.00,1.665)}
\multiput(24.535,25.20)(2.00,3.33){2}{
  \bezier{28}(0.00,0.00)(0.333,1.167)(1.00,1.665)}
\put(20.00,30.00){\makebox(0,0)[cc]{$\cdot $}}
\put(23.00,29.50){\makebox(0,0)[cc]{$\cdot $}}
\put(17.00,29.50){\makebox(0,0)[cc]{$\cdot $}}
\put(11.00,15.50){\makebox(0,0)[cc]{$\cdot $}}
\put(9.80,18.50){\makebox(0,0)[cc]{$\cdot $}}
\put(12.90,13.00){\makebox(0,0)[cc]{$\cdot $}}
\put(27.10,13.00){\makebox(0,0)[cc]{$\cdot $}}
\put(29.00,15.50){\makebox(0,0)[cc]{$\cdot $}}
\put(30.20,18.50){\makebox(0,0)[cc]{$\cdot $}}
\multiput(20.00,15.00)(0.00,-4.00){2}{
  \bezier{28}(0.00,0.00)(-1.00,-1.00)(0.00,-2.00)}
\multiput(20.00,13.00)(0.00,-4.00){2}{
  \bezier{28}(0.00,0.00)(1.00,-1.00)(0.00,-2.00)}
\end{picture}
\end{equation}
\vspace*{.35cm} \mbox{} \eol
where the auxiliary field $u$ is denoted again by the wavy line.

Now the path integral over $u(x)$ in \eq{usigmapartition}
is a typical saddle-point one:
the action $\sim N$ while the entropy $\sim 1$ since only one
integration over $u$ is left. The saddle-point equation 
for the nonlinear sigma model 
\be
\frac 1{g^2} -  N G\left( x,x;u_{\rm sp} \right)=0 
\label{uspnos}
\ee
while $G$ is defined by
\be
G \left( x,y;u \right)= \LA y 
\left| \frac{1}{-\partial_\mu^2 + u} \right| x \RA \,.
\label{uGres}
\ee

The coupling $g^2$ in \eq{uspnos} is $\sim 1/N$ as
is prescribed by the constraint~\rf{nconstraint} which involves
a sum over $N$ terms on the LHS. This guarantees that a solution to
\eq{uspnos} exists. Next orders of the $1/N$-expansion for the 
$2$-dimensional sigma model can be constructed analogously to
the previous Subsections.

The $1/N$-expansion of the $2$-dimensional nonlinear sigma model has
many advantages over perturbation theory, which is usually
constructed solving explicitly the constraint~\rf{nconstraint},
say, choosing
\be
n_N=\frac 1g \sqrt{1- g^2\sum_{a=1}^{N-1} n_a^2}
\ee
and expanding the square root in $g^2$. Only $N-1$ dynamical degrees
of freedom are left so that the O$(N)$-symmetry is broken in 
perturbation theory down to O$(N-1)$. The particles in
perturbation theory are massless (like Goldstone bosons) and
it suffers from infrared divergencies.

On the contrary, the solution to \eq{uspnos} has the form
\be
u_{\rm sp}=m^2_{\rm R}\equiv \Lambda^2 \e^{-\frac{4\pi}{Ng^2}} \,, 
\label{umR}
\ee
where $\Lambda$ is an ultraviolet cutoff. Therefore, 
all $N$ particles acquires the same mass $m_{\rm R}$ in the 
$1/N$-expansion so that the O$(N)$ symmetry is restored.
This appearance of mass is due to dimensional transmutation
which says in this case that the parameter $m_{\rm R}$ rather
than the renormalized coupling constant
$g^2_{\rm R}$ is observable. The emergence of the mass cures
the infrared problem.

To show that~\rf{umR} is a solution to \eq{uspnos}, 
let us look for a translationally invariant solution 
$
u_{\rm sp}(x) \equiv m^2_{\rm R} 
$.
Then \eq{uspnos} in the momentum space reads
\be
\frac{1}{g^2}=N \int^{\Lambda} \frac{d^2 k}{(2\pi)^2}
\frac{1}{k^2+m^2_{\rm R}}=\frac N{4\pi} \int_0^{\Lambda^2} 
\frac{d k^2}{k^2+m^2_{\rm R}}=\frac N{4\pi} 
\ln{\frac{\Lambda^2}{m^2_{\rm R}}} \,.
\label{2dsmR}
\ee
The exponentiation results in \eq{umR}.

Equation~\rf{2dsmR} relates the bare coupling $g^2$ and the cutoff
$\Lambda$ and allows us to calculate the Gell-Mann--Low function yielding
\be
\Beta(g^2)\equiv\frac {\Lambda\,d g^2}{d \Lambda}=
-\frac{N g^4}{2\pi}\,.
\ee
The analogous one-loop perturbation-theory formula for any $N$ 
reads~\cite{Pol75}
\be
\Beta(g^2)=-\frac{(N-2) g^4}{2\pi}\,.
\ee
Thus, the sigma-model is asymptotically free in $2$-dimensions for $N>2$
which is the origin of the dimensional transmutation. There is no
asymptotic freedom for $N=2$ since O$(2)$ is Abelian.

\subsection{Large-$N$ factorization in vector models}\label{ss:vfactor}

The fact that a path integral has a saddle point at large $N$
implies a very important feature of large-$N$ theories ---
the factorization. It is a general property of the large-$N$
limit and holds not only for the O$(N)$ vector models.
However, it is useful to illustrate it by these solvable examples.

The factorization at large $N$
holds for averages of {\it singlet\/} operators, for example
\be
\left\langle \,u\left( x_1\right) \ldots u\left( x_k\right) 
\,\right\rangle \equiv 
Z^{-1}\int_\uparrow Du\e^{-\frac N2{\rm Sp}\ln \left[ - \partial
_\mu  ^2+u\right] +\frac 1{2g^2}\int d^2x\,u}\,
u\left( x_1\right) \cdots
u\left( x_k\right) 
\ee
in the $2$-dimensional sigma model.

Since the path integral has a saddle point at some
$
u\left(x\right)=u_{\rm sp}\left(x\right)
$
which is, in fact, $x$-independent due to
translational invariance,
we get to the leading order in $1/N$:
\be
\left\langle\, u\left( x_1\right) \ldots u\left( x_k\right) 
\,\right\rangle   =
 u_{\rm sp}\left( x_1\right) \ldots 
u_{\rm sp}\left( x_k\right) +{\cal O}\left( \frac 1N\right) ,
\ee
which can be written in the factorized form
\be
\left\langle \,u\left( x_1\right) \ldots u\left( x_k\right) \,\right\rangle
  =
\left\langle\, u\left( x_1\right) \,\right\rangle \ldots \left\langle \,
u\left(x_k\right) \,\right\rangle +{\cal O}\left( \frac 1N\right) .
\label{vfactorization}
\ee

Therefore, $u$ becomes ``classical'' as $N\ra\infty$ in the sense of the 
$1/N$-expansion.
This is an analog of the WKB-expansion in $\hbar=1/N$. ``Quantum'' 
corrections are suppressed as $1/N$.

We shall return to discussing the large-$N$ factorization in the
next Section when considering the large-$N$ limit of QCD.


\section{Large-N QCD}
\label{s:mqcd}

The method of the $1/N$-expansion can be applied to QCD.
This was done by 't~Hooft~\cite{Hoo74} using 
the inverse number of colors for the gauge
group \/SU$(N_c)$ as an expansion parameter.

For a \/SU$(N_c)$ gauge theory without virtual quark loops, 
the expansion goes in $1/N^2_c$
and rearranges diagrams of perturbation theory according to their
topology. The leading order in $1/N_c^2$ is given by planar diagrams,
which have a topology of a sphere, while the expansion in $1/N^2_c$
plays the role of a topological expansion.
This reminds an expansion in the string coupling constant in string models 
of the strong interaction, which also has a topological character.

Virtual quark loops can be easily incorporated in the $1/N_c$-expansion.
One distinguishes between the 't~Hooft limit when the number of
quark flavors $N_f$ is fixed as $N_c\ra\infty$ and the Veneziano 
limit~\cite{Ven76} when the ratio $N_f/N_c$ is fixed as $N_c\ra\infty$.
Virtual quark loops are suppressed in the 't~Hooft limit as $1/N_c$ 
and lead in the Veneziano limit to the same topological expansion as 
dual-resonance models of strong interaction.

The simplification of QCD in the large-$N_c$ limit is due to the fact
that the number of planar graphs grows with the number of vertices 
only exponentially rather than factorially as do the total number of graphs.
Correlators of gauge invariant operators factorize in the large-$N_c$
limit which looks like the leading-order term of a ``semiclassical''
WKB-expansion in $1/N_c$.

We begin this Section with a description of the double-line representation
of diagrams of QCD perturbation theory and rearrange it as the topological
expansion in $1/N_c$. Then we discuss some properties of the
$1/N_c$-expansion for a generic matrix-valued field.

\subsection{Index or ribbon graphs} 

In order to describe the $1/N_c$-expansion of the Yang--Mills theory, 
it is convenient to represent the gauge field by a Hermitean
$N\times N$ matrix 
\be
A_\mu^{ij}(x) =g \sum_a A_\mu^a(x) [t^a]^{ij}. 
\label{calA}
\ee
Here $[t^a]^{ij}$ are the generators of the gauge group
($a=1,\ldots, N_c^2-1$ for SU$(N_c)$)
with the normalization
\be
\tr t^a t^b = \frac 12 \delta^{ab}\,.
\label{1/2}
\ee
The (Euclidean) action reads
\be
S[A]=\int d^d x\frac{1}{2g^2}\tr F_{\mu\nu}^2\,,
\label{QCDaction}
\ee
where
\be
F_{\mu\nu}=\partial_\mu A_\nu -\partial_\nu A_\mu -i [A_\mu,A_\nu]
\label{fieldstrenght}
\ee
is the non-Abelian field strength and $g$ is the coupling constant.

The propagator of the matrix field $A^{ij}(x)$ has  
the form 
\be
\left\langle \,A_\mu ^{ij}\left( x\right) \, A_\nu ^{kl}\left( y\right)
\,\right\rangle_{\rm Gauss} = \frac 12
\left(\delta ^{il}\delta ^{kj} -\frac 1{N_c}\delta ^{ij}\delta ^{kl}\right)
D_{\mu\nu} \left( x-y \right),
\label{matpropagator}
\ee
where we have assumed, as usual, a gauge fixing  
to define the propagator in perturbation theory. For instance, one has
\be
D_{\mu\nu} \left( x-y \right)=\frac{g^2}{4 \pi^2}
\frac{\delta_{\mu\nu}}{\left( x-y \right)^2}
\label{D(x-y)}
\ee
in the Feynman gauge.

Equation~\rf{matpropagator} can be immediately derived from 
the standard formula
\be
\left\langle  A_\mu ^a\left( x\right) \, A_\nu ^b\left( y\right)
\right\rangle_{\rm Gauss}  =  \delta ^{ab} D_{\mu\nu} \left( x-y \right)
\ee
multiplying by the generators of the SU$( N_c)$ gauge group according
to the definition~\rf{calA} and using the completeness condition
\be
\sum\limits_{a=1}^{N_c^2-1}
\left( t^a\right) ^{ij}\left( t^a\right) ^{kl}  =  
\frac12\left(\delta ^{il}\delta
^{kj}-\frac 1{N_c}\delta ^{ij}\delta ^{kl}\right) ~~~~
\fbox{for \ SU$(N_c)$}~,~~
\label{completeness}
\ee
where the factor of $1/2$ is due to the normalization~\rf{1/2} of 
the generators.

We concentrate in this Section only on the structure of diagrams
in the index space, \ie the space of the indices associated with the 
SU$(N_c)$ group. We shall not consider, in most cases, 
space-time structures of diagrams which are prescribed by Feynman's rules. 

Omitting at large $N_c$ the second
term in parentheses on the RHS of \eq{matpropagator},  we depict the
propagator by the double line
\be
\left\langle\, A_\mu ^{ij}\left( x\right) \,A_\nu ^{kl}\left( y\right)
\,\right\rangle_{\rm Gauss}  \propto  
g^2 \delta ^{il}\delta ^{kj} =
\unitlength=1mm
\linethickness{0.6pt}
\begin{picture}(31.00,4.00)(10,9)
\put(20.00,11.00){\vector(1,0){6.00}}
\put(26.00,11.00){\line(1,0){4.00}}
\put(30.00,9.00){\vector(-1,0){6.00}}
\put(24.00,9.00){\line(-1,0){4.00}}
\put(18.00,12.00){\makebox(0,0)[cc]{$i$}}
\put(18.00,8.00){\makebox(0,0)[cc]{$j$}}
\put(32.00,12.00){\makebox(0,0)[cc]{$l$}}
\put(32.00,8.00){\makebox(0,0)[cc]{$k$}}
\end{picture} .
\label{doubleline}
\ee
Each line represents the Kronecker delta-symbol and has orientation
which is indicated by arrows. This notation is obviously consistent
with the space-time structure of the propagator which describes
a propagation from $x$ to $y$.

The arrows are due to the fact that the matrix $A_\mu^{ij}$ is Hermitean
and its off-diagonal components are complex conjugate.
The independent fields are, say, the complex fields $A_\mu^{ij}$ for $i>j$ 
and the diagonal real fields $A_\mu^{ii}$. 
The arrow represents the
direction of the propagation of the indices
of the complex field $A_\mu^{ij}$ 
for $i>j$ while the complex-conjugate one, $
A_\mu^{ji}=(A_\mu^{ij})^*$, propagates in the opposite
direction. For the real fields $A_\mu^{ii}$, the arrows are not
essential.

The double-line notation
appears generically in all models describing {\it matrix\/} fields
in contrast to {\it vector\/} (in internal symmetry space) fields whose
propagators are depicted by single lines as in the previous Section. 

The three-gluon vertex, which is generated by the 
action \rf{QCDaction}, is depicted in the double-line notations as 
\be
\mbox{\beginpicture
\setcoordinatesystem units <0.50000cm,0.50000cm>
\unitlength=0.50000cm
\linethickness=1pt
\setplotsymbol ({\makebox(0,0)[l]{\tencirc\symbol{'160}}})
\setshadesymbol ({\thinlinefont .})
\setlinear
%
%
\linethickness= 0.500pt
\setplotsymbol ({\thinlinefont .})
\plot  5.080 21.273  6.191 20.637 /
%
%
\linethickness= 0.500pt
\setplotsymbol ({\thinlinefont .})
\putrule from  4.763 21.749 to  4.763 22.860
%
%
\linethickness= 0.500pt
\setplotsymbol ({\thinlinefont .})
\plot  4.763 21.749  3.651 21.114 /
%
%
\linethickness= 0.500pt
\setplotsymbol ({\thinlinefont .})
\putrule from  5.397 21.749 to  5.397 22.860
%
%
\linethickness= 0.500pt
\setplotsymbol ({\thinlinefont .})
\plot  5.397 21.749  6.509 21.114 /
%
%
\linethickness= 0.500pt
\setplotsymbol ({\thinlinefont .})
\plot  5.857 21.495  6.079 21.368 /
%
%
\plot  5.827 21.439  6.079 21.368  5.890 21.549 /
%
%
%
\linethickness= 0.500pt
\setplotsymbol ({\thinlinefont .})
\plot  4.174 21.416  4.396 21.543 /
%
%
\plot  4.207 21.362  4.396 21.543  4.144 21.473 /
%
%
%
\linethickness= 0.500pt
\setplotsymbol ({\thinlinefont .})
\plot  4.619 21.018  4.396 20.892 /
%
%
\plot  4.585 21.073  4.396 20.892  4.648 20.962 /
%
%
%
\linethickness= 0.500pt
\setplotsymbol ({\thinlinefont .})
\plot  5.730 20.908  5.508 21.035 /
%
%
\plot  5.760 20.965  5.508 21.035  5.697 20.854 /
%
%
%
\linethickness= 0.500pt
\setplotsymbol ({\thinlinefont .})
\putrule from  5.397 22.274 to  5.397 22.115
%
%
\plot  5.334 22.369  5.397 22.115  5.461 22.369 /
%
%
%
\linethickness= 0.500pt
\setplotsymbol ({\thinlinefont .})
\putrule from  4.763 22.384 to  4.763 22.543
%
%
\plot  4.826 22.289  4.763 22.543  4.699 22.289 /
%
%
%
\linethickness= 0.500pt
\setplotsymbol ({\thinlinefont .})
\putrule from  9.842 21.749 to  9.842 22.860
%
%
\linethickness= 0.500pt
\setplotsymbol ({\thinlinefont .})
\plot  9.842 21.749  8.731 21.114 /
%
%
\linethickness= 0.500pt
\setplotsymbol ({\thinlinefont .})
\plot 10.478 21.749 11.589 21.114 /
%
%
\linethickness= 0.500pt
\setplotsymbol ({\thinlinefont .})
\plot 10.160 21.273  9.049 20.637 /
%
%
\linethickness= 0.500pt
\setplotsymbol ({\thinlinefont .})
\putrule from  9.842 22.305 to  9.842 22.147
%
%
\plot  9.779 22.401  9.842 22.147  9.906 22.401 /
%
%
%
\linethickness= 0.500pt
\setplotsymbol ({\thinlinefont .})
\putrule from 10.478 22.337 to 10.478 22.496
%
%
\plot 10.541 22.242 10.478 22.496 10.414 22.242 /
%
%
%
\linethickness= 0.500pt
\setplotsymbol ({\thinlinefont .})
\plot  9.445 21.526  9.222 21.400 /
%
%
\plot  9.411 21.581  9.222 21.400  9.474 21.470 /
%
%
%
\linethickness= 0.500pt
\setplotsymbol ({\thinlinefont .})
\plot  5.080 21.273  3.969 20.637 /
%
%
\linethickness= 0.500pt
\setplotsymbol ({\thinlinefont .})
\putrule from 10.478 21.749 to 10.478 22.860
%
%
\put{$\propto g^{-2} \left( 
\delta^{i_1j_3}\delta^{i_2j_1}\delta^{i_3j_2} -
\delta^{i_1j_2}\delta^{i_3j_1}\delta^{i_2j_3}
\right)$} [lB] at 13.032 21.607
%
%
\linethickness= 0.500pt
\setplotsymbol ({\thinlinefont .})
\plot 11.081 21.416 10.859 21.543 /
%
%
\plot 11.111 21.473 10.859 21.543 11.048 21.362 /
%
%
%
\linethickness= 0.500pt
\setplotsymbol ({\thinlinefont .})
\plot  9.620 20.972  9.842 21.099 /
%
%
\plot  9.653 20.918  9.842 21.099  9.590 21.028 /
%
%
%
\linethickness= 0.500pt
\setplotsymbol ({\thinlinefont .})
\plot 10.160 21.273 11.271 20.637 /
%
%
\linethickness= 0.500pt
\setplotsymbol ({\thinlinefont .})
\plot 10.651 20.987 10.873 20.860 /
%
%
\plot 10.621 20.931 10.873 20.860 10.684 21.041 /
%
%
%
\put{\SetFigFont{6}{7.2}{rm}$i_1$} [B] at  5.715 22.860
%
%
\put{\SetFigFont{6}{7.2}{rm}$j_1$} [B] at  4.445 22.860
%
%
\put{\SetFigFont{6}{7.2}{rm}$i_3$} [B] at  6.032 20.161
%
%
\put{\SetFigFont{6}{7.2}{rm}$j_2$} [B] at  4.128 20.161
%
%
\put{\SetFigFont{6}{7.2}{rm}$i_2$} [B] at  3.334 21.273
%
%
\put{\SetFigFont{6}{7.2}{rm}$i_1$} [B] at  9.525 22.860
%
%
\put{\SetFigFont{6}{7.2}{rm}$j_1$} [B] at 10.795 22.860
%
%
\put{\SetFigFont{6}{7.2}{rm}$i_2$} [B] at  9.207 20.161
%
%
\put{\SetFigFont{6}{7.2}{rm}$j_2$} [B] at  8.414 21.273
%
%
\put{\SetFigFont{6}{7.2}{rm}$j_3$} [B] at  6.826 21.273
%
%
\put{\SetFigFont{6}{7.2}{rm}$j_3$} [B] at 11.113 20.161
%
%
\put{\SetFigFont{6}{7.2}{rm}$i_3$} [B] at 11.906 21.273
%
%
\put{$-$} [B] at  7.620 21.749
\linethickness=0pt
\putrectangle corners at  3.234 23.165 and 13.032 20.104
\endpicture}

\label{threegluon}
\ee
where the subscripts $1$, $2$ or $3$ refer to each of the three gluons.
The relative minus sign is due to the commutator in the cubic in $A$ term
in the action~\rf{QCDaction}. The color part of the three-vertex
is antisymmetric under interchanging the gluons. The space-time structure,
which reads in the momentum space as
\bea
\lefteqn{
\gamma_{\mu_1\mu_2\mu_3}\left(p_1,p_2,p_3 \right)}\non
&=&\delta_{\mu_1\mu_2}
\left(p_1-p_2 \right)_{\mu_3}+\delta_{\mu_2\mu_3}
\left(p_2-p_3 \right)_{\mu_1}+\delta_{\mu_1\mu_3}
\left(p_3-p_1 \right)_{\mu_2}\,,
\eea
is antisymmetric as well. We consider all three gluons as incoming
so that their momenta obey $p_1+p_2+p_3=0$.
The full vertex is symmetric as is prescribed by Bose statistics.

The color structure in \eq{threegluon} can alternatively be
obtained by multiplying
the standard vertex
\be
\Gamma^{a_1a_2a_3}_{\mu_1\mu_2\mu_3}\left(p_1,p_2,p_3 \right)=
f^{a_1a_2a_3}\gamma_{\mu_1\mu_2\mu_3}\left(p_1,p_2,p_3 \right)
\ee
by $\left(t^{a_1}\right)^{i_1j_1}\left(t^{a_2}\right)^{i_2j_2}
\left(t^{a_3}\right)^{i_3j_3}$, 
with $f^{abc}$ being the structure constants of the SU$(N_c)$ group,
and using the formula
\be
f^{a_1a_2a_3}
\left(t^{a_1}\right)^{i_1j_1}\left(t^{a_2}\right)^{i_2j_2}
\left(t^{a_3}\right)^{i_3j_3}=\frac12
\left( \delta^{i_1j_3}\delta^{i_2j_1}\delta^{i_3j_2} -
\delta^{i_1j_2}\delta^{i_3j_1}\delta^{i_2j_3}
\right),
\ee
which is a consequence of the completeness condition~\rf{completeness}.

The four-gluon vertex involves six terms --- each of them is depicted
by a cross ---
which differ by interchanging of the color indices. 
We depict the color structure of the four-gluon vertex for 
simplicity in the case when
$i_1=j_2=i$, $i_2=j_3=j$, $i_3=j_4=k$, $i_4=j_1=l$ but $i,j,k,l$ 
take on different values. Then only the following term is left
\be
\mbox{\beginpicture
\setcoordinatesystem units <0.50000cm,0.50000cm>
\unitlength=0.50000cm
\linethickness=1pt
\setplotsymbol ({\makebox(0,0)[l]{\tencirc\symbol{'160}}})
\setshadesymbol ({\thinlinefont .})
\setlinear
%
%
\linethickness= 0.500pt
\setplotsymbol ({\thinlinefont .})
\putrule from  5.397 21.749 to  5.397 22.860
%
%
\linethickness= 0.500pt
\setplotsymbol ({\thinlinefont .})
\putrule from  5.397 22.274 to  5.397 22.115
%
%
\plot  5.334 22.369  5.397 22.115  5.461 22.369 /
%
%
%
\linethickness= 0.500pt
\setplotsymbol ({\thinlinefont .})
\putrule from  4.763 22.384 to  4.763 22.543
%
%
\plot  4.826 22.289  4.763 22.543  4.699 22.289 /
%
%
%
\linethickness= 0.500pt
\setplotsymbol ({\thinlinefont .})
\putrule from  4.763 21.114 to  3.651 21.114
%
%
\linethickness= 0.500pt
\setplotsymbol ({\thinlinefont .})
\putrule from  4.763 21.749 to  3.651 21.749
%
%
\linethickness= 0.500pt
\setplotsymbol ({\thinlinefont .})
\putrule from  6.509 21.114 to  5.397 21.114
%
%
\linethickness= 0.500pt
\setplotsymbol ({\thinlinefont .})
\putrule from  6.509 21.749 to  5.397 21.749
%
%
\linethickness= 0.500pt
\setplotsymbol ({\thinlinefont .})
\putrule from  4.763 20.003 to  4.763 21.114
%
%
\linethickness= 0.500pt
\setplotsymbol ({\thinlinefont .})
\putrule from  5.397 20.003 to  5.397 21.114
%
%
\linethickness= 0.500pt
\setplotsymbol ({\thinlinefont .})
\putrule from  5.397 20.527 to  5.397 20.369
%
%
\plot  5.334 20.623  5.397 20.369  5.461 20.623 /
%
%
%
\linethickness= 0.500pt
\setplotsymbol ({\thinlinefont .})
\putrule from  4.763 20.637 to  4.763 20.796
%
%
\plot  4.826 20.542  4.763 20.796  4.699 20.542 /
%
%
%
\linethickness= 0.500pt
\setplotsymbol ({\thinlinefont .})
\putrule from  4.128 21.749 to  4.286 21.749
%
%
\plot  4.032 21.685  4.286 21.749  4.032 21.812 /
%
%
%
\linethickness= 0.500pt
\setplotsymbol ({\thinlinefont .})
\putrule from  4.763 21.749 to  4.763 22.860
%
%
\linethickness= 0.500pt
\setplotsymbol ({\thinlinefont .})
\putrule from  5.874 21.114 to  5.715 21.114
%
%
\plot  5.969 21.177  5.715 21.114  5.969 21.050 /
%
%
%
\put{$\propto g^{-2}$} [lB] at  7.938 21.431
%
%
\linethickness= 0.500pt
\setplotsymbol ({\thinlinefont .})
\putrule from  5.874 21.749 to  6.032 21.749
%
%
\plot  5.779 21.685  6.032 21.749  5.779 21.812 /
%
%
%
\linethickness= 0.500pt
\setplotsymbol ({\thinlinefont .})
\putrule from  4.128 21.114 to  3.969 21.114
%
%
\plot  4.223 21.177  3.969 21.114  4.223 21.050 /
%
%
%
\put{\SetFigFont{6}{7.2}{rm}$i$} [B] at  5.715 22.860
%
%
\put{\SetFigFont{6}{7.2}{rm}$l$} [B] at  4.445 22.860
%
%
\put{\SetFigFont{6}{7.2}{rm}$i$} [B] at  6.826 21.907
%
%
\put{\SetFigFont{6}{7.2}{rm}$k$} [B] at  4.445 19.844
%
%
\put{\SetFigFont{6}{7.2}{rm}$j$} [B] at  5.715 19.844
%
%
\put{\SetFigFont{6}{7.2}{rm}$l$} [B] at  3.334 21.907
%
%
\put{\SetFigFont{6}{7.2}{rm}$k$} [B] at  3.334 20.796
%
%
\put{\SetFigFont{6}{7.2}{rm}$j$} [B] at  6.826 20.796
\linethickness=0pt
\putrectangle corners at  3.109 23.089 and  7.938 19.768
\endpicture}

\label{fourgluon}
\ee
and there are no deltas on the RHS since the color structure is fixed.
In other words, we pick up only one color structure by 
equaling indices pairwise. 

\begin{figure}
\vspace*{3mm}
\centering{
\input{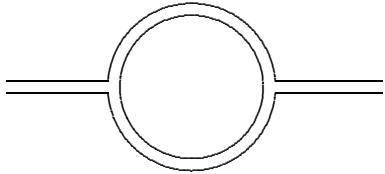}
}
\caption[Double-line representation of a one-loop diagram  
for gluon propagator]   
{
Double-line representation of a one-loop diagram for the 
gluon propagator. The sum over the $N_c$ indices is associated 
with the closed index line. The relative contribution of this diagram is
$\sim g^2 N_c\sim 1$.}
   \label{fi:1loop}
\end{figure}

The diagrams of perturbation theory can now be completely rewritten in
the double-line notation~\cite{Hoo74}.
The simplest one which describes the one-loop correction to the gluon
propagator is depicted in Figure~\ref{fi:1loop}.%
\footnote{Here and in the most figures below 
the arrows of the index lines are omitted for simplicity.}
This diagram involves two three-gluon vertices and a sum over 
the $N_c$ indices which is associated with the closed index line.
Therefore, the relative contribution of this diagram is 
$\sim g^2 N_c$.

In order for the large-$N_c$ limit to be nontrivial, 
the bare coupling constant $g^2$ should satisfy 
\be
g^2\sim\frac 1{N_c} \,.
\label{orderg}
\ee
This dependence on $N_c$
is similar to Eqs.~\rf{orderG} and \rf{uspnos} for the
vector models and is prescribed by the asymptotic-freedom formula
\be
g^2
  =  \frac {24\pi^2}{11N_c\ln \left( \Lambda/\Lambda _{QCD}\right)}
\label{NcAF}
\ee
of the pure SU$(N_c)$ gauge theory.

Thus, the relative
contribution of the diagram of Figure~\ref{fi:1loop} is of order
\be
\hbox{Figure~\protect{\ref{fi:1loop}}}~\sim~g^2 N_c~\sim~1
\ee
in the large-$N_c$ limit.

The double lines of the diagram in Figure~\ref{fi:1loop} can be viewed
as bounding a piece of a plane. Therefore, these lines represent
a two-dimensional object rather than a one-dimensional one as the single
lines do in vector models. These double-line graphs are often called
in mathematics the {\it ribbon\/} graphs or {\it fatgraphs}.
We shall see below their connection with Riemann surfaces.

\subsubsection*{Remark on the U$(N_c)$ gauge group}

As is said above, the second term in the parentheses
on the RHS of \eq{completeness} can be omitted at large $N_c$.
Such a completeness condition emerges for the U$(N_c)$ group
whose generators $T^A$ ($A=1,\ldots,N_c^2$) are
\be
T^A=\left(t^a,\case{{\rm I}}{\sqrt{2N}} \right) ,~~~~
\tr T^A T^B=\frac12 \delta^{AB}\,.
\ee
They obey the completeness condition 
\be
\sum\limits_{A=1}^{N_c^2}
\left( T^A\right) ^{ij}\left( T^A\right) ^{kl}  =  \frac12
\delta ^{il}\delta^{kj} ~~~~
\fbox{for \ U$(N_c)$}~.
\label{completenessU}
\ee
The point is that elements of both the SU$(N_c)$ group and the U$(N_c)$
group can be represented in the form
$
U=\exp{iB}
$, 
where $B$ is a general Hermitean matrix for U$(N_c)$ and 
a traceless Hermitean matrix for SU$(N_c)$.

Therefore, the double-line representation of 
perturbation-theory diagrams which is described in this Section holds,
strictly speaking, only for the U$(N_c)$ gauge group. However, 
the large-$N_c$ limit of both the U$(N_c)$ group and the SU$(N_c)$ group is 
the same.

\subsection{Planar and non-planar graphs}\label{ss:b.}

The double-line representation of perturbation theory diagrams 
in the index space is very convenient to estimate their orders in $1/N_c$.
Each gluon propagator contributes a factor of $g^2$
Each three- or four-gluon vertex contributes a factor of $g^{-2}$.
Each closed index line contributes a factor of $N_c$. 
The order of $g$ in $1/N_c$ is given by \eq{orderg}.

\begin{figure}
\vspace*{3mm}
\centering{
\input{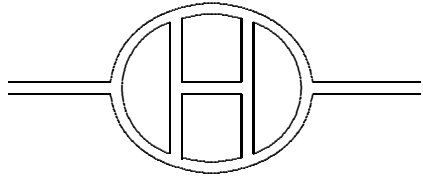}
}
\caption[Double-line representation of a four-loop diagram] 
{
Double-line representation of a four-loop diagram for the 
gluon propagator. The sum over the $N_c$ indices is associated 
with each of the four closed index lines whose number is equal
to the number of loops. The contribution of this diagram is
$\sim g^8 N_c^4\sim 1$.}
   \label{fi:4loop}
\end{figure}

Let us consider a typical diagram for the gluon propagator
depicted in Figure~\ref{fi:4loop}.
It has eight three-gluon vertices and four closed index lines
which coincides with the number of loops.
Therefore, the relative order of this diagram in $1/N_c$ is
\be
\hbox{Figure~\protect{\ref{fi:4loop}}}\sim\left(g^2 N_c\right)^4~\sim~1\,.
\ee

The diagrams of the type in Figure~\ref{fi:4loop}, which can be 
drawn on a sheet of a paper without crossing any lines, are called 
the {\it planar\/} diagrams. For such diagrams, an adding of a loop
inevitably results in adding of two three-gluon (or one four-gluon)
vertex. A planar diagram with $n_2$ loops has $n_2$ closed index lines.
It is of order
\be
n_2\hbox{-loop planar diagram}\sim\left(g^2 N_c\right)^{n_2}\sim 1 \,,
\ee
so that all planar diagrams survive in the large-$N_c$ limit.   

\begin{figure}
\vspace*{3mm}
\centering{
\input{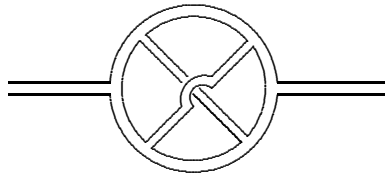}
}
\caption[Double-line representation of a non-planar diagram]   
{
Double-line representation of a three-loop non-planar diagram for the 
gluon propagator. The diagram has six three-gluon vertices but
only one closed index line (while three loops!).
The order of this diagram is $\sim g^6 N_c\sim 1/N_c^2$.}
   \label{fi:genus1}
\end{figure}

Let us now consider a non-planar diagram of the type depicted
in Figure~\ref{fi:genus1}.
This diagram is a three-loop one and has six three-gluon vertices.
The crossing of the two lines in the middle does not correspond
to a four-gluon vertex and is merely due to the fact that the 
diagram cannot be drawn on a sheet of a paper without crossing the lines. 
The diagram has only one closed index line.
The relative order of this diagram in $1/N_c$ is
\be
\hbox{Figure~\protect{\ref{fi:genus1}}}\sim g^6 N_c \sim\frac 1{N_c^2}\,.
\ee
It is therefore suppressed at large $N_c$ by $1/N_c^2$.

The non-planar diagram in Figure~\ref{fi:genus1} can be drawn without
line-crossing on a surface with one handle which is usually called
in mathematics a torus or the surface of genus one. A plane is
then equivalent to a sphere and has genus zero. Adding a handle to
a surface produces a {\em hole\/} according to mathematical terminology.
A general Riemann surface with $h$ holes has genus $h$. 

The above evaluations of the order of the diagrams in 
Figures~\ref{fi:1loop}--\ref{fi:genus1} can now be described by the unique
formula
\be
\hbox{genus-$h$ diagram}\sim
\left(\frac 1{N_c^2}\right)^{{\rm genus}} \,.
\label{genusexpansion}
\ee
Thus, the expansion in $1/N_c$ rearranges perturbation-theory diagrams
according to their topology~\cite{Hoo74}. For this reason, it is referred to 
as the {\it topological expansion\/} or the {\it genus expansion}.
The general proof of \eq{genusexpansion} for an arbitrary diagram 
is given in Subsection~\ref{ss:t.e.}.

Only planar diagrams, which are associated with genus zero, survive
in the large-$N_c$ limit. This class of diagrams is an analog of
the bubble graphs in the vector models. However, the problem of summing the 
planar graphs is much more complicated than that of summing 
the bubble graphs. Nevertheless, it is simpler than the problem of
summing all the graphs, since the number of the planar graphs with
$n_0$ vertices grows at large $n_0$ exponentially~\cite{Tut62,KNN77}
\be
\#_{\rm p}\left(n_0\right)\equiv\#~
\hbox{of planar graphs}\sim {\rm const}^{n_0}\,,
\label{noplanar}
\ee
while the number of all the graphs grows with $n_0$ factorially.
There is no dependence in \eq{noplanar} on the number of external
lines of a planar graph which is assumed to be much less than $n_0$.

It is instructive to see the difference between the planar diagrams
and, for instance, the ladder diagrams which describe $e^+e^-$ 
elastic scattering in QED. Let the ladder has $n$ rungs. Then there are
$n!$ ladder diagrams, but only one of them is planar. This simple
example shows why the number of planar graphs is much smaller than
the number of all graphs, most of which are non-planar.

We shall
discuss in these Lectures what is known about solving the problem of summing
the planar graphs.

Equation~\rf{genusexpansion} holds, strictly speaking, only for the
relative order while the contribution of tree diagrams to
a connected $n$-point Green's function is $\sim (g^2)^{n-1}$ which is its
natural order in $1/N_c$. 
In order to make contributions of all planar diagrams to be of the same
order $\sim 1$ in the large-$N_c$ limit, independently of the number of
external lines, it is convenient to contract 
the Kronecker deltas associated with external lines.
 
\begin{figure}
\vspace*{3mm}
\centering{
\input{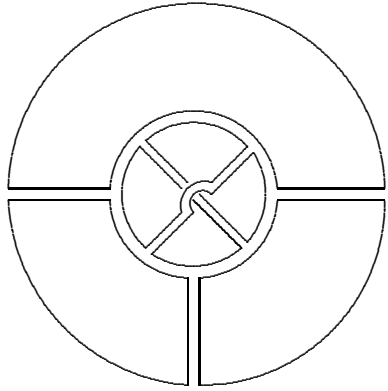}
}
\caption[Generic double-line index diagram]   
{
Generic index diagram with $n_0=10$ vertices, $n_1=10$ gluon
propagators, $n_2=4$ closed index lines, and $B=1$ boundary.
The color indices of the external lines are contracted 
by the Kronecker deltas (represented by the single lines) in a cyclic order.
The extra factor of $1/N_c$ is due to the normalization~\rf{trnormalization}.
Its order in $1/N_c$ is $\sim 1/N^2_c$ in accord with \eq{genusexpansion}.}
   \label{fi:generic}
\end{figure}

Let us do this in a cyclic order as is depicted in 
Figure~\ref{fi:generic} 
for a generic connected diagram with three external gluon lines.
The extra deltas which are added to contract the color indices are
depicted by the single lines. They can be viewed as a {\em boundary}\/ of 
the given diagram. The actual size of the boundary is not essential ---
it can be shrunk to a point. Then a bounded piece of a plane will be
topologically equivalent to a sphere with a puncture.
I shall prefer to draw planar diagrams in a plane with an extended
boundary (boundaries) rather than in a sphere with a puncture (punctures). 

It is clear from the graphic representation that the diagram in 
Figure~\ref{fi:generic} is associated with the trace over the color 
indices of the three-point Green's function
\be
G^{(3)}_{\mu_1\mu_2\mu_3}\left(x_1,x_2,x_3 \right) \equiv 
\frac {1}{N_c} \LA \tr \left[A_{\mu_1}\left(x_1 \right)
A_{\mu_2}\left(x_2 \right)A_{\mu_3}\left(x_3 \right) \right]\RA.
\label{trnormalization}
\ee
We have introduced here the factor $1/N_c$ to make $G_3$ of
${\cal O}(1)$ in the large-$N_c$ limit.
Therefore, the contribution
of the diagram in Figure~\ref{fi:generic} having one boundary
should be divided by $N_c$.

The extension of \eq{trnormalization} to multi-point Green's functions
is obvious:
\be
G^{(n)}_{\mu_1\cdots\mu_n}
\left(x_1,\ldots,x_n \right) \equiv \frac {1}{N_c} 
\LA\tr \left[A_{\mu_1}\left(x_1 \right) \cdots 
A_{\mu_n}\left(x_n \right)\right] \RA.
\label{multitrnormalization}
\ee
The factor $1/N_c$, which normalizes the trace, 
provides the natural normalization
\be
G^{(0)}=1
\ee 
of the averages.

\begin{figure}
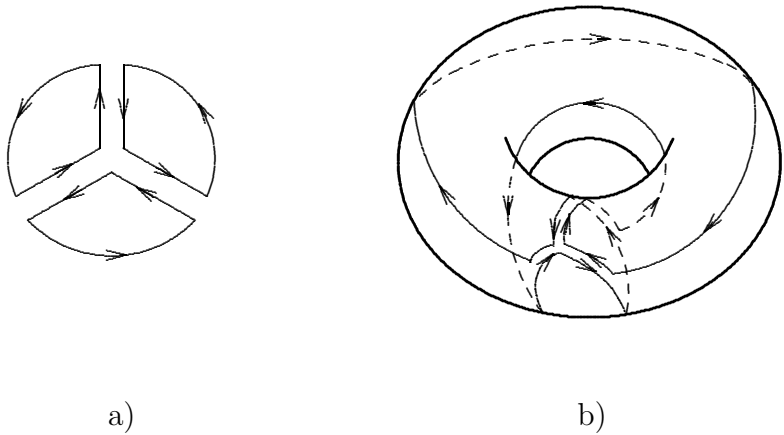

\vspace*{3mm}
\centering{
\input{mersed}
\hspace{2cm}
\input{torus}
}
\caption[Planar and non-planar parts of three-gluon vertex]   
{
Planar a) and non-planar b) contributions of the two color structures 
in \eq{threegluon} for three-gluon vertex to $G^{(3)}$ in the
lowest order of perturbation theory.}
   \label{fi:3vtorus}
\end{figure}

Though the two terms in the index-space representation~\rf{threegluon} 
of the three-gluon vertex 
look very similar, their fate in the topological expansion is quite
different. When the color indices are contracted anti-clockwise,
the first term leads to the planar contributions to $G^{(3)}$, the simplest
of which is depicted in Figure~\ref{fi:3vtorus}a. 
The anti-clockwise
contraction of the color indices in the second term leads to a non-planar
graph in Figure~\ref{fi:3vtorus}b which can be drawn without crossing
of lines only on a torus. Therefore, the two color structures of the 
three-gluon vertex contribute to different orders of the topological
expansion. The same is true for the four-gluon vertex.

\subsubsection*{Remark on oriented Riemann surfaces}

Each line of an index graph of the type depicted in Figure~\ref{fi:generic}
is oriented. This orientation continues along a closed index line while
the pairs of index lines of each double line have opposite orientations.
The overall orientation of the lines is prescribed by the orientation
of the external boundary which we choose to be, say, anti-clockwise
like in Figure~\ref{fi:3vtorus}a.
Since the lines are oriented, the faces of the Riemann surface associated
with a given graph are oriented too --- all in the same way 
--- anti-clockwise. 
Vice versa, such an orientation of the Riemann surfaces unambiguously
fixes the orientation of all the index lines. This is the reason why we
omit the arrows associated with the orientation of the index lines:
their directions are obvious.

\subsubsection*{Remark on cyclic-ordered Green's functions}

The cyclic-ordered Green's functions~\rf{multitrnormalization}
naturally arise in the expansion of the trace of the 
path-ordered non-Abelian
phase factor for a closed contour. One gets \label{p:c.o.}
\bea
\lefteqn{\LA\frac1{N_c}
\tr{}{\bf P}\e^{i\oint_\Gamma d x ^\mu A_\mu \left( x \right) } \RA} \non 
&=& \sum\limits_{n=0}^\infty\;i^n 
\oint\limits_\Gamma dx_1^{\mu_1}
\int\limits_{x_1}^{x_1} dx_2^{\mu_2} \ldots \!
\int\limits_{x_1}^{x_{n-1}} \!\!dx_n^{\mu_n}\;
G^{(n)}_{\mu_1\cdots\mu_n} \left(x_1,\ldots,x_n \right).
\hspace{1cm}
\label{Poexpansion}
\eea
The reason is because the ordering along a closed path implies
the cyclic-ordering in the index space. 

\subsubsection*{Remark on generating functionals for planar graphs}

\begin{figure}
\vspace*{3mm}
\centering{
\input{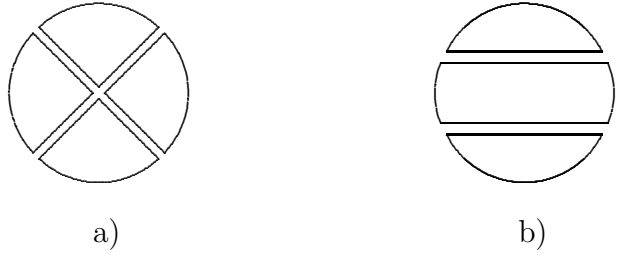}
}
\caption[Connected and disconnected planar graphs]   
{
Example of a) connected and b) disconnected planar graphs.}
   \label{fi:plcon}
\end{figure}

By connected or disconnected planar graphs we mean, respectively,  
the graphs which were connected or disconnected 
before the contraction of the color indices as is illustrated
by Figure~\ref{fi:plcon}.
The graph in Figure~\ref{fi:plcon}a is connected planar while the graph
in Figure~\ref{fi:plcon}b is disconnected planar.

The usual exponential relation between the generating 
functionals $W\left[ J \right]$ and $Z\left[ J \right]$ for
connected graphs and all graphs,
does not hold for the planar graphs.
The reason is that an exponentiation of such a connected planar diagram 
for the cyclic-ordered Green's functions~\rf{multitrnormalization} 
can give disconnected non-planar ones.

The generating functionals for all and connected planar graphs can
be constructed~\cite{Cvi81} by means of introducing non-commutative sources
$\j_\mu(x)$. ``Non-commutative'' means that there is no way to 
transform $\j_{\mu_1}(x_1)\j_{\mu_2}(x_2)$ into
$\j_{\mu_2}(x_2)\j_{\mu_1}(x_1)$. This non-commutativity of the sources
reflects the cyclic-ordered structure of the Green's 
functions~\rf{multitrnormalization} which possess only cyclic symmetry.

Using the short-hand notations 
\be
\j \circ \A\equiv\sum_\mu \int d^d x \;\j_\mu(x) \A_\mu(x)\,,
\ee
where the symbol $\circ$ includes the sum over the $d$-vector
(or whatever available) indices except for the color ones, 
we write down the definitions of the generating functionals for all
planar and connected planar graphs, respectively, as
\be
Z\left[\j \right]\equiv\sum_{n=0}^\infty \LA \frac1{N_c}\tr\left(
\j \circ \A \right)^n \RA
\label{planarall}
\ee
and
\be
W\left[\j \right]\equiv\sum_{n=0}^\infty \LA \frac1{N_c}\tr\left(
\j \circ \A \right)^n \RA_{\rm conn}\,.
\label{planarconnected}
\ee

The planar contribution to the
Green's functions~\rf{multitrnormalization} 
and their connected counterparts can be obtained, respectively, from the
generating functionals $Z\left[\j \right]$ and $W\left[\j \right]$
 applying the non-commutative derivative which is defined by
\be
\frac{\delta}{\delta \j_\mu(x)} \j_\nu(y)\,f\left(\j \right)
=\delta_{\mu\nu} \delta^{(d)}\left(x-y\right) f\left(\j \right)\,,
\ee
where $f$ is an arbitrary function of $\j$'s. In other words the derivative
picks up only the leftmost variable.

\begin{figure}
\vspace*{3mm}
\centering{
\hspace*{0cm}\epsfbox{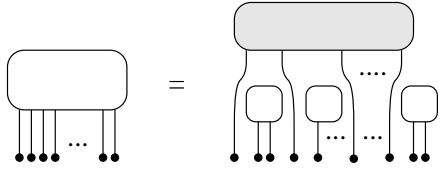}
 }
\caption[Graphic derivation of \protect{\eq{plconndisconn}}]   
{
Graphic derivation of \eq{plconndisconn}: 
$Z\left[\j \right]$ is denoted by an empty box, 
$W\left[\j \right]$ is denoted by a shadow box, $\j$ is denoted by 
a filled circle.
By picking a leftmost external line of a planar graph,
we end up with a connected planar graph,
whose remaining external lines are somewhere to the right 
interspersed by disconnected planar graphs. It is evident that
$\j Z\left[\j \right]$ plays the role of a new source for the 
connected planar graph. If we instead pick up the rightmost external line,
we get the inverse order $ Z\left[\j \right]\j$, which results
in \eq{cvicy}.}
   \label{fi:cvi}
\end{figure}

The relation which replaces the usual one for planar graphs 
is
\be
Z\left[\j \right]=W\left[\j Z\left[\j \right]\right],
\label{plconndisconn}
\ee
while the cyclic symmetry says
\be
W\left[\j Z\left[\j \right]\right]=W\left[ Z\left[\j \right]\j\right].
\label{cvicy}
\ee
A graphic derivation of Eqs.~\rf{plconndisconn} and \rf{cvicy}
is given in Figure~\ref{fi:cvi}.
In other words, given $W\left[\j \right]$, one should construct
an inverse function as the solution to the equation
\be
\j_\mu(x)=\J_\mu(x) W\left[\j \right],
\label{invfunct}
\ee
after which \eq{plconndisconn} says
\be
Z\left[\j \right]=W\left[\J \right].
\ee
More about this approach to the generating functionals for planar graphs
can be found in Ref.~\cite{CLS82}.

An iterative solution to
 \eq{invfunct} for the Gaussian case can be easily found.
In the Gaussian case, only $G^{(2)}$ is nonvanishing which yields
\be
W\left[\j \right]=1-g^2 \j\circ D \circ \j \,,
\ee
where the propagator $D$ is given by \eq{D(x-y)}.
Using \eq{plconndisconn}, we get explicitly
\be
Z\left[\j \right]=1-g^2 \int d^d x \,d^d y\, D_{\mu\nu}(x-y)\,
\j_\mu(x)\, Z\left[\j \right] \,\j_\nu(y) \,Z\left[\j \right].
\label{ccvi}
\ee
While this equation for $Z\left[\j \right]$ is quadratic, 
its solution can be written 
only as a continued fraction due to the non-commutative nature of 
the variables. In order to find it, we rewrite \eq{ccvi} as
\be
Z\left[\j \right]=\frac{1}{1+g^2 \int d^d x \,d^d y\, D_{\mu\nu}(x-y)\,
\j_\mu(x)\, Z\left[\j \right] \,\j_\nu(y)} \,,
\label{iccvi}
\ee
whose iterative solution reads~\cite{Cvi81}
\be
Z\left[\j \right]=\frac{1}{\displaystyle 1+g^2\j\frac{\circ D \circ}
{\displaystyle 1+g^2\j\frac{\circ D \circ}
{\displaystyle 1+g^2\j\frac{\circ D \circ}{\vdots}\j}\j}\j}\,.
\ee

\subsection{Topological expansion and quark loops}\label{ss:t.e.}

It is easy to incorporate quarks in the topological expansion.
A quark field belongs to the fundamental representation of the 
gauge group SU$(N_c)$ and its propagator is represented by 
a single line
\be
\left\langle  \psi _i\bar\psi _j\right\rangle  \propto  \delta _{ij} ~=
\unitlength=1mm
\linethickness{0.6pt}
\begin{picture}(31.00,4.00)(10,10)
\put(20.00,11.00){\vector(1,0){6.00}}
\put(26.00,11.00){\line(1,0){4.00}}
\put(18.00,11.00){\makebox(0,0)[cc]{$i$}}
\put(32.00,11.00){\makebox(0,0)[cc]{$j$}}
\end{picture}.
\label{singleline}
\ee
The arrow indicates, as usual, the direction of propagation of
a (complex) field $\psi$. We shall omit these arrows for simplicity.

\begin{figure}
\vspace*{3mm}
\centering{
\input{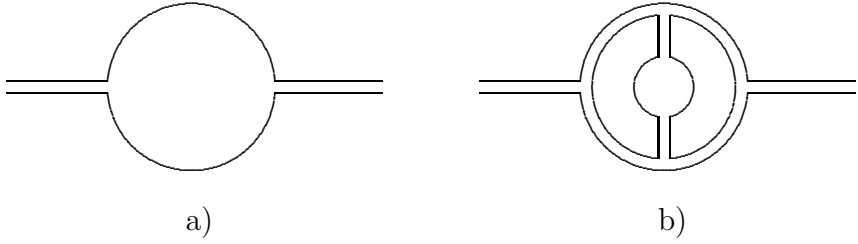}
}
\caption[Diagrams for gluon propagator which involve quark loop]   
{
Diagrams for the gluon propagator with a quark loop which is represented
by the single lines.
The diagram a) involves one quark loop and has no closed index lines 
so that its order is $\sim g^2\sim 1/N_c$. 
The diagram b) involves three loops one of which is a quark loop.  
Its relative order is $\sim g^6 N_c^2\sim 1/N_c$.}
   \label{fi:quarkl}
\end{figure}

The diagram for the gluon propagator which involves one quark loop
is depicted in Figure~\ref{fi:quarkl}a.
It has two three gluon vertices and no closed index lines so that its
order in $1/N_c$ is
\be
\hbox{Figure~\protect{\ref{fi:quarkl}}a}~\sim~g^2~\sim~\frac{1}{N_c} \,.
\ee
Analogously, the relative order of a more complicated tree-loop diagram in 
Figure~\ref{fi:quarkl}b, which involves one quark loop and two closed
index lines, is
\be
\hbox{Figure~\protect{\ref{fi:quarkl}}b}~\sim~g^6 N_c^2~\sim~\frac{1}{N_c} \,.
\ee

It is evident from this consideration that quark loops are not
accompanied by closed index lines. One should add a closed index line
for each quark loop in order for a given diagram with $L$ quark
loops to have the same double-line representation as for pure gluon diagrams.
Therefore, given \eq{genusexpansion}, diagrams with $L$ 
quark loops are suppressed at large $N_c$ by 
\be
L~\hbox{quark loops}~\sim~\left(\frac1{N_c}\right)^{L+2\cdot{\rm genus}}\,.
\label{Lqloops}
\ee

The single-line representation of the quark loops is similar to the one of 
the external boundary in Figure~\ref{fi:generic}. Moreover, such a diagram
emerges when one calculates perturbative gluon corrections to the
vacuum expectation value of the quark operator
\be
O_\Gamma=\frac 1{N_c} \bar \psi \Gamma \psi \,,
\label{OGamma}
\ee
where $\Gamma$ stands for one of the combinations of the gamma-matrices:
\be
\Gamma~=~{\rm I},\;\gamma_5,\;\gamma_\mu,\;i\gamma_\mu \gamma_5,\;
\Sigma_{\mu\nu}=\frac{1}{2i}[\gamma_\mu,\gamma_\nu]\;,\ldots\;.
\ee
The factor of $1/N_c$ is introduced in~\rf{OGamma}
to make it ${\cal O}(1)$ in 
the large-$N_c$ limit. Therefore, the external 
boundary can be viewed as a single line associated with valence quarks.
The difference between virtual quark loops and external boundaries is
that each of the latter comes along with the factor of $1/N_c$
due to the definitions~\rf{multitrnormalization} and \rf{OGamma}.  

In order to prove Eqs.~\rf{genusexpansion} and its quark 
counterpart~\rf{Lqloops},
let us consider a generic diagram in the index space which has
$n_0^{(3)}$ three-point vertices (either three-gluon or quark-gluon ones),
$n_0^{(4)}$ four-gluon vertices, $n_1$ propagators 
(either gluon or quark ones), 
$n_2$ closed index lines, $L$ virtual quark loops and $B$ external boundaries. 
Its order in $1/N_c$ is 
\be
\frac1{N_c^B}(g^2)^{n_1-n_0^{(3)}-n_0^{(4)}} N_c^{n_2}~\sim~
N_c^{n_2 -n_1+ n_0 -B}
\label{ordergeneric}
\ee
where the total number of vertices
$
n_0=n_0^{(3)}+ n_0^{(4)}
$
is introduced.
The extra factor of $1/N_c^B$ is due to 
the extra normalization factor of $1/N_c$ in operators associated with
external boundaries.

The exponent on the RHS of \eq{ordergeneric} can be expressed via the 
Euler characteristics $\chi$ of a given graph of genus $h$.
Let us first mention that a proper Riemann surface, which is
associated with a given graph, is open and has $B+L$ boundaries
(represented by single lines). This surface can be closed by attaching
a cap to each boundary. The single lines then become double lines
together with the lines of the boundary of each cap. We have already 
considered this procedure when deducing \eq{Lqloops} from \eq{genusexpansion}.

The number of faces for a
closed Riemann surface constructed in such a manner is \mbox{$n_2+L+B$}, 
while the number of edges and vertices are $n_1$ and $n_0$, respectively.
Euler's theorem says that
\be
\chi\equiv 2-2h=n_2+L+B-n_1+n_0\,.
\label{Euler}
\ee
Therefore the RHS of \eq{ordergeneric} can be rewritten as
\be
N_c^{n_2-n_1+n_0-B}=N_c^{2-2h-L-2B} \,.
\ee

We have thus proven that the order in $1/N_c$ of a generic graph
does not depend on its order in the coupling constant and is 
completely expressed via the genus $h$ and the number 
of virtual quark loops $L$ and external boundaries $B$ by
\be
\hbox{generic graph}~\sim~\left(\frac1{N_c}\right)^{2h+L+2(B-1)} \,.
\label{ordergenericee}
\ee 
For $B=1$, we recover Eqs.~\rf{genusexpansion} and \rf{Lqloops}.

\subsubsection*{Remark on the order of gauge action}

We see from \eq{multitrnormalization} that the natural variables for the 
large-$N_c$ limit are the 
matrices $A_\mu$ which include the 
factor of $g$ (see \eq{calA}).
In these variables, 
the action~\rf{QCDaction} is ${\cal O}(N_c^2)$ at large $N_c$, 
since $g^2$ is $\sim 1/N_c$ and the trace is 
$\sim N_c$.

This result can be anticipated from the free theory because the kinetic
part of the action involves the sum over $N_c^2-1$ free gluons.
Therefore, the non-Abelian field strength~\rf{fieldstrenght} is 
$\sim 1$ for $g^2\sim 1/N_c$.  

The fact that the action is ${\cal O}(N^2_c)$ in the large-$N_c$ limit
is a generic property of the models describing matrix fields.
It will be crucial for developing saddle-point approaches
at large $N_c$ which are considered  below.

\subsubsection*{Remark on phenomenology of multicolor QCD}

While $N_c=3$ in the real world, there are phenomenological indications that
$1/N_c$ may be considered as a small parameter. We have already mentioned
some of them in the text --- the simplest one is that the ratio of
the $\rho$-meson width to its mass, which is $\sim 1/N_c$, is small.
Considering $1/N_c$ as a small parameter immediately leads to qualitative
phenomenological consequences which are preserved by the planar diagrams
associated with multicolor QCD, but are violated by the
non-planar diagrams.

The most important consequence is the relation of the $1/N_c$-expansion
to the topological expansion in the dual-resonance model of hadrons.
Vast properties of hadrons are explained by the dual-resonance
model. A very clear physical picture behind this model
is that hadrons are excitations of a string with quarks at the ends. 

I shall briefly list some consequences of multicolor QCD:
\begin{itemize}
\item[1)]
The ``naive'' quark model of hadrons emerges at $N_c=\infty$.
Hadrons are built out of (valence or constituent) quark and antiquark 
$q\bar{q}$, while
exotic states like $qq\bar{q}\bar{q}$ do not appear.
\item[2)]
The partial width of decay of the $\phi$-meson, which is built out of
$s\bar{s}$ (the strange quark and antiquark), into $K^+K^-$ is $\sim1/N_c$,
while that into $\pi^+\pi^-\pi^0$ is $\sim 1/N_c^2$. This
explains Zweig's rule. The masses of the $\rho$- and $\omega$-mesons
are degenerate at $N_c=\infty$.
\item[3)]
The coupling constant of \label{m-mcoupl}
meson-meson interaction is small at large 
$N_c$. 
\item[4)]
The widths of glueballs are $\sim1/N_c^2$, \ie they should be even
narrower than mesons built out of quarks. The glueballs do not
interact or mix with mesons at $N_c=\infty$.
\end{itemize}

All these hadron properties (except the last one) approximately
agree with experiment, and were well-known
even before 1974 when multicolor QCD was introduced.
Glueballs are not yet detected experimentally (maybe because of
their property listed in the item~4).

\subsection{Large-$N_c$ factorization}\label{ss:mfac}

The vacuum expectation values of several colorless or white operators,
which are singlets with respect to the gauge group, factorize in
the large-$N_c$ limit of QCD (or other matrix models). This property
is similar to that already discussed in Subsection~\ref{ss:vfactor} for
the vector models.

The simplest gauge-invariant operator in a pure SU$(N_c)$ gauge theory
is the square of the non-Abelian field strength:
\be
O\left( x\right)=
\frac {1}{N_c}\,{\rm tr}\,F_{\mu \nu }^2\left( x\right).
\label{defO}
\ee
The normalizing factor is the same as in 
Eqs.~\rf{trnormalization}, \rf{multitrnormalization}, which 
provides the natural normalization
\be
\left\langle \frac {1}{N_c}\,{\rm tr}\,F_{\mu \nu }^2\left( x\right)
\right\rangle  =  \left\langle \frac {g^2}{2N_c}\,
F_{\mu \nu }^a\left( x\right) F_{\mu \nu }^a\left( x\right) \right\rangle 
 \sim 1\,. 
\label{estimateO}
\ee
The contribution of all planar graphs to the average on the LHS of
\eq{estimateO} is of order 1 in accord with the 
general formula~\rf{ordergenericee} for $B=1$.

In order to verify the factorization in the large-$N_c$ limit,
let us consider the index space diagrams for the average of
the product of
two colorless operators $O\left( x_1\right)$ and $O\left( x_2\right)$
given by~\rf{defO}. It involves a factorized part when
gluons are emitted and
absorbed by the same operators. The contribution of the 
factorized part is or order 1 as above. 

Alternatively, the connected correlator of the two operators
is associated with the 
general formula~\rf{ordergenericee} for two boundaries $B=2$.
Its contribution is suppressed by $1/N_c^2$ in the large-$N_c$ limit. 
For this correlator, at least one
gluon line is emitted and absorbed by different
operators $O\left( x_1\right)$ and $O\left( x_2\right)$.
Notice, that these graphs themselves are planar, while
the suppression comes from the number of boundaries.

This example illustrates the general property that
only (planar) diagrams with gluon lines emitted and absorbed by
the same operators survive as $N_c\ra\infty$. Since correlations
between the colorless operators $O\left( x_1\right)$ and 
$O\left( x_2\right)$ are of order $1/N_c^2$,  the 
{\em factorization}\/ property holds as $N_c\ra\infty$:
\begin{eqnarray}
\lefteqn{\left\langle \frac {1}{N_c}\,{\rm tr}\,F^2\left( x_1\right) 
\frac {1}{N_c}\,{\rm tr}\, F^2\left( x_2\right) \right\rangle }
\non & = & 
\left\langle \frac {1}{N_c}\,{\rm tr}\,
F^2\left( x_1\right) \right\rangle \left\langle \frac {1}{N_c}\,{\rm tr}\,
F^2\left( x_2\right) \right\rangle +{\cal O}\left( \frac 1{N_c^2}\right).
\label{FFFF}
\eea

For a general set of gauge-invariant operators $O_1$, \ldots, $O_n$, the
factorization property can be represented by
\be
\LA \,O_1\cdots O_n\,\RA=\LA\, O_1 \,\RA \cdots \LA\, O_n \,\RA
+{\cal O}\left( \frac 1{N_c^2}\right) \,.
\label{mfactorization}
\ee
This is analogous to \eq{vfactorization} for the vector models.

The factorization in large-$N_c$ QCD was first discovered by
A.A.~Migdal in the late seventies. An important observation that
the factorization implies a semiclassical nature of the
large-$N_c$ limit of QCD was done by Witten~\cite{Wit79}. We shall
discuss this in the next two Subsections.

The factorization property also holds for gauge-invariant
operators constructed from quarks like in \eq{OGamma}
as a consequence of \eq{ordergenericee}. 

\subsubsection*{Remark on factorization beyond perturbation theory}

The large-$N_c$ factorization~\rf{mfactorization}
has been shown above to all orders of perturbation theory.
It can be also verified at all
orders of the strong coupling expansion in the SU$(N_c)$ lattice
gauge theory. A non-perturbative proof 
of the factorization will be given in the next
Section by using quantum equations of motion (the loop equations). 

\subsection{The master field} \label{ss:m.f.} 

The large-$N_c$ factorization in QCD assumes that gauge-invariant
objects behave as $c$-numbers, rather than as operators.
Likewise the vector models, this suggests that the path integral 
is dominated by a saddle point.

We already saw in Subsection~\ref{ss:vfactor} that the factorization
in the vector models does not mean that the fundamental field itself, 
for instance $\vec{n}$ in the sigma-model, becomes ``classical''. 
It is the case, instead, for a singlet composite field.

We are now going to apply a similar idea to the Yang--Mills theory
whose partition function reads
\be
Z =  \int DA_\mu ^a\e^{-S} \,.
\label{YMpart}
\ee
The action, $\sim N_c^2$, is large as $N_c\ra\infty$, but the ``entropy'' 
is also $\sim N_c^2$ due to the $N_c^2-1$ integrations over $A_\mu^a$:
\be
 DA_\mu ^a  \sim  \e^{N_c^2} \,.
\ee
Consequently, the saddle-point equation of the large-$N_c$ Yang--Mills
theory is {\em not}\/ the classical one. 

The idea is to rewrite the path integral over $A_\mu$ 
for the Yang--Mills theory
as that over a colorless composite field $\Phi\left[A\right]$,
likewise it was done in Subsection~\ref{ss:n.s.m.} for the sigma-model.
The expected new path-integral representation of the partition
function~\rf{YMpart} would be something like
\be
Z  \propto  \int D\Phi \,\frac 1{ 
\frac{\partial \Phi \left[ A\right] }{\partial
A_\mu ^a}}\e^{-N_c^2S\left[ \Phi \right] } \,.
\label{YMpartp}
\ee
The Jacobian 
\be
\frac{\partial \Phi \left[ A\right] }{\partial A_\mu ^a}  \equiv  
\e^{-N_c^2J\left[ \Phi \right] } 
\label{YMJac}
\ee
in \eq{YMpartp} is related to the old entropy factor, so
that
$ 
J\left[ \Phi \right]\sim 1
$
in the large-$N_c$ limit.

The original partition function~\rf{YMpart} can be then rewritten as
\be
Z  \propto  \int D\Phi \e^{N_c^2J\left[ \Phi \right] -N_c^2S\left[ \Phi
\right] }, 
\label{newYMpart}
\ee
where $S\left[ \Phi \right]$ represents the Yang--Mills action in the
new variables. The new ``entropy'' factor $D\Phi$ is ${\cal O}(1)$ because
the variable $\Phi\left[ A \right] $ is a color singlet.
The large parameter $N_c$ enters \eq{newYMpart} only in the exponent.
Therefore, the saddle-point equation can be immediately written:
\be
\frac{\delta S}{\delta \Phi } = \frac{\delta J}{\delta \Phi } \,.
\label{YMsaddlepoint}
\ee

Remembering that $\Phi$ is a functional of $A_\mu$:
$
\Phi  \equiv  \Phi \left[ A\right], 
$
we rewrite the saddle-point equation~\rf{YMsaddlepoint} as 
\be
\frac{\delta S}{\delta A_\nu ^a}  =  
\left(\nabla _\mu F_{\mu \nu }\right)^a 
 =  \frac{\delta J}{\delta A_\nu ^a}\,. 
\label{master}
\ee
It differs from the classical Yang--Mills equation 
by the term on the RHS coming from the Jacobian~\rf{YMJac}.

Given $J\left[ \Phi \right]$ which depends on the precise
from of the variable $\Phi\left[ A \right]$, \eq{master} 
has a solution
\be
A_\mu(x)= A_\mu^{\rm cl}(x)\,.
\label{Acl}
\ee 
Let us first assume that there exists only one solution to \eq{master}.
Then the path integral is saturated by a single configuration~\rf{Acl},
so that the vacuum expectation values of gauge-invariant operators
are given by their values at this configuration:
\be
\LA\, O \, \RA=O\left( A_\mu^{\rm cl}(x) \right).
\label{atmaster}
\ee
The factorization property~\rf{mfactorization} will obviously be satisfied.

An existence of such a classical field configuration in multicolor QCD
was conjectured by Witten~\cite{Wit79}. It was discussed in the
lectures by Coleman~\cite{Col79} who called it the {\em master field}.
Equation~\rf{master} which determines the master field is often
referred to as the master-field equation.

A subtle point with the master field is that a solution to \eq{master}
is determined only up to a gauge transformation. To preserve gauge
invariance, it is more reasonable to speak about the whole
gauge orbit as a solution of \eq{master}. However, this will not change
\eq{atmaster} since the operator $O$ is gauge invariant.

The conjecture about an existence of the master field has surprisingly
rich consequences. Since vacuum expectation values are Poincar{\'e}
invariant, the RHS of \eq{atmaster} does. This implies that 
$A_\mu^{\rm cl}(x) $ must itself be Poincar{\'e} invariant up to a
gauge transformation: a change of $A_\mu^{\rm cl}(x) $ under 
translations or rotations can be compensated by a gauge transformation.
Moreover, there must exist a gauge in which $A_\mu^{\rm cl}(x) $ is
space-time independent: 
$
A_\mu^{\rm cl}(x)=A_\mu^{\rm cl}(0)\,.  
\label{Pindependent}
$
In this gauge, rotations must be equivalent to a global gauge 
transformation, so that $A_\mu^{\rm cl}(0)$ transforms as a 
Lorentz vector.

In fact, the idea about such a master field in multicolor QCD may be
incorrect as was pointed out by Haan~\cite{Haa81}. The conjecture
about an existence of only one solution to the master-field 
equation~\rf{master} seems to too strong. If several solutions
exists, one needs an additional averaging over these solutions.
This is a very delicate matter, since this additional averaging 
must still preserve the factorization property. One might better think
about this situation as if $A_\mu^{\rm cl}(0)$ would be an operator
in some Hilbert space rather than a $c$-valued function. 
Such an operator-valued master field is sometimes called the master
field in the {\em weak}\/ sense, while the above conjecture about
a single classical configuration of the gauge field, which saturates
the path integral, is called the master field in the {\em strong}\/ sense.

The concept of the master field is rather vague until a precise form
of the composite field $\Phi\left[A\right]$, and consequently 
the Jacobian $\Phi\left[A\right]$ that enters \eq{master}, is not defined. 
However, what is important is that the master field (in the weak sense)
is space-time independent. This looks like a simplification of
the problem of solving large-$N_c$ QCD. A Hilbert space, in which the 
operator $A_\mu^{\rm cl}(0)$ acts, should be specified by 
$\Phi\left[A\right]$. We shall consider in the next Subsection 
a realization of these ideas for the case
of $\Phi\left[A\right]$ given by the trace of the non-Abelian phase
factor for closed contours.
 
\subsubsection*{Remark on non-commutative probability theory}

An adequate mathematical language for describing the master
field in multicolor QCD (and, generically, in matrix models at large $N_c$)
was found by I.~Singer in 1994. It is based on the concept of free
random variables of non-commutative probability theory, introduced
by Voiculescu~\cite{VDN92}. How to describe the master field in this
language and some other applications of  
non-commutative free random variables to the problems of planar
quantum field theory are discussed in Refs.~\cite{Dou95,GG95}.

\subsection{$1/N_c$ as semiclassical expansion}\label{ss:semic}

A natural candidate for the composite operator $\Phi\left[A\right]$ 
from the previous Subsection is given by the trace of the non-Abelian phase
factor for closed contours --- the Wilson loop. It is labeled 
by the loop $C$ in the same sense as the field
$A_\mu \left( x \right)$ is labeled by the point $x$, 
so we shall use the notation
\be
\Phi\left( C \right) \equiv \Phi\left[A\right]
= \frac{1}{N_c} \tr {{\bf P}}\e^{i \oint_C dx^\mu A_\mu\left( x \right)}.
\label{closedWloop}
\ee

Nobody up to now managed to reformulate QCD at finite $N_c$ in terms of
$\Phi\left( C \right)$ in the language of path integral.
This is due to the fact that self-intersecting loops are not
independent (they are related by the so-called Mandelstam 
relations~\cite{Man79}), and the Jacobian is huge. 
The reformulation was done~\cite{MM79} 
in the language of Schwinger--Dyson or loop equations which will be 
described in the next Section.

Schwinger--Dyson equations are a convenient way of performing the 
semiclassical expansion, which is an alternative to the path integral.
Let us illustrate an idea how to do this by an example of the $\varphi^3$
theory.
The RHS of the Schwinger--Dyson equations
 is proportional to the Planck's constant
$\hbar$.
In the semiclassical limit $\hbar\ra0$, we get 
\be
\left( -\partial _1^2+m^2\right) \left\langle \varphi \left( x_1\right)
\ldots \varphi \left( x_n\right) \right\rangle +\frac \lambda 2\left\langle
\varphi ^2\left( x_1\right) \ldots \varphi \left( x_n\right) \right\rangle
 =  0 \,, 
\ee
whose solution is of the factorized form
\be
\left\langle \varphi \left( x_1\right) \ldots \varphi \left( x_n\right)
\right\rangle  =  \left\langle \varphi \left( x_1\right) \right\rangle
\ldots \left\langle \varphi \left( x_n\right) \right\rangle +
{\cal O}\left( \hbar\right) 
\ee
provided that 
\be
\left\langle \varphi \left( x\right) \right\rangle  \equiv  
\varphi _{\rm cl}\left(x\right) 
\ee
obeys 
\be
\left( -\partial ^2+m^2\right) \varphi _{\rm cl}\left( x\right) +\frac \lambda
2\varphi _{\rm cl}^2\left( x\right)  =  0 \,.
\label{classphi3}
\ee

Equation~\rf{classphi3} is nothing but the classical equation of 
motion for the $\varphi^3$ theory, which specifies extrema of the
action entering the path integral.
Thus, we have reproduced, using the Schwinger--Dyson equations,
the well-known fact that the path integral is dominated by
a classical solution as $\hbar\ra0$. It is also clear how to
perform the semiclassical expansion in $\hbar$ in the language
of the Schwinger--Dyson equations: one should solve them by
iterations.

The reformulation of multicolor QCD in terms of 
the loop functionals $\Phi\left( C \right) $ is, in a sense,
a realization of the idea of the master field in the weak sense,
when the master field acts as an operator in the space of loops. 

\subsubsection*{Remark on the large-$N_c$ limit as statistical averaging}

There is yet
another, pure statistical, explanation why the large-$N_c$
limit is a ``semiclassical'' limit for the collective variables
$\Phi\left( C \right) $. The matrix $U^{ij}\left[ C_{xx}\right]$,
that describes the parallel transport along a closed contour $C_{xx}$,
can be reduced by the gauge transformation to
\be
U\left[ C_{xx}\right]=\Omega\left[ C_{xx}\right]
{\rm diag}\left( \e^{ig \alpha_1(C)},\ldots, \e^{ig \alpha_{N_c}(C)}
\right) \Omega^\dagger\left[ C_{xx}\right].
\ee
Then $\Phi\left( C \right) $ reads
\be
\Phi\left( C \right)=\frac{1}{N_c} \sum_{j=1}^{N_c}\e^{ig \alpha_j(C)}.
\label{Udiag}
\ee
The phases $\alpha_j(C)$ are gauge invariant and normalized
so that $\alpha_j(C)\sim 1$ as $N_c \ra\infty$. For simplicity
we omit below all the indices (including space ones) except color.

The commutator of $\Phi$'s can be estimated using the 
representation \rf{Udiag}. Since
$
\left[ \alpha_i,\,\alpha_j \right]\propto\delta_{ij}\,,
$
one gets
\be
\left[ \Phi\left( C \right),\,\Phi\left( C^\prime \right) 
\right]\sim g^2 \frac{1}{N_c}\sim\frac{1}{N_c^2}
\label{6.4}
\ee
in the limit~\rf{orderg},
\ie the commutator can be neglected as $N_c\ra\infty$, and the field
$\Phi\left( C \right)$ becomes classical.

Note that the commutator~\rf{6.4} is of order $1/N_c^2$. One factor
$1/N_c$ is because of $g$ in the definition~\rf{Udiag} of 
$\Phi\left( C \right)$, while the other has a deep reason.
Let us image the summation over $j$ in \eq{Udiag} as some statistical
averaging. It is well-known in statistics that such averages weakly 
fluctuate as $N_c\ra\infty$, so that the dispersion is of order
$1/N_c$. It is the factor which emerges in the commutator~\rf{6.4}.

We see that the factorization is valid only for the gauge-invariant
quantities which involve the averaging over the color indices,
like that in \eq{Udiag}.
There is no reason to expect factorization for gauge invariants
which do not involve this averaging, for instance for the phases
$\alpha_j(C)$. Moreover, their commutator is $\sim 1$,
so that $\alpha_j(C)$'s strongly fluctuate even at $N_c=\infty$.
An explicit example of such strongly fluctuating gauge-invariant
quantities was first constructed in Ref.~\cite{Haa81}.

The r{\'e}sum{\'e} to this Remark is that the factorization
is due to the additional statistical averaging in the large-$N_c$
limit. There is no reason to assume an existence of the master field
in the strong sense in order to explain the factorization.


\section{QCD in Loop Space}\label{s:l.s.}

QCD can be entirely reformulated in terms of the colorless composite field
$\Phi\left(C\right)$ 
--- the trace of the Wilson loop for closed contours. 
This fact involves two main steps:
\begin{itemize}
\item[i)]
All the observables are expressed via $\Phi\left(C\right)$.
\item[ii)]
Dynamics is entirely reformulated in terms of $\Phi\left(C\right)$.
\end{itemize}

This approach is especially useful in the large-$N_c$ limit where
everything is expressed via the vacuum expectation value of 
$\Phi\left(C\right)$ --- the Wilson loop average. Observables are
given by summing the Wilson loop average
over paths with the same weight as in free theory.
The Wilson loop average obeys itself
a close functional equation --- the loop equation.

We begin this Section with presenting the formulas which relate
observables to the Wilson loops. Then we translate quantum
equation of motion of Yang--Mills theory into loop space.
We derive the closed equation for the Wilson loop average
as $N_c\ra\infty$ and discuss its various properties, including
a non-perturbative regularization. Finally, we briefly comment on
what is known about solutions of the loop equation. 

\subsection{Observables in terms of Wilson loops}

All observables in QCD can be  expressed via 
the Wilson loops $\Phi\left(C\right)$ defined by \eq{closedWloop}.
This property was first
advocated by Wilson~\cite{Wil74} on a lattice.
Calculation of QCD observables can be divided in two steps:
\begin{itemize}
\item[1)]
Calculation of the Wilson loop averages for arbitrary contours.
\item[2)]
Summation of the Wilson loop averages over the contours
with some weight depending on a given observable.
\end{itemize}

At finite $N_c$, observables are expressed via the $n$-loop averages
\be
W_n\left( C_1,\ldots,C_n\right)  =  
\LA\, \Phi\left(C_1\right) \cdots \Phi\left(C_n\right)\,\RA \,,
\label{n-loop}
\ee
which are analogous to the $n$-point Green functions
for $\varphi^3$ theory.
The appropriate formulas for 
the continuum theory can be found in Ref.~\cite{MM81a}.

Great simplifications occur in these formulas at $N_c=\infty$,
when all observables are expressed only via the one-loop average
\be
W\left( C\right)  =  \LA\, \Phi\left(C\right)\,\RA \equiv
 \left\langle \frac 1{N_c}\,{\rm tr}\,{\bf \,P}
\e^{i\oint_C dx^\mu \,A_\mu }\right\rangle\,. 
\label{one-loop}
\ee
This is associated with the quenched approximation.

For example, the average of the product of two colorless
quark vector currents \rf{OGamma} is given at large $N_c$ by
\be
\left\langle \bar \psi \gamma _\mu \psi \left( x_1\right) \bar \psi \gamma
_\nu \psi \left( x_2\right) \right\rangle  = 
\sum\limits_{C\ni x_1,x_2}
J_{\mu \nu }\left( C\right) \left\langle\, \Phi \left( C\right)\,
\right\rangle \,,
\label{twocurrents}
\ee
where the sum runs over contours $C$ passing through the points $x_1$ and 
$x_2$ as is depicted in Figure~\ref{fi:contours}a. 
\begin{figure}[t]
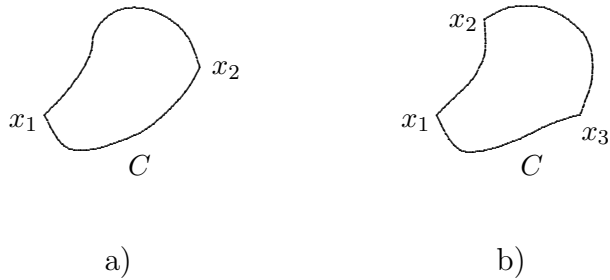

\vspace*{3mm}
\centering{
\input{2paths}\hspace*{2cm}\input{3paths}
}
\caption[Contours in sum over paths representing observables]   
{
Contours in the sum over paths representing observables:
a) in \eq{twocurrents} and b) in \eq{threecurrents}.
The contour a) passes two nailed points $x_1$ and $x_2$.
The contour b) passes three nailed points $x_1$, $x_2$, and $x_3$.
}
   \label{fi:contours}
\end{figure}
An analogous formula for the (connected)
correlators of three quark scalar currents reads
\be
\left\langle \bar \psi \psi \left( x_1\right) \bar \psi \psi \left(
x_2\right) \bar \psi \psi \left( x_3\right) \right\rangle _{\rm conn} = 
\sum\limits_{C\ni x_1,x_2,x_3}
J\left( C\right) 
\left\langle\,\Phi \left( C\right) \,\right\rangle\,, 
\label{threecurrents}
\ee
where the sum runs over contours $C$ passing through the three points 
$x_1$, $x_2$, and $x_3$ as is depicted in Figure~\ref{fi:contours}b.
A general (connected) correlator of $n$ quark currents is given
by a similar formula with $C$ passing through $n$ points $x_1$, \ldots,
$x_n$ (some of them may coincide).

The weights $J_{\mu \nu }\left( C\right)$ in \eq{twocurrents} and
$J\left( C\right)$ in \eq{threecurrents} 
are completely determined by free theory.
If quarks were scalars rather than spinors, then we would get
\be
J\left( C\right)  = \e^{-\frac12 m^2 \tau -\frac12 \int_0^\tau dt\,
\dot{z}_\mu^2(t) }
= \e^{-m L(C)} \qquad \fbox{scalar quarks}\,,
\ee
where $L(C)$ is the length of the (closed) contour $C$.
For spinor quarks, an additional disentangling of the gamma-matrices
is needed (ee, {\it e.g.}, Ref.~\cite{BNZ79}).  

\subsubsection*{Remark on renormalization of Wilson loops}

Perturbation theory for $W(C)$ can be obtained by expanding
the path-ordered exponential in the definition~\rf{one-loop} in
$g$ (see \eq{Poexpansion}) and averaging over the gluon field $A_\mu$.
Because of ultraviolet divergencies,
we need a (gauge invariant) regularization.
After such a regularization introduced, the Wilson loop average
for a smooth contour $C$ of the type in Figure~\ref{fi:smooth}a
\begin{figure}[t]
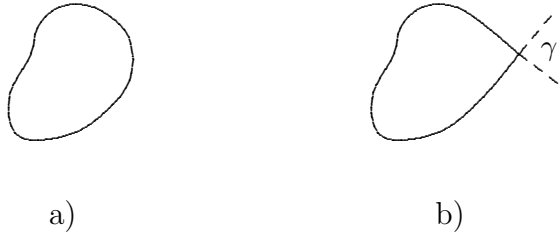

\vspace*{3mm}
\centering{
\input{smooth}\hspace*{3cm}\input{cusp}
}
\caption[Examples of smooth contour and contour with cusp]   
{
Examples of a) smooth contour and b) contour with a cusp.
The tangent vector to the contour jumps through angle $\gamma$
at the cusp.
}
   \label{fi:smooth}
\end{figure}
reads
\be
W\left( C \right)=\e^{-g^2\frac{(N_c^2-1)}{8\pi N_c} \frac {L(C)}a}
\, W_{\rm ren}\left( C \right),
\label{Wren}
\ee
where $a$ is the cutoff,
$L(C)$ is the length of $C$, and $W_{\rm ren}\left( C \right)$
is finite when expressed via the renormalized charge $g_{\rm R}$.
The exponential factor is due to the renormalization of the
mass of a heavy test quark. This factor does not emerge in
the dimensional regularization where $d=4-\eps$. The multiplicative
renormalization of the smooth
Wilson loop was shown in Refs.~\cite{GN80a,Pol80,DV80}.

If the contour $C$ has a cusp (or cusps) 
but no self-intersections as is illustrated by 
Figure~\ref{fi:smooth}b, then $W\left( C \right)$ is still
multiplicatively renormalizable~\cite{BNS81}:
\be
W\left( C \right)=Z \left( \gamma \right)
\, W_{\rm ren}\left( C \right),
\label{WrenZ}
\ee
while the (divergent) factor $Z\left( \gamma \right)$ depends on the cusp 
angle (or angles) $\gamma$ (or $\gamma$'s)
and $W_{\rm ren}\left( C \right)$
is finite when expressed via the renormalized charge $g_{\rm R}$.

\subsection{Schwinger--Dyson equations for Wilson loop}

Dynamics of (quantum) Yang--Mills theory is described by 
the quantum equation of motion
\be
\nabla_\mu^{ab} F_{\mu \nu }^b\left( x\right)  
\stackrel{\rm w.s.}{=}  \hbar \,\frac \delta {\delta
A_\nu^a \left( x\right) } 
\label{YMqeq}
\ee
which is
understood in the weak sense, \ie for the averages
\be
\Big\langle  \,\nabla_\mu^{ab} F_{\mu \nu }^b\left( x\right)
 Q\left[A\right]\,\Big\rangle  
=  \hbar \LA \frac \delta {\delta
A_\nu^a \left( x\right)}  Q\left[A\right] \RA \,.
\label{YMSD}
\ee 
The standard set of Schwinger--Dyson equations of Yang--Mills 
theory emerges when the functional $ Q\left[A\right]$ is chosen in the form
of the product of $A$'s as in \eq{multitrnormalization}.

Strictly speaking, the last statement is incorrect, since we have not
added, in  
Eqs.~\rf{YMqeq} and \rf{YMSD}, contributions coming from the variation of
 gauge-fixing and ghost terms in the Yang--Mills
action. However, these two contributions are mutually cancelled for
gauge-invariant functionals $Q[A]$. We shall deal below only with 
such gauge-invariant functionals (the Wilson loops). This is why we have
not considered the contribution of the  gauge-fixing and ghost terms. 

It is also convenient to use the matrix notation~\rf{calA}, when
\eq{YMqeq} for the Wilson loop takes on the form
\be
\left\langle \frac 1{N_c}\,{\rm tr}{\bf \,P}\,
\nabla _\mu F_{\mu \nu }\left( x\right) \e^{i\oint_Cd\xi ^\mu A_\mu
}\right\rangle  
 =  \left\langle
\frac {g^2}{2N_c}\,{\rm tr}\,
\frac \delta {\delta A_\nu \left( x\right) }{\bf P}
\e^{i\oint_Cd\xi ^\mu A_\mu }\right\rangle \,,
\label{SDle}
\ee
where we have restored the units with $\hbar=1$.

The variational derivative on the RHS can be calculated by virtue 
of the formula 
\be
\frac{\delta A_\mu ^{ij}\left( y\right) }{\delta A_\nu ^{kl}\left( x\right) }
 =  \delta _{\mu \nu }\,\delta^{(d)} \left( x-y\right) 
\left( \delta ^{il}\delta^{kj}-\frac 1{N_c}\delta ^{ij}\delta ^{kl}\right) 
\label{caldAdA}
\ee
which is a consequence of 
\be
\frac{\delta A_\mu ^a\left( y\right) }{\delta A_\nu ^b\left( x\right) }  =
 \delta _{\mu \nu }\,\delta^{(d)} \left( x-y\right) \delta ^{ab} \,.
\label{dAdA}
\ee
The second term in the parentheses in \eq{caldAdA} --- same as in 
\eq{completeness} --- is because ${A}_\mu$ is a matrix from the adjoint
representation of SU$(N_c)$.

By using \eq{caldAdA}, we get for the variational derivative on 
RHS of \eq{SDle}:
\begin{eqnarray}
\lefteqn{
{\rm tr}\,
\frac \delta {\delta A_\nu \left( x\right) }{\bf P}
\e^{i\oint_Cd\xi ^\mu \,{A}_\mu } ~ = ~ 
i \oint\limits_C d y_\nu \,
\delta^{(d)} \left( x-y\right)\,\times} \non  & & \!\!\!\!\!\!
  \left[ \frac 1{N_c} \,
{\rm tr}\,{\bf P}\e^{i\int_{C_{yx}}d\xi ^\mu \,{A}_\mu }
\frac 1{N_c}{\rm tr
}\,{\bf P}\e^{i\int_{C_{xy}}
d\xi ^\mu \,{A}_\mu }-\frac 1{N_c^3}{\rm tr}\,
{\bf P}\e^{i\int_Cd\xi ^\mu \,{A}_\mu }\right].~~~~ ~~~~
\eea
The contours $C_{yx}$ and $C_{xy}$, 
which are depicted in Figure~\ref{fi:twoparts}, 
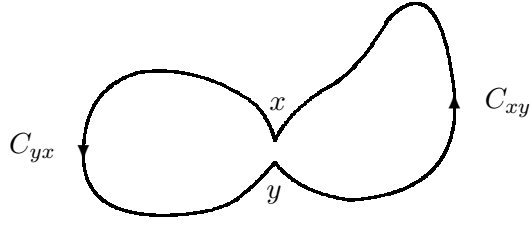
\begin{figure}[tbp]
\unitlength=1.00mm
\begin{picture}(44.0,33.00)(-20,98)
\thicklines
\bezier{60}(19.00,119.00)(27.00,120.00)(34.00,115.00)
\bezier{80}(11.50,108.00)(11.50,117.00)(19.00,119.00)
\bezier{80}(11.50,108.00)(11.50,100.00)(22.00,100.00)
\bezier{40}(22.00,100.00)(30.00,100.00)(33.00,103.00)
\bezier{64}(44.00,117.00)(48.00,119.00)(52.00,126.00)
\bezier{108}(52.00,126.00)(58.67,133.00)(60.67,117.50)
\bezier{148}(47.00,102.00)(62.67,103.00)(60.67,117.50)
\bezier{40}(44.00,117.00)(39.00,114.00)(37.00,110.00)
\bezier{64}(47.00,102.00)(40.00,103.00)(37.00,107.00)
\bezier{32}(34.00,115.00)(36.00,113.33)(37.00,110.00)
\bezier{32}(37.00,107.00)(34.67,104.33)(33.00,103.00)
\put(37.50,114.00){\makebox(0,0)[cb]{$x$}}
\put(37.00,104.00){\makebox(0,0)[ct]{$y$}}
\put(65.00,115.00){\makebox(0,0)[lc]{$C_{xy}$}}
\put(8.0,109.00){\makebox(0,0)[rc]{$C_{yx}$}}
\put(61.00,115.30){\vector(0,1){1.00}}
\put(11.65,108.20){\vector(0,-1){1.00}}
\end{picture} 
\caption[Contours $C_{yx}$ and $C_{xy}$ entering loop equation]
   {
   Contours $C_{yx}$ and $C_{xy}$ which enter the RHS's 
   of Eqs.~\rf{SDle} and \rf{le}. }
\label{fi:twoparts}
\end{figure}
are the parts of
the loop $C$: from $x$ to $y$ and from $y$ to $x$, respectively.
They are always closed
due to the presence of the delta-function. It implies that 
$x$ and $y$ should be the same points of {\it space}\/ but not necessarily
of the {\it contour} (\ie they may be associated with different values of the
parameter $\sigma$).  

We finally rewrite \eq{SDle} as
\begin{eqnarray}
\lefteqn{
\left\langle \frac 1{N_c}\,{\rm tr}{\bf \,P}\,
\nabla _\mu F_{\mu \nu }\left( x\right)\e^{i\oint_Cd\xi ^\mu A_\mu
} \right\rangle } \non & = &
i\lambda \oint\limits_C dy_\nu \,
\delta^{(d)} \left( x-y\right) \left[ \left\langle
\,\Phi \left( C_{yx}\right)
 \Phi \left( C_{xy}\right)\, \right\rangle -\frac
1{N^2}\left\langle \,\Phi \left( C\right) \,\right\rangle \right]~~~~~~ 
\label{le}
\end{eqnarray}
where we have introduced 
\be
\lambda=\frac{g^2 N_c}{2} \,.
\label{lvsg}
\ee
Notice that the RHS of \eq{le}
is completely represented via the (closed) Wilson loops.

\subsection{Path and area derivatives}

As we already mentioned,
the RHS of \eq{le} is completely represented via the (closed) Wilson loops.
It is crucial for the loop-space formulation of QCD that the
LHS of \eq{le} can also be represented in loop space as some operator
applied to the Wilson loop. To do this we need to develop
a differential calculus in loop space.

Loop space consists of arbitrary continuous closed loops, $C$.
They can be described in a parametric form by the functions 
$x_\mu(\sigma)\in L_2$,%
\footnote{Let us remind that $L_2$ stands for the Hilbert space of functions
$x_\mu(\s)$ whose square is integrable over the Lebesgue measure: 
\mbox{$\int_{\s_0}^{\s_f} d \s x^2_\mu(\s) < \infty $}.}
where \mbox{$\sigma_0\leq\sigma\leq\sigma_f$ and $\mu=1,\ldots,d$}, 
which take on values 
in a $d$-dimensional Euclidean space.
The functions $x_\mu(\sigma)$ can be 
discontinuous, generally speaking, for an arbitrary choice of the parameter 
$\s$. The continuity of the loop $C$ implies a continuous dependence 
on parameters of the type of proper length.

The functions $x_\mu(\s)\in L_2$ which are associated with the elements 
of loop space obey the following restrictions:
\begin{itemize}
\item[i)] The points $\s=\s_0$ and $\s=\s_f$ are identified:  
$x_\mu(\s_0)=x_\mu(\s_f)$
--- the loops are closed.
\item[ii)] 
The functions $x_\mu(\s)$ and $\Lambda_{\mu\nu}x_\nu(\s)+\alpha_\mu$, with 
$\Lambda_{\mu\nu}$ and $\alpha_\mu$ independent of $\s$, represent the 
same element of the loop space --- rotational and translational invariance.
\item[iii)] 
The functions $x_\mu(\s)$ and $x_\mu(\s^\p)$ with $\s^\p=f(\s),~f^\p(\s)\geq0$
describe the same loop --- reparametrization invariance.
\end{itemize}
An example of functionals which are defined on the elements of loop space
is the Wilson loop average~\rf{one-loop} or, more generally,
the $n$-loop average~\rf{n-loop}. 

The differential calculus in loop space is built out of the path
and area derivatives.

The {\em area derivative}\/ of a functional ${\cal F}\left(C\right)$
is defined by the difference
\vskip 0 cm
\unitlength=.6mm
\begin{equation}
\frac{\delta {\cal F}(C)}{ \delta \sigma_{\mu\nu}(x)}  \equiv
\frac{1}{\left| \delta \sigma_{\mu\nu}\right|}
\left[ \;
{\cal F} \left( \hbox{
\begin{picture}(44.0,13.00)(10,112)
\thicklines
\bezier{80}(11.00,108.00)(11.00,119.00)(19.00,119.00)
\bezier{64}(19.00,119.00)(28.00,121.00)(32.00,126.00)
\bezier{108}(32.00,126.00)(39.00,133.00)(41.00,118.00)
\bezier{44}(41.00,118.00)(44.00,122.00)(45.00,118.00)
\bezier{36}(41.00,117.00)(44.50,113.50)(45.00,118.00)
\bezier{148}(21.00,100.00)(43.00,101.00)(41.00,117.00)
\bezier{80}(11.00,108.00)(11.00,100.00)(21.00,100.00)
\put(39.00,117.50){\makebox(0,0)[cc]{$x$}}
\thinlines
\put(47.50,115.00){\vector(1,0){8.00}}
\put(47.50,115.00){\vector(4,3){6.0}}
\put(52.50,111.0){\makebox(0,0)[cb]{{\small $\mu$}}}
\put(50.00,118.00){\makebox(0,0)[rb]{{\small $\nu$}}}
\end{picture} }
\right)
~-~ {\cal F} \left( \hbox{
\begin{picture}(33.00,13.00)(10,112)
\thicklines
\bezier{80}(11.00,108.00)(11.00,119.00)(19.00,119.00)
\bezier{64}(19.00,119.00)(28.00,121.00)(32.00,126.00)
\bezier{108}(32.00,126.00)(39.00,133.00)(41.00,117.50)
\bezier{148}(21.00,100.00)(43.00,101.00)(41.00,117.50)
\bezier{80}(11.00,108.00)(11.00,100.00)(21.00,100.00)
\put(39.00,117.50){\makebox(0,0)[cc]{$x$}}
\end{picture} } \right) 
\right]
\label{aa}
\end{equation}
\vskip .2 cm \noindent
where an infinitesimal loop 
$\delta C_{\mu\nu}(x)$ is attached to a given loop at the point $x$ in the 
$\mu\nu$-plane and
${\left| \delta \sigma_{\mu\nu}\right|}$ stands for the area 
enclosed by the $\delta C_{\mu\nu}(x)$.
For a rectangular loop $\delta C_{\mu\nu}(x)$, one gets
$
\delta \sigma _{\mu \nu } = dx_\mu \wedge dx_\nu 
$,
where the symbol $\wedge$ implies antisymmetrization.

Analogously, the {\em path derivative}\/ is defined by
\vskip 0 cm
\unitlength=.6mm
\begin{equation}
\partial_\mu^x \, {{\cal F}(C_{xx})}  \equiv
\frac{1}{\left| \delta x_{\mu}\right|}
\left[ \;
{\cal F} \left( \hbox{
\begin{picture}(44.0,13.00)(10,112)
\thicklines
\bezier{80}(11.00,108.00)(11.00,119.00)(19.00,119.00)
\bezier{64}(19.00,119.00)(28.00,121.00)(32.00,126.00)
\bezier{108}(32.00,126.00)(39.00,133.00)(41.00,118.00)
\bezier{148}(21.00,100.00)(43.00,101.00)(41.00,117.00)
\bezier{80}(11.00,108.00)(11.00,100.00)(21.00,100.00)
\put(39.00,117.50){\makebox(0,0)[cc]{$x$}}
\put(46.20,117.50){\circle*{1.25}}
\bezier{16}(41.00,118.00)(43.50,117.00)(46.00,118.00)
\bezier{16}(41.00,117.00)(43.50,116.00)(46.00,117.00)
\thinlines
\put(47.50,115.00){\vector(1,0){8.00}}
\put(52.50,111.00){\makebox(0,0)[cb]{{\small $\mu$}}}
\end{picture} }
\right)
~-~ {\cal F} \left( \hbox{
\begin{picture}(33.00,13.00)(10,112)
\thicklines
\bezier{80}(11.00,108.00)(11.00,119.00)(19.00,119.00)
\bezier{64}(19.00,119.00)(28.00,121.00)(32.00,126.00)
\bezier{108}(32.00,126.00)(39.00,133.00)(41.00,117.50)
\bezier{148}(21.00,100.00)(43.00,101.00)(41.00,117.50)
\bezier{80}(11.00,108.00)(11.00,100.00)(21.00,100.00)
\put(38.80,117.50){\makebox(0,0)[cc]{$x$}}
\put(41.20,117.50){\circle*{1.25}}
\end{picture} } \right) 
\right]
\label{pp}
\end{equation}
\vskip .2 cm \noindent
where $\delta x_\mu$  is an infinitesimal path                
along which the point $x$ is shifted from the loop and 
${\left| \delta x_{\mu}\right|}$ stands for the length of the 
$\delta x_\mu$. 

These two differential
operations are well-defined for so-called functionals of the Stokes type which 
satisfy the backtracking condition --- they do not change when an appendix 
passing back and forth is added to the loop at some point $x$: 
\vskip 0 cm
\unitlength=.6mm
\begin{equation}
{\cal F} \left( \hbox{
\begin{picture}(44.0,13.00)(10,112)
\thicklines
\bezier{80}(11.00,108.00)(11.00,119.00)(19.00,119.00)
\bezier{64}(19.00,119.00)(28.00,121.00)(32.00,126.00)
\bezier{108}(32.00,126.00)(39.00,133.00)(41.00,118.00)
\bezier{148}(21.00,100.00)(43.00,101.00)(41.00,116.00)
\bezier{80}(11.00,108.00)(11.00,100.00)(21.00,100.00)
\bezier{49}(41.00,118.00)(49.00,118.00)(49.00,124.00)
\bezier{49}(41.00,116.00)(51.00,116.00)(51.00,124.00)
\put(50.2,124.00){\oval(2.0,2.0)[t]}
\put(39.00,117.00){\makebox(0,0)[cc]{$x$}}
\end{picture} }
\right)
~=~ {\cal F} \left( \hbox{
\begin{picture}(33.00,13.00)(10,112)
\thicklines
\bezier{80}(11.00,108.00)(11.00,119.00)(19.00,119.00)
\bezier{64}(19.00,119.00)(28.00,121.00)(32.00,126.00)
\bezier{108}(32.00,126.00)(39.00,133.00)(41.00,117.50)
\bezier{148}(21.00,100.00)(43.00,101.00)(41.00,117.50)
\bezier{80}(11.00,108.00)(11.00,100.00)(21.00,100.00)
\end{picture} } \right).
\label{btr}
\end{equation}
\vskip .2 cm \noindent
This condition is equivalent to the
Bianchi identity of Yang--Mills theory and is obviously satisfied by 
the Wilson loop~\rf{one-loop} due to the properties of the non-Abelian
phase factor. Such functionals are 
known in mathematics as Chen integrals.

A simple example of the Stokes functional is the area of the minimal
surface, $A_{\rm min}(C)$. It obviously satisfies \eq{btr}. Otherwise, 
the length $L(C)$ of the loop $C$ is not a Stokes functional,
since the lengths of contours on the LHS and RHS of \eq{btr}
are different.

For the Stokes functionals, the variation on the RHS of \eq{aa} 
is proportional to the area enclosed by the infinitesimally small loop 
$\delta C_{\mu\nu}(x)$ and does not depend on its shape. 
Analogously, the variation on the RHS of \eq{pp} 
is proportional to the length of the infinitesimal path $\delta 
x_{\mu}$ and does not depend on its shape. 

If $x$ is a regular point (like any point of the contour for the 
functional~\rf{one-loop}), the RHS of \eq{pp} vanishes due to the backtracking 
condition~\rf{btr}. In order for the result to be nonvanishing, the point $x$
should be a {\it marked}\/ (or irregular) point. 
A simple example of the functional with a marked point $x$ is 
\be
\Phi^a[C_{xx}] \equiv \frac{1}{N_c}\tr{} \Big( t^a  {\bf \,P} 
\e^{ i \int _{C_{xx}} d\xi^\mu A_\mu(\xi)} \Big)
\label{Wlmarked}
\ee
with the SU$(N_c)$ generator $t^a$  inserted in the 
path-ordered product at the point $x$.

The area derivative of the Wilson loop is given by
the Mandelstam formula
\be
\frac \delta {\delta \sigma _{\mu \nu }\left( x\right) }\frac 1{N_c}\,{\rm tr}
\,{\bf P}\e^{i \oint_Cd\xi ^\mu \,{A}_\mu } 
 =  \frac i{N_c}\,{\rm tr}\,
{\bf P}\, F_{\mu \nu }\left( x\right) 
\e^{i \oint_C d\xi ^\mu \,{A}_\mu }\,.
\label{Mandelstamformula}
\ee
In order to prove it, it is convenient to choose 
$\delta C_{\mu\nu}(x)$ to be a rectangle in the $\mu\nu$-plane 
and straightforwardly use the definition~\rf{aa}. 
The sense of \eq{Mandelstamformula} is very simple: $\F_{\mu\nu}$
is a curvature associated with the connection $\A_\mu$.

The functional on the RHS of \eq{Mandelstamformula} has a marked point $x$,
and is of the type in \eq{Wlmarked}. 
When the path derivative acts on such a functional
according to the definition~\rf{pp}, the result reads
\be
\partial _\mu ^x\frac 1{N_c}\,{\rm tr}\,{\bf P}\,B\left( x\right)
\e^{i\oint_Cd\xi ^\mu \,{A}_\mu } = \frac 1{N_c}\,{\rm tr}\,{\bf P}
\,\nabla _\mu B\left( x\right) \e^{i\oint_Cd\xi ^\mu \,{A}_\mu }\,,
\label{pathderivative}
\ee
where
\be
\nabla _\mu B = \partial _\mu B-i\left[ \A_\mu ,B\right] 
\ee
is the covariant derivative in the adjoint representation.

Combining Eqs.~\rf{Mandelstamformula} and \rf{pathderivative},
we finally represent the expression on the LHS of \eq{SDle} (or \eq{le}) as
\be
\frac 1{N_c}\,{\rm tr}\,{\bf P}\,\nabla _\mu \F_{\mu \nu }\left( x\right)
\e^{i\oint_Cd\xi ^\mu \,{A}_\mu } = \partial _\mu ^x\frac \delta
{\delta \sigma _{\mu \nu }\left( x\right) }\frac i{N_c}\,{\rm tr}\,{\bf P}
\e^{i\oint_Cd\xi ^\mu \,{A}_\mu }, 
\label{L_nu}
\ee
\ie via the action of the path and area derivatives on the Wilson loop.
It is therefore rewritten in loop space.

A r{\'e}sum{\'e} of the results of this subsection is presented
in Table~\ref{t:2} as a vocabulary for
translation of Yang--Mills theory
from the language of ordinary space in the language of
loop space.
\begin{table}[t] 
\vspace*{.2cm}
\centerline{
\begin{tabular}{||r|l||r|l||} \hline 
\multicolumn{2}{||c||} {Ordinary space}   & 
\multicolumn{2}{c||} {Loop space}  \\ \hline\hline  
\mbox{}& & & \\
$\Phi\left[ A \right]$ & phase factor &  $\Phi\left( C \right)$  
 & loop functional \\ 
\mbox{} & & & \\\hline 
\mbox{} & & & \\
$F_{\mu\nu}\left(x\right)$& field strength & 
$\frac{\delta}{\delta \sigma_{\mu\nu}\left(x\right)}$ & 
 area derivative \\ 
\mbox{} & & & \\\hline 
\mbox{} & & & \\
$\nabla_\mu^x$& covariant derivative& 
 $\partial_\mu^x$ &  path derivative \\ 
\mbox{} & & & \\\hline 
\mbox{} & & & \\
$\nabla \wedge F =0$&  Bianchi identity & 
  &  Stokes functionals \\ 
\mbox{} & & & \\\hline 
\mbox{} & & & \\
$\nabla_\mu F_{\mu\nu}$\hspace*{.5cm} &Schwinger-Dyson & 
  &  Loop   \\ 
$=\delta/\delta A_\nu$&  equations & 
  &   equations \\ \mbox{} & & & \\\hline
\end{tabular} }
\caption[Vocabulary from ordinary space in loop space]   
{  
   Vocabulary for translation of Yang--Mills theory
from ordinary space in loop space.
     }
\label{t:2}
\end{table}

\subsubsection*{Remark on Bianchi identity for Stokes functionals}

The backtracking relation~\rf{btr} can be equivalently represented as
\begin{eqnarray}
\epsilon _{\mu \nu \lambda \rho }\,\partial _\mu ^x
\,\frac \delta {\delta \sigma
_{\nu \lambda }\left( x\right) }\Phi \left( C\right) ~ = ~ 0 \,,
\label{LBianchi}
\end{eqnarray}
by choosing the appendix in \eq{btr} to be an infinitesimal
straight line in the $\rho$-direction and geometrically applying 
the Stokes theorem.
Using Eqs.~\rf{Mandelstamformula} and \rf{pathderivative},
\eq{LBianchi} can in turn be rewritten as 
\begin{eqnarray}
\epsilon _{\mu \nu \lambda \rho }\frac 1{N_c}\,{\rm tr}\,{\bf P}\,
\nabla_\mu \F_{\nu
\lambda }\left( x\right) \e^{i\oint_Cd\xi ^\mu \,{A}_\mu } ~ = ~ 0 \,.
\end{eqnarray}
Therefore, \eq{LBianchi} represents the Bianchi identity
in loop space. 

\subsubsection*{Remark on relation to variational derivative}

The standard variational derivative, 
$\delta/\delta x_\mu(\s)$, can be expressed via the path and area derivatives 
by the formula
\be
\frac{\delta}{\delta x_\mu(\s)} = \dot{x}_\nu(\s) 
\frac{\delta}{\delta\s_{\mu\nu}( x(\s) )} +
\sum_{i=1}^m \d_\mu^{x_i} \delta(\s-\s_i) \,,
\label{varder}
\ee
where the sum on the RHS is present for the case of a functional having 
$m$ marked (irregular) points $x_i\equiv x(\s_i)$. A simplest example of 
the functional with $m$ marked points is just a function of $m$ variables 
$x_1,\ldots,x_m$. 

By using \eq{varder}, the path derivative can be calculated as the 
limiting procedure
\be
\d_\mu^{x(\s)}= \int\limits_{\s-0}^{\s+0} d \s^\p
\frac{\delta}{\delta x_\mu(\s^\p)}  \,.
\label{vpath}
\ee
The result is obviously nonvanishing only when $\d_\mu^x$ is applied to a 
functional with $x(\s)$ being a marked point.

It is nontrivial that the area derivative can also be expressed via
the variational derivative~\cite{Pol80}:
\be
\frac{\delta}{\delta\s_{\mu\nu}( x(\s) )} =
 \int\limits_{\s-0}^{\s+0} d \s^\p (\s^\p -\s)
\frac{\delta}{\delta x_\mu(\s^\p)}  
\frac{\delta}{\delta x_\nu(\s)}  \,.
\label{varea}
\ee
The point is that the six-component quantity,
${\delta}/{\delta\s_{\mu\nu}( x(\s) )} $,
is expressed via the four-component one,
${\delta}/{\delta x_{\mu}(\s)} $, which is possible because
the components of
${\delta}/{\delta\s_{\mu\nu}( x(\s) )} $ are dependent
due to the loop-space Bianchi identity~\rf{LBianchi}.

\subsection{Loop equations}

By virtue of \eq{L_nu}, \eq{le} can be represented completely
in loop space:
\begin{eqnarray}
\lefteqn{ 
\partial _\mu ^x\frac \delta {\delta \sigma _{\mu \nu }\left( x\right) }
\Big\langle\,\Phi \left( C\right)\,\Big\rangle} \non
&{=} & 
\l \oint\limits_Cdy_\nu \,\delta^{(d)} \left( x-y\right)
\LA\left[ \Phi \left( C_{yx}\right) \Phi \left( C_{xy}\right)
 -\frac 1{N_c^2}\Phi \left( C\right) \right] \RA \,,
\end{eqnarray}
or, using the definitions~\rf{n-loop} and \rf{one-loop}
of the loop averages, as
\be
\partial _\mu ^x\frac \delta {\delta \sigma _{\mu \nu }\left( x\right) }
W \left( C\right) =
\l \oint\limits_Cdy_\nu \,\delta^{(d)} \left( x-y\right)
\left[ W_2 \left( C_{yx}, C_{xy}\right)
 -\frac 1{N_c^2}W \left( C\right) \right] \,.
\label{12le}
\ee

This equation is not closed. Having started from $W(C)$, we obtain another
quantity, $W_2\left( C_1, C_2\right)$, so that  \eq{12le} connects
the one-loop average with a two-loop one.  This is similar to 
the case of the (quantum) $\varphi^3$-theory, whose Schwinger--Dyson 
equations connect the $n$-point Green functions with
different $n$. We shall derive this complete set of equations for
the $n$-loop averages in this Subsection later on.

However, the two-loop average factorizes in the large-$N_c$ limit:
\be
W_2\left( C_1, C_2\right)=W\left( C_1\right) W\left( C_2\right)
+{\cal O}\left(\frac 1{N_c^2} \right),
\label{12fac}
\ee
as was discussed in Subsection~\ref{ss:mfac}. Keeping the constant
$\lambda$ (defined by \eq{lvsg}) fixed in the large-$N_c$ limit
as is prescribed by \eq{orderg}, we get~\cite{MM79} 
\be
\partial _\mu ^x\frac \delta {\delta \sigma _{\mu \nu }\left( x\right)
}W\left( C\right) = \l\oint\limits_C dy_\nu \,
\delta^{(d)} \left( x-y\right)
W\left( C_{yx}\right) W\left( C_{xy}\right) 
\label{MM}
\ee
as $N_c\ra\infty$.

Equation~\rf{MM} is a closed equation for the Wilson loop average in the 
large-$N_c$ limit. It is referred to as the {\em loop equation}.

To find $W(C)$, 
\eq{MM} should be solved in the class of Stokes functionals with the initial 
condition 
\be
W(0)=1
\label{ic}
\ee
for loops which are shrunk to points. 

The factorization~\rf{12fac} can itself be derived 
from the chain of loop equations. Proceeding as before, we get
\begin{eqnarray}
\lefteqn{\frac 1{\l}\partial _\mu ^x\frac \delta {\delta \sigma _{\mu \nu }
\left( x\right) } W_n\left( C_1,\ldots ,C_n\right) }\non
& = & \oint\limits_{C_1}dy_\nu \,
\delta^{(d)}  \left( x-y\right) \left[ W_{n+1}\left( C_{xy},C_{yx},\ldots
,C_n\right) -\frac 1{N_c^2}W_n\left( C_1,\ldots ,C_n\right) \right] 
\nonumber\\ 
&  & +\sum\limits_{j\geq 2}\frac 1{N_c^2}\oint\limits_{C_j}dy_\nu \,
\delta^{(d)}  \left(
x-y\right) \Big[ W_{n-1}\left( C_1C_j,\ldots ,\underline{C_j},\ldots
,C_n\right) \non  & &
\hspace*{4.5cm}  -W_n\left( C_1,\ldots ,C_n\right) \Big] \,.
\label{MMn}
\end{eqnarray}
Here $x$ belongs to $C_1$; $C_1C_j$ stands for the joining
of $C_1$ and $C_j$; $\underline{C_j}$ means that $C_j$ is omitted.

Equation~\rf{MMn} looks like the Schwinger--Dyson
equation for the $\varphi^3$-theory.
Moreover, the number of colors $N_c$ enters \eq{MMn} simply as
a scalar factor $N_c^{-2}$, likewise Plank's constant $\hbar$ enters
in the $\varphi^3$-theory. It is the major advantage of the use of loop space.
What is said in Subsection~\ref{ss:semic} about the ``semiclassical''
nature of the $1/N_c$-expansion of QCD is explicitly realized in
\eq{MMn}. Its expansion in $1/N_c$ is straightforward. 

At $N_c = \infty$, \eq{MMn} is simplified to
\be
\partial _\mu ^x\frac \delta {\delta \sigma _{\mu \nu }\left( x\right)
}W_n\left( C_1,\ldots ,C_n\right)=
\l\oint\limits_{C_1}dy_\nu \,
\delta^{(d)} ( x-y) W_{n+1}( C_{yx},C_{xy},\ldots,C_n). 
\ee
This equation possesses a factorized solution
\be
W_n\left( C_1,\ldots ,C_n\right)= 
W\left( C_1\right) \cdots W\left( C_n\right)
+{\cal O}\left( \frac 1{N_c^2}\right) 
\label{facsolution}
\ee
provided $W\left( C\right)$ obeys \eq{MM} which plays the role of
a ``classical'' equation in the large-$N_c$ limit.
Thus, we have given a non-perturbative proof of the
large-$N_c$ factorization of the Wilson loops.

\subsection{Relation to planar diagrams} 

The perturbation-theory expansion of the Wilson loop average 
can be calculated from \eq{Poexpansion} which we represent in 
the form
\bea
\lefteqn{W\left( C \right) = 1
+ \sum\limits_{n=2}^\infty\; i^n
\oint\limits_C dx_1^{\mu_1}
\oint\limits_C dx_2^{\mu_2} \ldots 
\oint\limits_C dx_n^{\mu_n} } \non & &
\times \theta_{\rm c}(1,2,\ldots,n)\,
G^{(n)}_{\mu_1\mu_2\cdots\mu_n} \left(x_1,x_2,\ldots,x_n \right),
\label{cPoexpansion}
\eea
where $\theta_{\rm c}(1,2,\ldots,n)$ orders the points $x_1$, $\ldots$, $x_n$
along contour in the cyclic order
and $G^{(n)}_{\mu_1\cdots\mu_n}$ is given by \eq{multitrnormalization}.
This $\theta$-function has the meaning of the propagator of
a test heavy particle which lives in the contour $C$. 

We assume, for definitiveness, the dimensional regularization
throughout this Subsection to make all the integrals well-defined.

Each term on the RHS of \eq{cPoexpansion}  
can be conveniently represented by the diagram in
Figure~\ref{fi:diagen},
\begin{figure}
\vspace*{3mm}
\centering{
\input{diagen}
 }
\caption[Graphic representation of \protect{\eq{cPoexpansion}}]   
{
Graphic representation of the terms on the RHS of \eq{cPoexpansion}.
 }
   \label{fi:diagen}
\end{figure} 
where the integration over the contour $C$ is associated 
with each point $x_i$ lying in the contour $C$. 

These diagrams are analogous to those discussed in Subsection~\ref{ss:b.} 
with one external boundary --- the Wilson loop in the given case.
In the large-$N_c$ limit, only planar diagrams survive.
Some of them, which are of the lowest order in $\lambda$, 
are depicted in Figure~\ref{fi:Feypl}. 
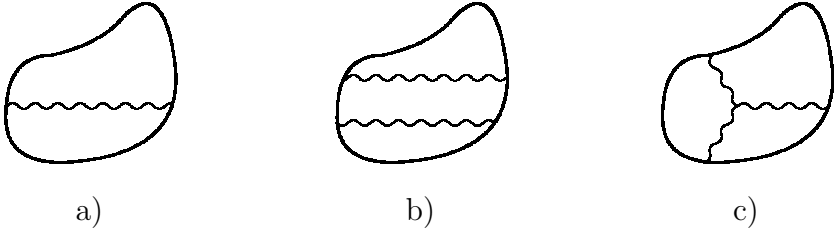
\begin{figure}[tb]
\hspace*{-.2cm}
\unitlength=.75mm
\begin{picture}(33.00,50.00)(5,85)
\thicklines
\bezier{80}(11.00,108.00)(11.00,119.00)(19.00,119.00)
\bezier{64}(19.00,119.00)(28.00,121.00)(32.00,126.00)
\bezier{108}(32.00,126.00)(39.00,133.00)(41.00,117.50)
\bezier{148}(21.00,100.00)(43.00,101.00)(41.00,117.50)
\bezier{80}(11.00,108.00)(11.00,100.00)(21.00,100.00)
\thinlines
\multiput(11.30,110.00)(4.00,0.00){7}{
  \bezier{28}(0.00,0.00)(1.00,1.00)(2.00,0.00)}
\multiput(13.30,110.00)(4.00,0.00){7}{
  \bezier{28}(0.00,0.00)(1.00,-1.00)(2.00,0.00)}
\bezier{28}(39.30,110.00)(39.90,110.50)(40.50,110.50)
\put(26.00,94.00){\makebox(0,0)[ct]{{\large a)}}}
\end{picture}
\begin{picture}(33.00,50.00)(-19,85)
\thicklines
\bezier{80}(11.00,108.00)(11.00,119.00)(19.00,119.00)
\bezier{64}(19.00,119.00)(28.00,121.00)(32.00,126.00)
\bezier{108}(32.00,126.00)(39.00,133.00)(41.00,117.50)
\bezier{148}(21.00,100.00)(43.00,101.00)(41.00,117.50)
\bezier{80}(11.00,108.00)(11.00,100.00)(21.00,100.00)
\thinlines
\bezier{28}(12.00,114.50)(12.50,114.50)(13.00,115.00)
\multiput(13.00,115.00)(4.00,0.00){7}{
  \bezier{28}(0.00,0.00)(1.00,1.00)(2.00,0.00)}
\multiput(15.00,115.00)(4.00,0.00){7}{
  \bezier{28}(0.00,0.00)(1.00,-1.00)(2.00,0.00)}
\multiput(11.00,107.00)(4.00,0.00){7}{
  \bezier{28}(0.00,0.00)(1.00,-1.00)(2.00,0.00)}
\multiput(13.00,107.00)(4.00,0.00){7}{
  \bezier{28}(0.00,0.00)(1.00,1.00)(2.00,0.00)}
\put(26.00,94.00){\makebox(0,0)[ct]{{\large b)}}}
\end{picture} 
\begin{picture}(33.00,50.00)(-42,85)
\thicklines
\bezier{80}(11.00,108.00)(11.00,119.00)(19.00,119.00)
\bezier{64}(19.00,119.00)(28.00,121.00)(32.00,126.00)
\bezier{108}(32.00,126.00)(39.00,133.00)(41.00,117.50)
\bezier{148}(21.00,100.00)(43.00,101.00)(41.00,117.50)
\bezier{80}(11.00,108.00)(11.00,100.00)(21.00,100.00)
\thinlines
\multiput(24.50,110.00)(4.00,0.00){4}{
  \bezier{28}(0.00,0.00)(1.00,1.00)(2.00,0.00)}
\multiput(26.50,110.00)(4.00,0.00){4}{
  \bezier{28}(0.00,0.00)(1.00,-1.00)(2.00,0.00)}
\multiput(24.50,110.00)(-2.00,-3.33){3}{
  \bezier{28}(0.00,0.00)(-1.333,-0.333)(-1.00,-1.665)}
\multiput(23.50,108.335)(-2.00,-3.33){3}{
  \bezier{28}(0.00,0.00)(0.333,-1.167)(-1.00,-1.665)}
\multiput(24.50,110.00)(-2.00,3.33){3}{
  \bezier{28}(0.00,0.00)(-1.333,0.333)(-1.00,1.665)}
\multiput(23.50,111.665)(-2.00,3.33){2}{
  \bezier{28}(0.00,0.00)(0.333,1.167)(-1.00,1.665)}
\bezier{28}(19.50,118.25)(19.60,118.80)(20.00,119.30)
\put(26.00,94.00){\makebox(0,0)[ct]{{\large c)}}}
\end{picture} 
\caption[Planar diagrams for $W(C)$ to order $\lambda^2$]   
{
   Planar diagrams for $W(C)$: a) of order $\lambda$ with gluon
   propagator, and of order $\lambda^2$ b) with two noninteracting 
   gluons and c) with the three-gluon vertex. Diagrams of order 
   $\lambda^2$ with one-loop insertions to gluon propagator are
   not drawn. 
 }
\label{fi:Feypl}
\end{figure}  

The large-$N_c$ loop equation~\rf{MM} describes the sum of the planar
diagrams. Its iterative solution in $\lambda$ reproduces the set of
planar diagrams for $W\left( C\right)$ provided the initial 
condition~\rf{ic} and some boundary conditions
for asymptotically large contours are imposed.

Equation~\rf{cPoexpansion} can be viewed as an ansatz 
for $W\left( C\right)$ with some unknown functions 
$G^{(n)}_{\mu_1\cdots\mu_n} \left(x_1,\ldots,x_n \right)$  
to be determined by the substitution 
into the loop equation. To preserve symmetry properties of 
$W\left( C\right)$, the functions $G^{(n)}$ must be symmetric
under a cyclic permutation of the points $1$, $\ldots$, $n$
and depend only on $x_i-x_j$ (translational invariance).
A main advantage of this ansatz is
that it automatically corresponds to a Stokes functional,
due to the properties of vector integrals, and the initial
condition~\rf{ic} is satisfied. 

The action of the area and path derivatives on the ansatz~\rf{cPoexpansion}
is easily calculable. For instance, the area derivative reads
\bea
\lefteqn{\frac{\delta W\left( C \right)}{\delta \sigma_{\mu\nu}(z)} ~=~ 
 \sum\limits_{n=1}^\infty\; i^n
\oint\limits_C dx_1^{\mu_1}  \ldots 
\oint\limits_C dx_n^{\mu_n}  \,\theta_{\rm c}(1,2,\ldots,n)}\non & &
\times \left[ \left( \d^z_\mu \delta_{\nu\alpha}
- \d^z_\nu \delta_{\mu\alpha}\right)
G^{(n+1)}_{\alpha\mu_1\cdots\mu_n} \left(z,x_1,\ldots,x_n \right) 
\right.\non & &
~~~+\left.\left( \delta_{\mu\beta} \delta_{\nu\alpha}
- \delta_{\mu\alpha}\delta_{\nu\beta} \right)
G^{(n+2)}_{\alpha\beta\mu_1\cdots\mu_n} \left(z,z,x_1,\ldots,x_n \right)
\right].
\eea
The analogy with the Mandelstam formula~\rf{Mandelstamformula}
is obvious.

More about solving the loop equation by the ansatz~\rf{cPoexpansion}
can be found in Refs.~\cite{MM81a,BGSN82,Mig83}.

\subsection{Loop-space Laplacian and regularization}

The loop equation~\rf{MM}
is {\it not}\/ yet entirely formulated in loop space. It 
is a $d$-vector equation whose both sides depend explicitly on the point $x$ 
which does not belong to loop space. The fact that we have a $d$-vector 
equation for a scalar quantity means, in particular, that \eq{MM} is 
overspecified. 

A practical difficulty in solving \eq{MM} is that the 
area and path derivatives,
$ {\delta}/{\delta\s_{\mu\nu}( x)}$ and $\d_\mu^x$, 
which enter the LHS are complicated, generally speaking, non-commutative
operators. They are intimately related to the Yang--Mills perturbation theory 
where they correspond to the non-Abelian field strength $F_{\mu\nu}$
and the covariant derivative $\nabla_\mu$.
However, it is not easy to apply these operators to a generic functional 
$W(C)$ which is defined on elements of loop space.

A much more convenient form of the loop equation can be obtained by 
integrating 
both sides of \eq{MM} over $dx_\nu$ along the same contour $C$, which
yields
\be
\oint\limits_C dx_\nu \,\d_\mu^x \,\frac{\delta}{\delta\s_{\mu\nu}( x)} W(C)=
\lambda \oint\limits_C dx_\mu \oint\limits_C d y_\mu\, \delta^{(d)}(x-y)\, 
W(C_{yx}) W(C_{xy})\,.
\label{ile}
\ee
Now both the operator on the LHS and the functional on the RHS are 
scalars without labeled points 
and are well-defined in loop space. The operator on the LHS 
of \eq{ile} can be interpreted as an infinitesimal variation of elements 
of loop space.

Equations \rf{MM} and \rf{ile} are completely equivalent.
A proof of equivalence of scalar \eq{ile} and original $d$-vector 
\eq{MM} is based on the important property of \eq{MM} whose both sides are 
identically annihilated by the operator $\d_\nu^x$.
It is a consequence of the identity 
\be
\nabla_\mu \nabla_\nu \,\F_{\mu\nu}=
-\frac i2 \left[\F_{\mu\nu},\F_{\mu\nu} \right]=0
\ee
in the ordinary space.
Due to this property, the vanishing of the contour integral of some vector 
is equivalent to vanishing of the vector itself, so that \eq{MM} can in turn 
be deduced from \eq{ile}.

Equation~\rf{ile} is associated with the so-called second-order 
Schwinger--Dyson equation
\be
\int d^dx\,\nabla _\mu F_{\mu \nu }^a\left( x\right) \frac \delta {\delta
A_\nu ^a\left( x\right) }
\stackrel{\rm w.s.}{=}  
\hbar \int d^dx\,d^dy\,\delta^{(d)} 
\left(
x-y\right) \frac \delta {\delta A_\nu ^a\left( y\right) }\frac \delta
{\delta A_\nu ^a\left( x\right) } 
\label{2ndYMSD}
\ee
in the same sense as \eq{MM} is associated with \eq{YMqeq}.
It is called ``second order'' since the RHS involves two variational
derivatives with respect to $A_\nu$.

The operator on the LHS of \eq{ile} is a well-defined object in loop space.
When applied to regular functionals which do not have marked points, 
it can be represented, using 
Eqs.~\rf{vpath} and \rf{varea}, in an equivalent form
\be
\Delta \equiv
\oint\limits_C dx_\nu \,\d_\mu^x \,\frac{\delta}{\delta\s_{\mu\nu}( x)}
 =\int\limits_{\s_0}^{\s_f} d\s  \int\limits_{\s-0}^{\s+0} d\s^\p 
\frac{\delta}{\delta x_\mu(\s^\p)} \frac{\delta}{\delta x_\mu(\s)} \,.
\label{defDelta}
\ee
As was first pointed out by Gervais and Neveu~\cite{GN79b}, 
this operator is nothing but
a functional extension of the Laplace operator, which is
known in mathematics as the Levy operator.%
\footnote{See the book by Levy \cite{Lev51} and a review~\cite{Fel86}.} 
Equation~\rf{ile} can be 
represented in turn as an (inhomogeneous) functional Laplace equation
\be
\Delta W(C) 
= \lambda \oint\limits_C dx_\mu \oint\limits_C d y_\mu \, 
\delta^{(d)}(x-y) \, W(C_{yx}) W(C_{xy})\,.
\label{Le}
\ee
We shall refer to this equation as the loop-space Laplace equation.

The form~\rf{Le} of the loop equation is convenient for a
non-perturbative ultraviolet regularization.

The idea is to 
start from the regularized version of \eq{2ndYMSD}, replacing the
delta-function on the RHS by the kernel of the regularizing operator: 
\be
\delta^{ab}\delta^{(d)}(x-y) \stackrel{\rm Reg.}{\Longrightarrow}
\LA y\left|\bbox{R}^{ab} \right| x \RA =
\bbox{R}^{ab}\,\delta^{(d)}(x-y)
\label{defRkern}
\ee
with
\be
\bbox{R}^{ab}  
= \left( \e^{{a^2 \nabla^2}/{2}}  \right)^{ab} \,,
\label{defR}
\ee
where $\nabla_\mu$ is the covariant derivative in the adjoint representation.
The regularized version of \eq{2ndYMSD} is
\be
\int d^dx\,\nabla _\mu F_{\mu \nu }^a\left( x\right) \frac \delta {\delta
A_\nu ^a\left( x\right) }  
\stackrel{\rm w.s.}{=} 
\hbar \int d^dx\,d^dy\, 
\LA y\left|\bbox{R}^{ab} \right| x \RA
 \frac \delta {\delta A_\nu ^b\left( y\right) }\frac \delta
{\delta A_\nu ^a\left( x\right) } \,.
\label{rs-oSD}
\ee
 
To translate \eq{rs-oSD} in loop space, we use the path-integral 
representation 
\be
\LA y\left|\bbox{R}^{ab} \right| x \RA = 
\int\limits_{r(0)=x \atop r(a^{2})=y} D r(t)
\e^{-\frac 12 \int_0^{a^{2}} d t \,\dot{r}^2(t)} \;
2\tr{}\left[ t^a U(r_{yx})t^b  U(r_{xy})
   \right] 
\label{rpi} 
\ee
with
\be
U(r_{yx}) =
{\bf P} \e^{i\int_x^y  d r_\mu \A_\mu(r)}\,,
\ee
where the integration is over regulator paths $r_\mu(t)$ from $x$ to $y$
whose typical length is $\sim a$. 
 
Calculating the variational derivatives on the RHS of \eq{rs-oSD},
using \eq{rpi} and the completeness condition~\rf{completeness},
we get as $N\ra\infty$:
\bea
\lefteqn{\int d^dx\,d^dy\, 
\LA y\left|\bbox{R}^{ab} \right| x \RA
 \frac \delta {\delta A_\nu ^b\left( y\right) }\frac \delta
{\delta A_\nu ^a\left( x\right) } \Phi\left( C \right) ~=~
 \l \oint\limits_C dx_\mu \oint\limits_C dy_\mu } \non & & \times
 \!\!\!\int\limits_{r(0)=x \atop r(a^{2})=y}\!\!\! {\cal D} r(t)
\e^{-\frac 12 \int_0^{a^{2}} d t \dot{r}^2(t)}\,
\Phi\left(C_{yx}r_{xy}\right) \Phi\left(C_{xy}r_{yx}\right)\,,\hspace*{1cm}
\label{AnLe}
\eea
where the contours $C_{yx}r_{xy}$ and $C_{xy}r_{yx}$ are depicted in
Figure~\ref{fi:Fig.2}.
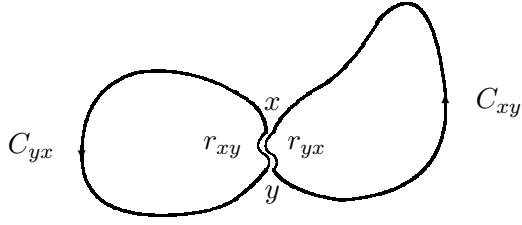
\begin{figure}[tbp]
\unitlength=1.00mm
\begin{picture}(44.0,33.00)(-20,98)
\thicklines
\bezier{60}(20.00,119.00)(28.00,120.00)(35.00,115.00)
\bezier{80}(12.50,108.00)(12.50,117.00)(20.00,119.00)
\bezier{80}(12.50,108.00)(12.50,100.00)(23.00,100.00)
\bezier{40}(23.00,100.00)(31.00,100.00)(34.00,103.00)
\bezier{64}(44.00,117.00)(48.00,119.00)(52.00,126.00)
\bezier{108}(52.00,126.00)(58.67,133.00)(60.67,117.50)
\bezier{148}(47.00,102.00)(62.67,103.00)(60.67,117.50)
\bezier{40}(44.00,117.00)(39.00,114.00)(38.00,111.00)
\bezier{64}(47.00,102.00)(40.00,103.00)(38.00,106.00)
\bezier{32}(35.00,115.00)(37.00,113.33)(37.00,111.00)
\bezier{32}(37.00,106.00)(35.67,104.33)(34.00,103.00)
\put(38.0,114.00){\makebox(0,0)[cb]{$x$}}
\put(38.0,104.00){\makebox(0,0)[ct]{$y$}}
\put(65.00,115.00){\makebox(0,0)[lc]{$C_{xy}$}}
\put(9.0,109.00){\makebox(0,0)[rc]{$C_{y x}$}}
\thinlines
\bezier{28}(38.00,111.00)(36.00,109.00)(38.00,108.00)
\bezier{28}(38.00,108.00)(39.00,107.00)(38.00,106.00)
\bezier{28}(37.00,111.00)(35.00,109.00)(37.00,108.00)
\bezier{28}(37.00,108.00)(38.00,107.00)(37.00,106.00)
\put(34.00,109.00){\makebox(0,0)[rc]{$r_{xy}$}}
\put(40.00,109.00){\makebox(0,0)[lc]{$r_{yx}$}}
\put(61.00,115.30){\vector(0,1){1.00}}
\put(12.65,108.20){\vector(0,-1){1.00}}
\end{picture}
\caption[Contours $C_{y x}r_{xy}$ and $C_{xy}r_{yx}$ in the regularized
loop equation]   {
   Contours $C_{y x}r_{xy}$ and $C_{xy}r_{yx}$  which enter the RHS's
   of Eqs.~\rf{AnLe} and \rf{rLe}. }
\label{fi:Fig.2}
\end{figure}
Averaging over the gauge field and using the large-$N_c$ factorization, 
we arrive at the regularized loop-space 
Laplace equation~\cite{HM89}
\be
\Delta W(C) =  \l \oint\limits_C dx_\mu \oint\limits_C dy_\mu
 \!\!\! \int\limits_{r(0)=x \atop r(a^{2})=y} \!\!\! D r(t)
\e^{-\frac 12 \int_0^{a^{2}} d t \,\dot{r}^2(t)}\,
W(C_{yx}r_{xy}) W(C_{xy}r_{yx}) 
\label{rLe}
\ee
which manifestly recovers \eq{Le} when $a\ra 0$.

The constructed regularization is non-perturbative while perturbatively 
reproduces regularized Feynman diagrams. An 
advantage of this regularization of the loop equation is that the contours
$C_{y x}r_{xy}$ and $C_{xy}r_{yx}$ on the RHS of \eq{rLe} both are closed
and do not have marked points if $C$ does not have.
Therefore, \eq{rLe} is written entirely in loop space.

\subsubsection*{Remark on functional Laplacian}

It is worth noting that the representation of the functional
Laplacian on the RHS of \eq{defDelta}, which involves the standard 
variational derivatives, is 
defined for a wider class of functionals than Stokes functionals.
It is easier to deal with the whole operator
$\Delta$, rather than separately with the area and path derivatives.

The functional Laplacian is parametric invariant and
possesses a number of remarkable properties.
While a finite-dimensional Laplacian is an operator of the second order, 
the functional Laplacian is that of the first order and satisfies the 
Leibnitz rule
\be
\Delta \left( U V \right) = 
\Delta \left( U \right) V + U \Delta \left( V \right) \,.
\label{Leibnitz}
\ee

The functional Laplacian can be approximated~\cite{Mak88} in loop space
by a (second-order) partial differential operator in such a way to preserve
these properties in the continuum limit. 
This loop-space Laplacian can be inverted to determine a Green function
$G\left( C,C^\p\right)$ in the form of a sum over surfaces $S_{C,C^\p}$
connecting two loops
which is analogous to the sum-over-path representation of the
Green function of the ordinary Laplacian. The standard perturbation
theory can then be recovered by iterating \eq{Le} 
(or its regularized version~\rf{rLe})
in $\lambda$ with the Green function of the loop-space Laplacian.

\subsection{Survey of non-perturbative solutions}

While the loop equations were proposed long ago, not much is
known about their non-perturbative solutions.
We briefly list some of the results.

It was shown in Ref.~\cite{MM80} that area law
\be
W \left( C\right) \equiv 
\left\langle \Phi \left( C\right) \right\rangle  \propto  \e^{-K\cdot
A_{\min }\left( C\right) } 
\label{a.l.as}
\ee
satisfies the large-$N_c$ loop equation for asymptotically
large $C$. However, a self-consistency equation for $K$,
which should relate it to the bare charge and the cutoff, was not
investigated. In order to do this, one needs more detailed information
about the behavior of $W \left( C\right) $ for intermediate loops.

The {\em free}\/ bosonic Nambu--Goto string which is defined as a sum over
surfaces spanned by $C$
\be
W \left( C\right)= \sum\limits_{S:\partial S=C} \e^{-K\cdot
A\left( S\right) }\,, 
\label{Nambu-Goto}
\ee
with the action being the area $A\left( S\right) $ of the surface $S$,
is {\em not}\/ a solution for intermediate loops. Consequently, 
QCD does not reduce to this kind of string, as was originally
expected in Refs.~\cite{GN79a,Nam79,Pol79,Foe79,Egu79}.
Roughly speaking, the ansatz~\rf{Nambu-Goto} is not consistent with the
factorized structure on the RHS of \eq{MM}.

Nevertheless, it was shown that if a free string satisfies \eq{MM},
then the same interacting string satisfies the loop equations for
finite $N_c$. Here ``free string'' means, as usual in string theory,
that only surfaces of
genus zero are present in the sum over surfaces, while surfaces 
or higher genera are associated with a string interaction.
The coupling constant of this interaction is 
${\cal O}\left(N_c^{-2} \right)$.

A formal solution of \eq{MM} for all loops was found by Migdal~\cite{Mig81}
in the form of a fermionic string
\be
W \left( C\right)   =  \sum\limits_{S:
\partial S=C}\int D\psi \e^{-\int d^2\xi \left[ \bar \psi \sigma _k\partial
_k\psi +\bar \psi \psi m^4\sqrt{g}\right] }, 
\label{elves}
\ee
where the world sheet of the string is parametrized by the coordinates
$\xi_1$ and $\xi_2$ for which the $2$-dimensional metric is
conformal, \ie diagonal. The field $\psi(\xi)$ describes $2$-dimensional
elementary fermions (elves) living in the surface $S$, and $m$
stands for their mass. Elves were introduced to provide factorization
which now holds due to some remarkable properties of $2$-dimensional
fermions. For large loops, the internal fermionic structure becomes
frozen, so that the empty string behavior~\rf{a.l.as} is recovered.
For small loops, the elves are necessary for asymptotic freedom.
However, it is unclear whether or not the string solution~\rf{elves}
is practically useful for a study of multicolor QCD, since the methods
of dealing with the string theory in four dimensions are not
yet developed.

A very interesting solution of the large-$N_c$ loop equation on a lattice
was found by Eguchi and Kawai~\cite{EK82}. They showed that
the SU$(N_c)$ gauge theory on an infinite lattice reduces at $N_c=\infty$
to the model on a hypercube.
The equivalence is possible only at $N_c=\infty$, when the space-time
dependence is absorbed by the internal symmetry group. 
More about this large-$N_c$ reduction will be said in the next Section.

\subsection{Wilson loops in QCD$_2$}

Two-dimensional QCD is popular since the paper by 't~Hooft~\cite{Hoo74b}
as a simplified model of QCD$_4$. 

One can always choose the axial gauge 
$
A_1=0
$,
so that the commutator in the non-Abelian field 
strength~\rf{fieldstrenght} vanishes in two dimensions. 
Therefore, there is no gluon
self-interaction in this gauge and the theory looks,
at the first glance, like the Abelian one.

The Wilson loop average in QCD$_2$ can be straightforwardly
calculated via the expansion~\rf{cPoexpansion} where only
disconnected (free) parts of the correlators $G^{(n)}$ for even $n$
should be left, since there is no interaction.
Only the planar structure of color indices contributes at $N_c=\infty$.
Diagrammatically, the diagrams of the type depicted in
Figure~\ref{fi:Feypl}a and Figure~\ref{fi:Feypl}b are relevant 
for contours without self-intersections,
while that in Figure~\ref{fi:Feypl}c should be omitted in two dimensions.

The color structure of the relevant planar
diagrams can be reduced by using
 the completeness 
condition~\rf{completeness} at large $N_c$. We have
\bea
\lefteqn{W\left( C \right) ~=~ 1
+ \sum\limits_{k}^\infty\; \left(-\lambda\right)^{k} 
\oint\limits_C dx_1^{\mu_1}
\oint\limits_C dx_2^{\nu_1} \cdots 
\oint\limits_C dx_{2k-1}^{\mu_k}
\oint\limits_C dx_{2k}^{\nu_k} } \non & &
\times \theta_{\rm c}(1,2,\ldots,2k)\,
D_{\mu_1\nu_1} \left(x_1-x_2\right)\cdots
D_{\mu_k\nu_k} \left(x_{2k-1}-x_{2k}\right),
\label{cPoexpansion2d}
\eea
where the points $x_1$, $\ldots$, $x_{2k}$ are still
cyclic ordered along the contour. We can exponentiate the
RHS of \eq{cPoexpansion2d} to get finally
\be
W\left( C \right)=\e^{-\frac \lambda 2 \oint_C dx^{\mu}
\oint_C dy^{\nu} D_{\mu\nu} \left(x-y\right)}\,.
\label{W2d}
\ee
This is the same formula as in the Abelian case
if $\lambda$ stands for $e^2$.

The propagator $D_{\mu\nu} \left(x,y\right)$ 
is, strictly speaking, the one in the
gauge $A_1=0$ which reads
\be
D_{\mu\nu} \left(x-y\right)=\frac 12 \delta_{\mu2}\delta_{\nu2}
\, |x_1-y_1|\,\delta^{(1)}\left(x_2-y_2\right) \,.
\label{axialpro}
\ee 
However, the contour integral on the RHS of \eq{W2d} is
gauge invariant, and we can simply choose instead
\be
D_{\mu\nu} \left(x-y\right)=\delta_{\mu\nu}\, 
\frac{1}{4\pi} \ln{\frac{\ell^2}{\left(x-y\right)^2}}\,,
\label{pppggg} 
\ee
where $\ell$ is an arbitrary parameter of the dimension
of length. Nothing depends on it because the contour integral
of a constant vanishes. 

The contour integral in the exponent 
on the RHS of \eq{W2d} can be graphically represented as is
depicted in Figure~\ref{fi:axial1}, 
\begin{figure}[tb]
\vspace*{3mm}
\centering{
\input{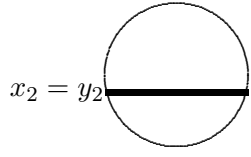}}
\caption[Contour integral in the axial gauge]   
{
   Graphic representation of the contour integral on the LHS of
   \eq{2gcontour} in the axial gauge. The bold line represents
   the gluon propagator~\rf{axialpro} with
   $x_2=y_2$ due to the delta-function.}
   \label{fi:axial1}
   \end{figure}
where $x_2=y_2$ due to the delta-function in \eq{axialpro} and the bold
line represents $|x_1-y_1|$. This  gives
\be
\oint\limits_C dx^{\mu}\oint\limits_C dy^{\nu} D_{\mu\nu} \left(x-y\right)
=A\left( C \right)
\label{2gcontour}
\ee
where $A\left( C \right)$ is the area enclosed by the contour $C$. 
We get finally
\be
W\left( C \right)=\e^{-\frac \lambda 2 A\left( C \right)}
\label{W2dfin}
\ee
for the contours without self-intersections.
 
Therefore, area law holds in two dimensions both
in the non-Abelian and Abelian cases. This is, roughly speaking, because
of the form of the two-dimensional propagator~\rf{pppggg} 
which falls down with the distance only logarithmically in the Feynman
gauge.

The difference between the Abelian and non-Abelian cases shows up
for the contours with self-intersections.

We first note that the simple formula~\rf{2gcontour} does {\em not}\/
hold for contours with arbitrary self-intersections. 

The simplest
contours with one self-intersection are depicted in Figure~\ref{fi:selfint1}.
\begin{figure}[tb]
\vspace*{3mm}
\centering{
\input{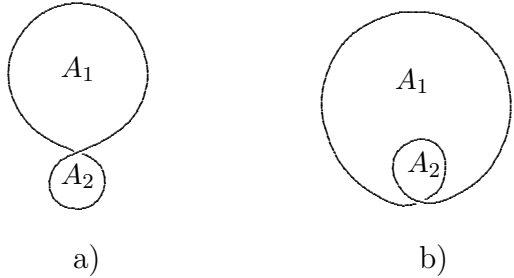}}
\caption[Contours with one self-intersection]   
{
  Contours with one self-intersection:   
  $A_1$ and $A_2$ stand for the areas of the proper windows.
  The total area enclosed by the contour in Figure~a) is $A_1+A_2$.
  The areas enclosed by the exterior and interior loops in Figure~b) 
  are  $A_1+A_2$ and $A_2$, respectively,
  while the total area of the surface with the folding is $A_1+2A_2$.}
   \label{fi:selfint1}
   \end{figure}
There is nothing special about the contour in Figure~\ref{fi:selfint1}a.
Equation~\rf{2gcontour} still holds in this case with $A\left(C\right)$
being the total area $A\left(C\right)=A_1+A_2$. 

The Wilson loop average for the contour in Figure~\ref{fi:selfint1}a
coincides both for the Abelian and non-Abelian cases and equals
\bea
W\left( C \right)~=~\e^{-\frac \lambda 2 \left( A_1+A_2 \right)}\,.
\label{W2dfina}
\eea
This is nothing but the exponential of the total area.

For the contour in Figure~\ref{fi:selfint1}b, we get
\be
\oint\limits_C dx^{\mu}\oint\limits_C dy^{\nu} D_{\mu\nu} \left(x-y\right)
=A_1+4A_2\,.
\label{2gcontour2}
\ee
This is easy to understand in the axial gauge where the ends of the propagator
line can lie both on the exterior and interior loops, or one end at
the exterior loop and the other end on the interior loop.
These cases are illustrated by Figure~\ref{fi:intext}.
\begin{figure}[tb]
\vspace*{3mm}
\centering{
\input{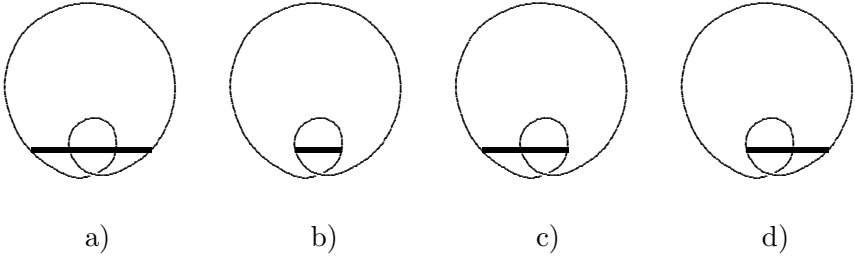}}
\caption[Three type of contribution in \protect{\eq{2gcontour2}}]   
{
 Three type of contribution in \eq{2gcontour2}
The ends of the propagator
line lie both on a) exterior and b) interior loops, or c), d) one end on
the exterior loop and another end on the interior loop.}
   \label{fi:intext}
   \end{figure}
The contributions of the diagrams in Figure~\ref{fi:intext}a,b,c,d are
$A_1+A_2$, $A_2$, $A_2$, and $A_2$, respectively.
The result given by \eq{2gcontour2} is obtained by summing 
over all four diagrams. 

For the contour in Figure~\ref{fi:selfint1}b, the Wilson loop average is 
\be
W\left( C \right)=\e^{-\frac \lambda 2 \left( A_1+4A_2 \right)}
\label{W2dfinba}
\ee
in the Abelian case and
\be
W\left( C \right)=
\left(1-\lambda A_2\right)
\e^{-\frac \lambda 2 \left( A_1+2A_2 \right)}
\label{W2dfinbna}
\ee
in the non-Abelian case at $N_c=\infty$. They coincide only to 
the order $\lambda$ as they should. 
The difference to the next orders is because only the diagrams 
with one propagator line connecting the interior and exterior loops
are planar and, therefore, contribute in the non-Abelian case. 
Otherwise, the diagram is non-planar and vanishes as $N_c\ra\infty$.
Notice, that the exponential of the total area $A\left(C\right)=A_1+2A_2$
of the surface with the folding, which is enclosed by the contour $C$,
appears in the exponent for the non-Abelian case.
The additional pre-exponential factor could be associated with an entropy
of foldings of the surface.

The Wilson loop averages~\rf{W2dfina} and \rf{W2dfinbna} 
in QCD$_2$ at large $N_c$ as well as
the ones for contours with arbitrary self-intersections,
which have a generic form
\be
W\left( C \right)= P
\left( A_1,\ldots, A_n\right)
\e^{-\frac \lambda 2 Area}
\ee 
where $P$ is a polynomial of the areas of individual windows
and $Area$ is the total area of the surface with foldings,
were
first calculated in Ref.~\cite{KK80} by solving the two-dimensional loop
equation and in Ref.~\cite{Bra80b} by applying the non-Abelian
Stokes' theorem. 

\subsubsection*{Remark on the string representation}

A nice property of QCD$_2$ at large $N_c$ is that the 
exponential of the area enclosed by the contour $C$ emerges%
\footnote{This is not true, as is already discussed, 
in the Abelian case for contours with self-intersections.} 
for the Wilson loop average $W\left(C\right)$.
This is as it should for the Nambu--Goto string~\rf{Nambu-Goto}.
However, the additional pre-exponential factors (like that in \eq{W2dfinbna})
are very difficult to interpret in the stringy language.
They may become negative for large loops which is impossible for
a bosonic string. This explicitly demonstrates in $d=2$ the statement
of the previous subsection that the Nambu--Goto string is not a solution
of the large-$N_c$ loop equation.


\section{Large-$N$ Reduction}

The large-$N_c$ reduction was first discovered by Eguchi and
Kawai~\cite{EK82} who showed that the Wilson lattice gauge theory on a
$d$-dimensional hypercubic lattice is equivalent at $N_c=\infty$ to the one
on a hypercube with periodic boundary conditions. This construction is based
on an extra $(Z_{N_c})^d$-symmetry which the reduced theory possesses to each
order of the strong coupling expansion.

Soon after it was recognized that a phase transition occurs in
the reduced model with decreasing the coupling constant, so that
this symmetry is broken in the weak coupling regime. 
To cure the construction at weak coupling, the quenching
prescription was proposed by Bhanot, Heller and Neuberger~\cite{BHN82} and
elaborated by many authors. 
An elegant
alternative reduction procedure based on twisting prescription was advocated
by Gonzalez-Arroyo and Okawa~\cite{GAO83}. Each of these prescriptions
results in the reduced model which is fully equivalent to multicolor
QCD, both on the lattice and in the continuum.

While the reduced models look as a great simplification, since
the space-time is reduced to a point, they still involve an integration
over $d$ infinite matrices which is in fact a continual path integral.
It is not clear at the moment whether or not this is a real simplification
of the original theory which can make it solvable. 
Nevertheless, the reduced models are useful and elegant representations of the
original theory at large $N_c$.

We shall start this Section by a simplest example of 
a pure matrix scalar theory. The quenched reduced model for
this case was proposed by Parisi \cite{Par82} on the lattice
end elaborated by Gross and Kitazawa \cite{GK82} in the continuum, 
while the twisted reduced model was advocated by Eguchi and
Nakayama~\cite{EN83}. Then we concentrate on the  Eguchi--Kawai
reduction of Yang--Mills theory.

\subsection{Reduction of scalar field}\label{ss:rscalar}

Let us begin with a simplest example of 
a pure matrix scalar theory on a lattice 
whose partition function is defined by the path
integral
 \be
Z=\int  \prod_x\prod_{i\geq j} d\varphi^{ij}_x
\e^{\sum_x N_c \tr{\left(-V[\varphi_x]+
\sum_{\mu}\varphi_x\varphi_{x+a\hat\mu}\right)}}\;.
\label{scalar}
\ee
Here $\varphi_x$ is a $N_c\times N_c$ Hermitean matrix field 
with $x$ running over sites of a hypercubic lattice and 
$V[\varphi]$ is some interaction potential, say
\bea
V[\varphi]~=~\frac {M}2 \varphi^2 +\frac {\lambda_3}3 \varphi^3 +
\frac {\lambda_4}4 \varphi^4\,.   
\eea

The prescription of the large-$N_c$ reduction is formulated as follows. We
substitute 
\be 
\varphi_x~\ra~ S_x \Phi S^\dagger_x \,,
\label{substitution}
\ee
where
\be
[S_x]^{kj} = \e^{ ip_k^\mu x_\mu} \delta^{kj} \non
= \hbox{diag}\left(\e^{ ip_1^\mu x_\mu},\ldots,
\e^{ip_{N_c}^\mu x_\mu} \right)
\label{S}
\ee
is a diagonal unitary matrix which eats the coordinate dependence,
so that $\Phi$ does {\em not}\/ depend on $x$.

The averaging of a functional $F[\varphi_x]$ which is defined
with the same weight as in \eq{scalar},
\be
\Big\langle F[\varphi_x]  \Big\rangle  
 \equiv \frac 1Z \int  \prod_x d\varphi_x
\e^{\sum_x N_c \tr{\left(-V[\varphi_x]+
\sum_{\mu}\varphi(x)\varphi(x+a\hat\mu)\right)}}F[\varphi_x]  \,,
\ee 
can be calculated at $N_c=\infty$ by
\be
\Big\langle F[\varphi_x]  \Big\rangle  \ra a^{N_c d}
\int_{-\frac\pi{a}}^\frac\pi{a} \prod_{\mu=1}^d \prod_{i=1}^{N_c}  
\frac{dp_i^\mu}{2\pi}
\Big\langle F[S_x\Phi S^\dagger_x]  \Big\rangle_{\rm Reduced}
\label{averages}
\ee
where the average on the RHS is calculated~\cite{Par82} 
for the {\it quenched reduced model\/} whose averages are defined by 
\bea
\lefteqn{
\Big\langle F[\Phi]  \Big\rangle _{\rm Reduced}
 \equiv \frac 1{Z_{\rm Reduced}} } \non & & \times 
\int \prod_{i\geq j}d\Phi_{ij} \e^{-N_c\tr{V[\Phi]}+
N_c\sum_{ij} \left|\Phi_{ij}\right|^2
\sum_\mu\cos{\left((p_i^\mu-p_j^\mu)a\right)} } 
F[\Phi]  \,.~~~~~~
\eea 
The partition function of the reduced model reads
\be
Z_{\rm Reduced}=\int \prod_{i\geq j}d\Phi_{ij} \e^{-N_c\tr{V[\Phi]}+
N_c\sum_{ij} \left|\Phi_{ij}\right|^2
\sum_\mu\cos{\left((p_i^\mu-p_j^\mu)a\right)} }
\label{QRM}
\ee
which can be deduced, modulo the volume factor, from the partition function 
\rf{scalar} by the substitution \rf{substitution}.

Notice that the integration over the momenta $p^\mu_i$ 
on the RHS of \eq{averages} is taken
{\em after}\/ the calculation of averages in the reduced model.
Such variables are usually called {\em quenched}\/ in statistical mechanics
which clarifies the terminology.  

Since $N_c\ra\infty$ it is not necessary to integrate over the quenched
 momenta in \eq{averages}. The integral should be recovered if
$p_i^\mu$'s would be uniformly distributed in a $d$-dimensional hypercube.
Moreover, a similar property holds for the matrix integral over $\Phi$ as
well, which can be substituted by its value at the saddle point configuration
$\Phi_s$:  
\bea 
\Big\langle F[\varphi_x]  \Big\rangle ~ \ra~ F[S_x\Phi_s
S^\dagger_x] \;,
\label{masterfield} 
\eea 
where the momenta $p^\mu_i$ are uniformly distributed in the
hypercube. 
Therefore, this saddle point configuration
plays the role of a master field in the sense of 
Subsection~\ref{ss:m.f.}.

In order to show how \eq{averages} works, let us demonstrate how
the planar diagrams of perturbation theory for the scalar matrix
theory~\rf{scalar} are recovered in the quenched reduced model.

The quenched reduced model~\rf{QRM} is of the general type
discussed in Section~\ref{s:mqcd}. 
The propagator is given by
\bea
\Big\langle \Phi_{ij} \Phi_{kl}  \Big\rangle _{\rm Gauss}
~=~\frac 1{N_c} G\left( p_i-p_j \right)\delta_{il} \delta_{kj}
\eea
with
\bea
G\left( p_i-p_j \right)~=~
\frac{1}{M-\sum_\mu\cos{\left((p_i^\mu-p_j^\mu)a\right)} }\,.
\label{Gpij}
\eea

It is convenient to associate the momenta $p_i$ and $p_j$ in \eq{Gpij}
with each of the two index lines representing the propagator and
carrying, respectively, indices $i$ and $j$. Remember, that these lines
are oriented for a Hermitean matrix $\Phi$ and their orientation
can be naturally associated with the direction of the flow of the momentum.
The total momentum carried by the double line is $p_i-p_j$.

The simplest diagram which represents the correction of the 
second order in $\lambda_3$ to the propagator is depicted 
Figure~\ref{fi:quench3}.
\begin{figure}[tb]
\vspace*{3mm}
\centering{
\input{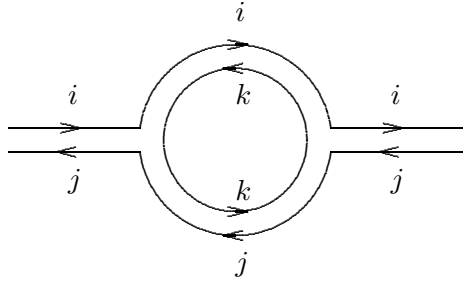}}
\caption[Simplest diagram for the quenched reduced model]   
{
   Simplest planar diagram of the second order in $\lambda_3$ for 
   the propagator in the quenched reduced model~\rf{QRM}.
   The momentum $p_i$ flows along the index line $i$. The momentum
   $p_i-p_j$ is associated with the double line $ij$.
   }
   \label{fi:quench3}
   \end{figure}
The momenta $p_i$ and $p_j$ flows along the index lines $i$ and $j$
while the momentum $p_k$ circulates along the index line $k$.
The contribution of the diagram in Figure~\ref{fi:quench3} reads
\bea
\frac{\lambda_3^2}{N_c^2} G\left( p_i-p_j \right)^2 
\sum_k G\left( p_i-p_k \right) G\left( p_k-p_j \right),
\label{Gpijk}
\eea
where the summation over the index $k$ is just a standard one
over indices forming a closed loop. 
   
In order to show that the quenched-model result~\rf{Gpijk} reproduces
the correction to the propagator in the original theory on an
infinite lattice, we pass to the variables of the total momenta
flowing along the double lines:
\bea
p_i-p_j=p;~~~p_k-p_j=q;~~~p_i-p_k=p-q\,,
\label{assignement}
\eea
which is obviously consistent with the momentum conservation
at each of the two vertices of the diagram in Figure~\ref{fi:quench3}.
Since $p_k$'s are uniformly distributed in the
hypercube, the summation over $k$ can be substituted as
$N_c\ra\infty$ by the integral
\bea
\frac{1}{N_c} \sum_k f\left( p_k\right) \Rightarrow a^d
\int_{-\frac \pi a}^{\frac \pi a} \frac{d^d q}{(2\pi)^d }
f\left( q\right). 
\label{sumint}
\eea 
The prescription~\rf{averages} then gives the correct
expression 
\bea
 a^d \frac{\lambda_3^2}{N_c} G\left( p \right)^2  
\int_{-\frac \pi a}^{\frac \pi a} \frac{d^d q}{(2\pi)^d }
 G\left( q \right) G\left( p-q \right)
\eea
for the second-order contribution of the 
perturbation theory for the propagator on the lattice.

It is now clear how a generic planar diagram is recovered
by the reduced model. We first represent the diagram by
the double lines and associate the momentum $p^\mu_i$ with
an index line carrying the index $i$. Then we write down
the expression for the diagram in the reduced model with the
propagator~\rf{Gpij}. Passing to the momenta
flowing along the double lines, similar to \eq{assignement},
we get an expression which coincides with the integrand of
the Feynman diagram for the theory on the whole lattice.
It is crucial that such a change of variables can always
be done for a planar diagram consistently with the momentum
conservation at each vertex. The last step is that the
summation over indices of closed index lines reproduces
the integration over momenta associated with each of the loops
according to \eq{sumint}. It is assumed that the number of
loops is much less than $N_c$ which is always true for
a given diagram since $N_c$ is infinite. 

We thus have shown how planar diagrams of the lattice theory defined
by the partition function~\rf{scalar} are recovered by the 
reduced model~\rf{QRM}. The lattice was needed only 
as a regularization to make
all integrals well-defined and was not crucial in the consideration.
This construction can be
formulated directly for the continuum theory~\cite{GK82,DW82}
where the propagator turns into
\bea
G\left( p_i-p_j \right)~=~
\frac{1}{\left(p_i-p_j\right)^2 +m^2 }
\label{Gpijcont}
\eea
and a Lorentz-invariant regularization can be achieved by choosing
$p^2 < \Lambda^2$.

\subsubsection*{Remark on the twisted reduced model}

An alternative reduction procedure is based on the twisting prescription
\cite{GAO83}.
We again perform the unitary transformation \rf{substitution} with the
matrices $S_x$  being expressed via a set of $d$ (unitary)
$N_c\times N_c$ matrices $\Gamma_\mu$ by 
\be
S_x=\Gamma_1^{x_1/a}\Gamma_2^{x_2/a}\Gamma_3^{x_3/a}\Gamma_4^{x_4/a}
\label{Stwi}
\ee
where the coordinates of the (lattice) vector $x_\mu$ are measured in the
lattice units.
The matrices
$\Gamma_\mu$ are explicitly constructed in Ref.~\cite{GAO83} and commute by
\be
\Gamma_\mu \Gamma_\nu = Z_{\mu\nu} \Gamma_\nu \Gamma_\mu
\label{commutator}
\ee
with $Z_{\mu\nu}=Z_{\nu\mu}^\dagger$ being elements of $Z_{N_c}$.

For the twisting reduction prescription, \eq{averages} is valid providing
the average on the RHS is calculated for the {\it twisted reduced
model\/} which is defined by the partition function~\cite{EN83}
\be
Z_{\rm TRM}=\int d\Phi \e^{-N_c\tr{V[\Phi]}+
N_c\sum_\mu \tr{\Gamma_\mu \Phi \Gamma_\mu^\dagger \Phi}}\;.
\label{TRM}
\ee

We can change the order of $\Gamma$'s in \eq{Stwi} defining 
a more general path-dependent factor
\be
S_x= {\bf P}\prod_{l\in C_{x\infty}} \Gamma_\mu\; .
\ee
The path-ordered product in this formula runs over all links $l=(z,\mu)$
forming a path $C_{x\infty}$ from infinity to the point $x$. 

Due to \eq{commutator}, changing the form of the path multiplies $S_x$ by
the Abelian factor
\be
Z(C)=\prod_{\Box\in S:\partial S=C} Z_{\mu\nu}(\Box)
\label{factor}
\ee
where $(\mu,\nu)$ is the orientation of the plaquette $\Box$.
The product runs over any surface spanned by the closed loop $C$ which is
obtained by passing the original path forward and the new path backward.
Due to the Bianchi identity
\be
\prod_{\Box\in cube} Z_{\mu\nu}(\Box)=1
\ee
where the product goes over six plaquettes forming a 3-dimensional cube on
the lattice, the product on the RHS of \eq{factor} does not depend on
the form of the surface $S$ and is a functional of the loop $C$.

It is now easy to see that under this change of the path we get
\be
[S_x]_{ij}[S^\dagger_x]_{kl}\ra
\left|Z(C)\right|^2[S_x]_{ij}[S^\dagger_x]_{kl}
\ee
and the path-dependence is canceled because $\left|Z(C)\right|^2=1$.
This is a general property which holds for the twisting reduction
prescription of any even (\ie invariant under the center $Z_{N_c}$)
representation of $SU(N_c)$.

\subsection{Reduction of Yang--Mills field}\label{ss:rYM}

The statement of the Eguchi--Kawai reduction of the Yang--Mills field
says that the theory on a $d$-dimensional space-time 
is equivalent at $N_c=\infty$ to the reduced model which is 
nothing but its reduction to a point. 
The action of the reduced model is given by
\be
S_{\rm EK} = \frac{1}{2g^2 \Lambda^d} \tr [A_\mu,A_\nu]^2\,,
\label{defSEK}
\ee
where $A_\mu$ are $d$ space-independent matrices
and $\Lambda$ is a dimensionful parameter.

A naive statement of the Eguchi--Kawai reduction
is that the averages coincide in both
theories, for example,
\be
\left\langle  \frac{1}{N_c} \tr {{\bf P}} \e^{i\oint d\xi^\mu A_\mu(\xi)}
\right\rangle_{\rm d-dim}=
\left\langle  \frac{1}{N_c} \tr {{\bf P}} \e^{i\oint d\xi^\mu A_\mu}
\right\rangle_{\rm EK}
\label{forexample}
\ee
where the LHS is calculated with the action~\rf{QCDaction} and
the RHS is calculated with the reduced action~\rf{defSEK}.
Strictly speaking, this naive statement is valid only in
$d=2$ or supersymmetric case for the reason which will be 
explained in a moment.

The precise equivalence is valid only if the average of open 
Wilson loops vanish in the reduced model:
\be
\label{openloop}
\left\langle  \frac{1}{N_c} \tr {{\bf P}} \e^{i\int_{C_{yx}} d\xi^\mu A_\mu}
\right\rangle_{\rm EK} =0 \,,
\ee
as it does in the $d$-dimensional theory due to the local
gauge invariance under which
\be
\left({\bf P} \e^{i\int_{C_{yx}} d\xi^\mu A_\mu(\xi)}\right)_{ij}
\rightarrow 
\left(\Omega^\dagger(y){\bf P} 
\e^{i\int_{C_{yx}} d\xi^\mu A_\mu(\xi)}\Omega (x)\right)_{ij}.
\ee
The point is that this gauge invariance transforms 
in the reduced model into (global)
rotation of the reduced field by constant matrices $\Omega$:
\be
A_\mu\rightarrow \Omega^\dagger A_\mu\Omega \,.
\label{rgauge}
\ee
which does {\em not}\/ guarantee 
such vanishing in the reduced model.

There exists, however, a symmetry of the reduced action~\rf{defSEK} under
the shift of $A_\mu$ by a unit matrix%
\footnote{This symmetry is rigorously defined on a lattice
where it is associated with a direction-dependent Z${}_{N_c}$ transformation.}:
\be
A_\mu^{ij}\rightarrow A_\mu^{ij}+ a_\mu \delta^{ij}\,,
\label{Rd}
\ee 
which is often called the R${}^d$-symmetry.
Under the transformation~\rf{Rd}, we get
\be
\left({\bf P} \e^{i\int_{C_{yx}} d\xi^\mu A_\mu}\right)_{ij}
\rightarrow 
\e^{i(y^\mu-x^\mu)a_\mu}
\left({\bf P} \e^{i\int_{C_{yx}} d\xi^\mu A_\mu}\right)_{ij}
\ee
which guarantees, if the symmetry is not broken, the vanishing
of the open Wilson loops
\be
W_{\rm EK}\left( C_{yx}\right) \equiv
\left\langle  \frac{1}{N_c} \tr {{\bf P}} \e^{i\int_{C_{yx}} d\xi^\mu A_\mu}
\right\rangle_{\rm EK}=0
\label{vanishing}
\ee
in the reduced model.

The equivalence of the two theories can then be shown
using the loop equation which reads for the reduced model
\begin{eqnarray}
\partial _\mu ^x\frac \delta {\delta \sigma _{\mu \nu }\left( x\right)
}W_{\rm EK}\left( C\right)&=&
\left\langle  \frac{1}{N_c} \tr {{\bf P}} [A_\mu,[A_\mu,A_\nu]]
\e^{i\oint_{C_{xx}} d\xi^\mu A_\mu}
\right\rangle_{\rm EK} \non & = &
\l \Lambda^d 
\left\langle  \frac{1}{N_c} \tr {{\bf P}} \frac{\partial}{\partial A_\nu}
\e^{i\oint_{C_{xx}} d\xi^\mu A_\mu}
\right\rangle_{\rm EK}
\non & = &\l \Lambda^d\oint\limits_C dy_\nu \,
W_{\rm EK}\left( C_{yx}\right) W_{\rm EK}\left( C_{xy}\right) \,.
\label{MMreduced}
\end{eqnarray}
The RHS is pretty much similar to the one in \eq{MM} while
$\delta^{(d)}(x-y)$ is missing.

This delta function can be recovered if the R${}^d$ symmetry is not
broken since
\be
W_{\rm EK}\left( C_{yx}\right) \sim 
\frac{\delta^{(d)}(x-y)}{\delta^{(d)}(0)}W_{\rm EK}\left( C_{yx}\right) 
\label{EKdeltas}
\ee
due to \eq{vanishing} for the open loops.

This is not a rigorous argument since a regularization is needed.
What actually happens is the following.
If we smear the delta function introducing
\be
\delta^{(d)}_\Lambda(x) = 
\left(\frac{\Lambda}{\sqrt{2\pi}}\right)^d \e^{-x^2\Lambda^2/2} \,,
\ee
then
\be
\frac{1}{\delta^{(d)}_\Lambda(0)}
\left(\delta^{(d)}_\Lambda(0)\right)^2 \propto 
\Lambda^d \e^{-x^2\Lambda^2} \rightarrow \delta^{(d)}(x)\,,
\ee
reproducing the delta function.

\subsection{R$^d$-symmetry in perturbation theory}

Since $N_c$ is infinite, the R$^d$-symmetry can
be broken spontaneously. The point is that the large-$N_c$
limit plays the role of a statistical averaging, as is mentioned
already in Subsection~\ref{ss:semic}, and phase transitions
are possible for infinite number of degrees of freedom.
This phenomenon occurs in perturbation theory of 
the reduced model for $d\geq3$.

The perturbation theory can be constructed expanding the fields
around solutions of the classical equation
\be
[A_\mu,[A_\mu,A_\nu]]=0 \,.
\label{rcleq}
\ee
Any diagonal matrix 
\be
A_\mu^{\rm cl}\equiv p_\mu = \hbox{diag} \left\{  
p_\mu^{(1)}, \ldots , p_\mu^{(N_c)}\right\}
\label{rAcl}
\ee
is a solution to \eq{rcleq}.

The perturbation theory of the reduced model can be constructed
expanding around the classical solution~\rf{rAcl}:
\be
A_\mu = A_\mu^{\rm cl}+ g A_\mu^{\rm q}\,,
\label{rAq}
\ee
where $A_\mu^{\rm q}$ is off-diagonal. 

Substituting~\rf{rAq} into the action~\rf{defSEK}, we get
\be
S_{\rm EK} = \tr \left\{ \frac 12 [p_\mu, A_\nu^{\rm q} ]^2
- \frac 12 [p_\mu, A_\mu^{\rm q} ]^2   \right\}
+ \hbox{higher orders}~.
\label{Sexpa}
\ee
To fix the gauge symmetry~\rf{rgauge},
it is convenient to add
\be
S_{\rm g.f.} = \tr \left\{ 
 \frac 12 [p_\mu, A_\mu^{\rm q} ]^2 +[p_\mu,b] [p_\mu,c] \right\} \,,
\label{Sgf}
\ee
where $b$ and $c$ are ghosts.

The sum of~\rf{Sexpa} and \rf{Sgf} gives
\be
S_2=\tr \left\{ \frac 12 [p_\mu, A_\nu^{\rm q} ]^2
+[p_\mu,b] [p_\mu,c]  \right\}
\ee
up to quadratic order in $A_\mu^{\rm q}$.

Doing the Gaussian integral over $A_\nu^{\rm q}$, we get at the one-loop
level:
\be
\int dp_\mu d A_\mu^{\rm q} \e^{-S_2} \ldots =
\int \prod_{k=1}^N dp_\mu^{(k)}
\prod_{i<j} \left[ (p_\mu^{(i)}-p_\mu^{(j)})^2  \right]^{1-d/2} \ldots\,,
\label{Vandermonde} 
\ee
where the integration over $p_\mu$ accounts for equivalent
classical solutions. 

For $d=1$ the product on the RHS of \eq{Vandermonde}  reproduces
the Vandermonde determinant. For $d=2$ it vanishes and does not
affect dynamics. For $d\geq3$ the measure is singular and the 
eigenvalues collapse. This leads us to a
spontaneous breakdown of the R$^d$ in perturbation theory.

The equivalence between the $N_c=\infty$ Yang--Mills theory on a 
whole space and the reduced model can be provided~\cite{BHN82}
introducing a quenching prescription similar
to the one described in Subsection~\ref{ss:rscalar}.
Then no collapse of eigenvalues happens and $d$-dimensional
planar graphs are reproduced by the reduced model.
More about the quenching prescription
in Yang--Mills theory can be found in the 
reviews~\cite{Mig83,Das87} and cited there original papers.

\subsubsection*{Remark on supersymmetric case}

In a supersymmetric gauge theory, there is an extra contribution 
from fermions to the exponent on the RHS of \eq{Vandermonde}.
Since the integration over fermions results in the 
extra factor $[ (p_\mu^{(i)}-p_\mu^{(j)})^2 ]^{\tr I/2}$,
this yields finally the exponent
$
1 - d/2 + {\tr I}/2 
$.
It vanishes in $d=4$ for either Majorana or Weyl fermions and
in $d=10$ for the Majorana--Weyl fermions. Therefore, 
the R$^d$-symmetry is not broken and no
quenching is needed in the supersymmetric case~\cite{MK83,IKKT97}.

\subsection{Twisted reduced model}

The continuum version of the twisted reduced model
can be constructed~\cite{GAKA83} by substituting 
$A_\mu \ra A_\mu -\gamma_\mu$ 
into the action~\rf{defSEK}, where the matrices $\gamma_\mu$
obey the commutation relation
\be
\left[\gamma_\mu, \gamma_\nu \right] =  B_{\mu\nu} I \,,
\label{icommutator}
\ee
where $ B_{\mu\nu}$ is an antisymmetric tensor and $d$ is even.
This is possible only for infinite Hermitean matrices (operators).
An example of such matrices is  $x$ and $p$ operators
in quantum mechanics.
\eq{icommutator} is a continuum version of \eq{commutator}.

The Wilson loop averages in the twisted reduced model are defined by
\be
W_{\rm TEK}\left( C_{yx}\right) =
\left\langle \frac{1}{N_c} 
\tr {{\bf P}} \e^{-i\int_{C_{yx}} d\xi^\mu \gamma_\mu} 
\frac{1}{N_c} \tr {{\bf P}} \e^{i\int_{C_{yx}} d\xi^\mu A_\mu}
\right\rangle_{\rm TEK}\,.
\label{WTEK}
\ee
They vanish for open loops which
is provided by the vanishing of the trace of
the path-ordered exponential of $\gamma_\mu$
in this definition.
For closed loops this factor does not vanish and 
is needed to provide 
the equivalence with $d$-dimensional Yang--Mills perturbation theory,
since the classical extrema of the twisted reduced model
are $A_\mu^{\rm cl}=\gamma_\mu$ and the perturbation theory
is constructed expanding around this classical solution.

The proof of the equivalence can be done using the loop equation
quite similarly to that of Subsection~\ref{ss:rYM} for the
Eguchi--Kawai model with an unbroken R$^d$ symmetry.


\acknowledgements 

I would like to thank the organizers for the warm hospitality
in Iceland.
This work is supported in part by the grant  RFFI                        
97--02--17927.

\end{document}